\documentclass[11pt,a4paper]{article}
\usepackage{jheppub}
\usepackage{amsmath}
\usepackage{flexisym}
\usepackage{breqn}
\usepackage{mathtools}
\usepackage{amssymb,bm,bbold}
\usepackage{enumerate}
\usepackage{comment}
\usepackage{esvect}
\usepackage{physics}
\usepackage{caption}
\usepackage{slashed}
\usepackage{import}
\usepackage[toc,page]{appendix}
\usepackage{subcaption}
\usepackage{shuffle}
\usepackage{flexisym}
\usepackage{breqn}
\usepackage{fancyhdr}
\usepackage{tikz}
	 \usetikzlibrary{arrows,decorations.markings}
	 \usetikzlibrary{calc}
	 \usetikzlibrary{3d}
    \usetikzlibrary{arrows.meta}
\usepackage{tikz-feynman}
\tikzfeynmanset{compat=1.1.0}
\tikzfeynmanset{warn luatex=false}
\usepackage{nicematrix}
\usepackage{pgfplots}
\pgfplotsset{compat=newest}
\usepgfplotslibrary{fillbetween}
\newif\ifshowcomments
\showcommentstrue   % change to \showcommentstrue to show comments

\ifshowcomments
  \newcommand{\yorgos}[1]{{\color{red}[\textbf{G.P:} #1]}}
  \newcommand{\rigers}[1]{{\color{orange}[\textbf{Rigers:} #1]}}
  \newcommand{\gab}[1]{{\color{blue}[\textbf{GD:} #1]}}
\else
  \newcommand{\yorgos}[1]{}
  \newcommand{\rigers}[1]{}
  \newcommand{\gab}[1]{}
\fi

\def\be{\begin{equation}}
\def\ee{\end{equation}}

\newcommand{\dlog}{d\text{log}}

\newcommand{\br}[1]{\langle #1 \rangle }

\def\cN{{\cal N}}

\newcommand{\cY}{\mathcal{Y}}

\def\centerarc[#1](#2)(#3:#4:#5){ \draw[#1] ($(#2)+({#5*cos(#3)},{#5*sin(#3)})$) arc (#3:#4:#5); }
\title{Novel cluster-algebraic letters for 5- and 6-point QCD processes}
\author[a]{Rigers Aliaj,}
\author[b]{Gabriele Dian}
\author[c,b,d]{and Georgios Papathanasiou}
\affiliation[a]{ II. Institut f{\"u}r Theoretische Physik, Universit{\"a}t Hamburg, Luruper Chaussee 149, 22607 Hamburg,
Germany}
\affiliation[b]{Deutsches Elektronen-Synchrotron DESY, Notkestr. 85, 22607 Hamburg, Germany}
\affiliation[c]{Department of Nuclear and Particle Physics,
Faculty of Physics, National and Kapodistrian University of Athens, Athens 15784, Greece}
\affiliation[d]{Department of Mathematics, City, University of London,
Northampton Square, EC1V 0HB, London, UK}
\emailAdd{rigers.aliaj@desy.de}
\emailAdd{gabriele.dian@desy.de}
\emailAdd{g.papathanasiou@phys.uoa.gr}
\abstract{
By breaking dual conformal invariance, we transform cluster-algebraic predictions for the alphabet of 9-point amplitudes in $\mathcal{N}=4$ super Yang-Mills theory to analogous predictions for 5- and 6-point processes in 
QCD. We start by obtaining, for the first time, candidate letters for 6-point processes with one massive external leg, and discover that they surprisingly also contain nested square roots. We confirm that our results essentially contain the alphabet of all 1-loop integrals with these kinematics, and in their massless limit also the recently computed alphabet of finite, planar 2-loop amplitudes for 6-point massless QCD processes. In the latter case, we additionally find 162 letters that may appear at higher loops. We similarly produce candidate letters for 5-point 2-mass processes, whose comparison with the literature reveals a nontrivial overlap that also includes new letters.

}

\begin{document}
\maketitle

\section{Introduction}\label{sec:Intro}
Maximally supersymmetric or $\cN=4$ super Yang-Mills (SYM) theory~\cite{Brink:1976bc,Gliozzi:1976qd} has proven to be an ideal theoretical laboratory for developing novel paradigms and computational methods, that ultimately find far-reaching applications in Quantum Chromodynamics (QCD) and in precision phenomenology for collider experiments. Celebrated examples of this highly successful strategy include \emph{generalised unitarity}~\cite{Bern:1994zx}, that trivialised the determination of 1-loop scattering amplitudes in generic gauge theories; and the \emph{canonical form}~\cite{Henn:2013pwa} of differential equations~\cite{Kotikov:1990kg,Remiddi:1997ny,Gehrmann:1999as} governing any basis~\cite{Chetyrkin:1981qh} of multi-loop Feynman integrals, that tremendously simplifies their evaluation. 

To be concrete, for the bases of integrals $\mathbf{f}$ evaluating to the often sufficient class of \emph{multiple polylogarithms}, which will be the focus of this paper, their canonical differential equations take the form
\begin{equation}\label{eq:CDE_intro}
    d\mathbf{f}(\vec{z};\epsilon)=\epsilon \left[\sum_{i}\mathbf{A}_{i} \,\dlog\alpha_i(\vec{z})\right]\mathbf{f}(\vec{z};\epsilon)\,,
\end{equation}
where $\vec{z}$ denotes all the kinematic variables the integrals depend on  (such as external momenta and internal masses), $\epsilon=(4-D)/2$ is  the dimensional regulator, $d=\sum_{j} \dd z_{j} \partial_j$ is the total differential, and $\mathbf{A}_i$ are constant matrices. Last but not least, each $\alpha_i$ is an algebraic function of the $\vec{z}$ components known as a \emph{letter} (of the symbol~\cite{Goncharov:2010jf}), with their entire set similarly denoted as the (symbol) \emph{alphabet}. 

Due to their symbolic, functional form, letters are arguably the most complicated elements of the differential equations, and obstruct a more efficient, fully numeric construction of a canonical basis. Yet mounting evidence suggests that this obstruction can be lifted once again with knowledge imported from $\cN=4$ SYM theory: This pertains to \emph{cluster algebras}~\cite{1021.16017}, mathematical objects which have been observed to describe the alphabet of six- and seven-particle amplitudes in the planar limit of the former theory~\cite{Golden:2013xva}, also enabling powerful bootstrap methods for their calculation as reviewed in~\cite{Travaglini:2022uwo}; closely related generalisations apply to higher-point amplitudes of the same theory as well~\cite{Drummond:2019cxm,Arkani-Hamed:2019rds,Henke:2019hve,Herderschee:2021dez,Henke:2021ity,Ren:2021ztg}. Most importantly, cluster algebras have also been discovered to underlie the analytic structure of a host of generic Feynman integrals as well as QCD processes they contribute to, most notably two-~\cite{Chicherin:2020umh} and three-loop~\cite{Aliaj:2024zgp} contributions to Higgs+jet amplitudes in the heavy-top limit. More examples of integrals where the presence of cluster algebras or their generalisations has been confirmed include~\cite{He:2021esx,He:2021fwf,He:2021non,He:2021mme,He:2021zuv,He:2021eec,He:2022ctv,He:2022tph,Bossinger:2022eiy,2638233,Zhao:2023okw,Pokraka:2025ali,Bossinger:2025rhf}, see also~\cite{Bossinger:2024apm} for related work.

Of course, the real test for every new method is not its ability to explain or postdict existing calculations, but to make new predictions and to aid future calculations. It is fair to say that all references of the previous paragraph have essentially done the former, perhaps with the exception of revealing \emph{adjacency relations}~\cite{Drummond:2017ssj} of the form $\mathbf{A}_{i}\cdot\mathbf{A}_{j}=0$ for certain matrices in eq.~\eqref{eq:CDE_intro}~\cite{Gehrmann:2024tds}, or certain absences of letters, see~\cite{Abreu:2021asb,Badger:2022ncb} and references therein, in QCD amplitudes. 

In this work, we take a bold step in this direction: We obtain, for the first time, cluster-algebraic candidate letters for planar six-point Feynman integrals with one external offshell or massive leg, and all remaining external and internal particles massless. These are relevant, for example, for the calculation of QCD corrections to the production of a massive vector boson in association with three jets/photons at the Large Hadron Collider. As we discuss in the following paragraphs, we arrive at these predictions starting from the proposed alphabet of nine-particle amplitudes in $\cN=4$ SYM theory~\cite{Henke:2021ity}, see also~\cite{Ren:2021ztg} for related partial results. A striking feature is the appearance of nested square root letters~\cite{FebresCordero:2023pww, Badger:2024fgb,Becchetti:2025oyb}, which so far had only been observed in integrals with massive internal propagators that integrate to more complicated function spaces. We also perform various tests of our obtained six-point one-mass letters, including the limit where the massive leg becomes massless. The relevant alphabet for planar two-loop six-point amplitudes in massless QCD has been obtained in the recent publications~\cite{Abreu:2024fei,Henn:2025xrc}, and remarkably we find that this is essentially contained in our limit after cyclic symmetrisation, along with additional letters that might appear at higher loops. As six-point one-mass master integrals also contain five-point two-mass integrals as a subsector, we similarly obtain candidate letters for the latter, as well as compare with the direct computation of~\cite{Abreu:2024yit}.

We provide a more detailed summary of our main findings in subsection~\ref{subsec:results_summary}, but before that let us also outline the method for obtaining them.

\subsection{Outline of the method}\label{subsec:MethodOutline}
\begin{figure}
\begin{center}
\begin{tikzpicture}[scale=0.7,baseline]
    % Define the number of sides and the radius
    \def\n{4}
    \def\radius{1.5}

    % Draw the nonagon
    \foreach \i in {1,...,\n} {
        \coordinate (P\i) at ({(\i-1/2)*360/\n}:\radius);
    }
    \draw[thick] (P1)-- (P2) node [midway,above] {$x_1$} -- (P3) node [midway,left] {$x_7$} -- (P4) node [midway,below] {$x_5$} -- cycle node [midway,right] {$x_3$} ;
  \foreach \i in {1,...,\n} {
        \coordinate (Q\i) at ({\i*360/\n}:2);
    }
\draw[thick] (P1)--({(1/2-0.1)*360/4}:2);
\draw[thick] (P1)--({(1/2+0.1)*360/4}:2);
\draw[thick] (P2)--({(3/2-0.1)*360/4}:2);
\draw[thick] (P2)--({(3/2+0.1)*360/4}:2);
\draw[thick] (P3)--({(5/2-0.1)*360/4}:2);
\draw[thick] (P3)--({(5/2+0.1)*360/4}:2);
\draw[thick] (P4)--({(7/2-0.1)*360/4}:2);
\draw[thick] (P4)--({(7/2+0.1)*360/4}:2);

\node (0,0) {DCI};
\node at (0,-2.5) {$\Phi(u,v)$};
\end{tikzpicture}$\quad \overset{x_7\to \infty}{\xrightarrow{\hspace*{40pt}}}\quad $\begin{tikzpicture}[scale=0.7,baseline]
    % Define the number of sides and the radius
    \def\n{4}
    \def\radius{1.5}

    % Draw the nonagon
    \foreach \i in {1,...,\n} {
        \coordinate (P\i) at ({(\i-1/2)*360/\n}:\radius);
    }
    \draw[thick] (P1)-- (-\radius,0) node [midway,above] {$x_1$} -- (P4) node [midway,below] {$x_5$} -- cycle node [midway,right] {$x_3$} ;
  \foreach \i in {1,...,\n} {
        \coordinate (Q\i) at ({\i*360/\n}:2);
    }
\draw[thick] (P1)--({(1/2-0.1)*360/4}:2);
\draw[thick] (P1)--({(1/2+0.1)*360/4}:2);
\draw[thick] (P4)--({(7/2-0.1)*360/4}:2);
\draw[thick] (P4)--({(7/2+0.1)*360/4}:2);

\draw[thick] (-\radius,0)--({(2+0.2)*360/4}:2);
\draw[thick] (-\radius,0)--({(2-0.07)*360/4}:2);
\draw[thick] (-\radius,0)--({(2+0.07)*360/4}:2);
\draw[thick] (-\radius,0)--({(2-0.2)*360/4}:2);

\node (0,0) {LI};
\node at (0,-2.5) {$\Phi(\hat{u},\hat{v})$};
\end{tikzpicture}
\end{center}
\caption{The dual conformal 4-mass box integral reduces to the Lorentz-invariant 3-mass triangle integral in the reference frame where $x_7\to \infty$. In the rest of the paper, we will use similar figures to represent the (precise sequence of massive and massless legs in the) external kinematics, irrespective of the loop order.}\label{fig:4to3pt}
\end{figure}
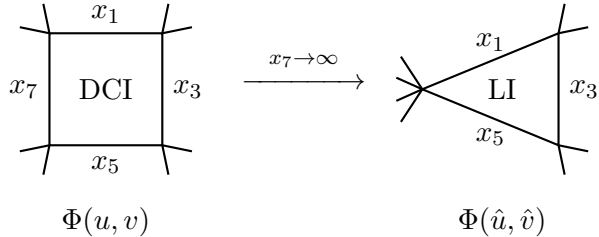

We will rely on a central property of appropriately normalised, planar $\cN=4$ SYM amplitudes as well as of certain Feynman integrals, \emph{dual conformal invariance} (DCI)~\cite{BROADHURST1993132,Drummond:2006rz,Alday:2007hr,Drummond:2007aua,Drummond:2007au}. Dual conformal  invariance implies that the quantities in question do not depend on the full set of momenta $p_i$ of the $n$ external particles, but rather on conformal cross ratios
\begin{equation}
u_{ijkl}:=\frac{x^2_{ij}x^2_{kl}}{x^2_{il}x^2_{kj}}\,,\label{eq:CrossRatios}
\end{equation}
of the \emph{dual position variables},
\begin{equation}\label{eq:x_coords_def}
p_{i}=: x_{i+1}-x_{i}:=x_{i+1 i}\,,
\end{equation}
where particle indices are cyclically identified, $x_{i+n}\equiv x_i$.

Since the cross ratios~\eqref{eq:CrossRatios} are invariant under conformal transformations of dual position variables, we can pick a reference frame where one of them goes to infinity, say $x_m\to \infty$. Restricting to the subset of transformations of the dual conformal group that do not move $x_m$, then breaks the latter to the Poincar\'e group times scale transformations with respect to these variables~\cite{Dirac:1936fq}, or equivalently to the Lorentz group times scale transformations with respect to the momenta $p_i$. 

We may illustrate this procedure in the example of the DCI 4-mass 1-loop box integral shown in the left-hand side of figure~\ref{fig:4to3pt}, which in the appropriate normalisation is a function of the two independent cross ratios\footnote{Note that here we form massive or offshell external legs $k_i=p_{2i-1}+p_{2i}$ as a sum of two massless ones, to make contact with the fact that all fields of $\cN=4$ SYM theory are massless, $p_i^2=x^2_{ii+1}=0$.}
\be\label{eq:BoxCrossRatios}
u=\frac{x^2_{13}x^2_{57}}{x^2_{15}x^2_{37}}\,,\quad v=\frac{x^2_{35}x^2_{17}}{x^2_{15}x^2_{37}}\,.
\ee
In the reference frame where, for example, $x_7\to \infty$, the above cross ratios reduce to
\be\label{eq:DU_CR_limits}
u\to \hat u = \frac{x^2_{13}}{x^2_{15}}
=\frac{k_1^2}{(k_1+k_2)^2}\,,\quad v\to \hat v=\frac{x^2_{35}}{x^2_{15}}
=\frac{k_2^2}{(k_1+k_2)^2}\,, 
\ee
with $k_1,k_2,k_3,k_4$ representing the massive legs of the 4-mass box, and the point at infinity is no longer visible, as shown in the right-hand side of figure~\ref{fig:4to3pt}. We have thus shown that, at least in strictly four dimensions, the Lorentz-invariant (LI) \emph{3-mass 1-loop triangle integral is identical to the 4-mass DCI box integral evaluated in a particular reference frame}! That the two integrals are equal to the same dilogarithmic function, 
\be\label{eq:DU_1loop}
\Phi(u,v)=2 \text{Li}_2(-\rho u)+2 \text{Li}_2(-\rho v)+\log\frac{v}{u}\log\frac{1+\rho v}{1+\rho u}+\log(\rho u)\log(\rho v)+\tfrac{1}{3}\pi^2\,
\ee
with
\be\label{eq:Delta_4mb}
\rho=\frac{2}{1-u-v+\Delta}\,,\quad \Delta=\sqrt{\left(1-u-v\right)^2-4uv}\,,
\ee
when appropriately normalised and upon a particular identification of their kinematic variables, was initially observed by direct calculation~\cite{Usyukina:1992jd}, and was only later understood as a consequence of conformal symmetry~\cite{BROADHURST1993132}.

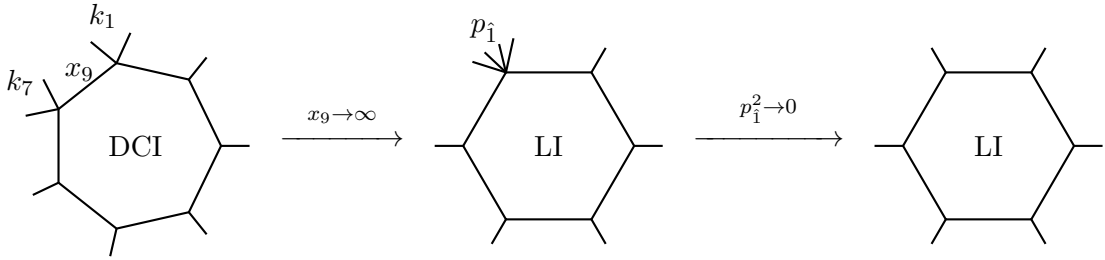
\begin{figure}
\begin{center}
\begin{tikzpicture}[scale=0.75,baseline]
    % Define the number of sides and the radius
    \def\n{7}
    \def\radius{1.5}
    \def\outer{2}
    \def\delta{10} % small angular offset in degrees
    % Draw the nonagon
    \foreach \i in {1,...,\n} {
        \coordinate (P\i) at ({\i*360/\n}:\radius);
    }
    \draw[thick] (P1)-- (P2)-- (P3) node[pos=0.2,left] {$x_9$} -- (P4)-- (P5) -- (P6)-- (P7)-- cycle ;
  \foreach \i in {1,...,\n} {
        \coordinate (Q\i) at ({\i*360/\n}:2);
    }
\draw[thick] (P1)--(Q1) ;
\draw[thick] (P4)--(Q4) ;
\draw[thick] (P5)--(Q5);
\draw[thick] (P6)--(Q6);
\draw[thick] (P7)--(Q7);

\draw[thick] (P2)--({(2-0.2)*360/7}:2) node[midway,above left,pos=0.8]{$k_1$};
\draw[thick] (P2)--({(2+0.2)*360/7}:2);
\draw[thick] (P3)--({(3+0.2)*360/7}:2);
\draw[thick] (P3)--({(3-0.2)*360/7}:2)node[midway,left,pos=0.9]{$k_7$};
    \node at (0,0) {DCI};

\end{tikzpicture}$\quad \overset{x_9\to \infty}{\xrightarrow{\hspace*{40pt}}}\quad $\begin{tikzpicture}[scale=0.75,baseline]
    % Define the number of sides and the radius
    \def\n{6}
    \def\radius{1.5}
    \def\outer{2}
    \def\delta{10} % small angular offset in degrees

    % Draw the hexagon
    \foreach \i in {1,...,\n} {
        \coordinate (P\i) at ({\i*360/\n}:\radius);
    }
    \draw[thick] (P1)--(P2)--(P3)--(P4)--(P5)--(P6)--cycle;

    % Outer points
    \foreach \i in {1,...,\n} {
        \coordinate (Q\i) at ({\i*360/\n}:\outer);
    }

    % Normal radial lines
    \draw[thick] (P1)--(Q1);
    \draw[thick] (P3)--(Q3);
    \draw[thick] (P4)--(Q4);
    \draw[thick] (P5)--(Q5);
    \draw[thick] (P6)--(Q6);

    \draw[thick] (P2) --({(2+0.2)*360/6}:2) ;
    \draw[thick] (P2) --({(2-0.07)*360/6}:2);
    \draw[thick] (P2) --({(2+0.07)*360/6}:2) node[above]{$p_{\hat{1}}$};
    \draw[thick] (P2) --({(2-0.2)*360/6}:2) ;
   
    \node at (0,0) {LI};
\end{tikzpicture}$\quad \overset{p_{\hat{1}}^2\to 0}{\xrightarrow{\hspace*{50pt}}}\quad $\begin{tikzpicture}[scale=0.75,baseline]
    % Define the number of sides and the radius
    \def\n{6}
    \def\radius{1.5}
    \def\outer{2}
    \def\delta{10} % small angular offset in degrees

    % Draw the hexagon
    \foreach \i in {1,...,\n} {
        \coordinate (P\i) at ({\i*360/\n}:\radius);
    }
    \draw[thick] (P1)--(P2)--(P3)--(P4)--(P5)--(P6)--cycle;

    % Outer points
    \foreach \i in {1,...,\n} {
        \coordinate (Q\i) at ({\i*360/\n}:\outer);
    }

    % Normal radial lines
    \draw[thick] (P1)--(Q1);
    \draw[thick] (P2)--(Q2);
    \draw[thick] (P3)--(Q3);
    \draw[thick] (P4)--(Q4);
    \draw[thick] (P5)--(Q5);
    \draw[thick] (P6)--(Q6);
    \node at (0,0) {LI};
\end{tikzpicture}
\end{center}
\caption{Breaking the dual conformal invariance of the DCI subalphabet of~\cite{Henke:2021ity} with seven points and two adjacent masses, we obtain for the first time candidate letters for 6-point 1-mass processes in QCD. Their massless limit, after cyclic symmetrisation, essentially encompasses the recent direct results of ~\cite{Abreu:2024fei,Henn:2025xrc}, and includes new predictions.}\label{fig:7to6pt}
\end{figure}

This relation between DCI and LI integrals, which we call breaking of dual conformal invariance, is of course not restricted to the above example. Precisely as a consequence of symmetry, it also applies to \emph{any} DCI planar integral with at least two massive adjacent legs.\footnote{Taking a dual coordinate $x_m$ to infinity implies that $x_{im}^2\to \infty$ for any other $x_i$, which cannot be satisfied when an adjacent leg is massless,  $p^2=x_{im}^2=0$ for $i=m-1$ or $m+1$.} Furthermore, since they are described by the same function, analogous to eq.~\eqref{eq:DU_1loop}, \emph{DCI and LI integrals related by the breaking of dual conformal invariance should also have the same alphabet}, cf. eq.~\eqref{eq:CDE_intro}, up to an identification of kinematic variables analogous to eq.~\eqref{eq:DU_CR_limits}.

\begin{figure}
\centering
\begin{tikzpicture}[scale=1.4]
  % Define a style for the nodes
  box/.style={
    draw,               % Draw border
    rectangle,          % Make it a rectangle
    rounded corners,    % Rounded corners
    minimum width=0.8cm,  % Minimum width of the box
    minimum height=0.5cm, % Minimum height of the box
    text centered,      % Center text
    font=\large         % Large font size
  }]
  \coordinate (A) at (0,1);
  \coordinate (B) at (-4,0);
  \coordinate (C) at (4,0);
 \draw (A) node {\begin{tikzpicture}[scale=0.8,baseline]
    % Define the number of sides and the radius
    \def\n{6}
    \def\radius{1.5}

    % Draw the nonagon
    \foreach \i in {1,...,\n} {
        \coordinate (P\i) at ({\i*360/\n}:\radius);
    }
    \draw[thick] (P1)-- (P2)-- (P3) -- (P4) -- (P5) -- (P6)-- cycle ;
    \draw[thick] (P2) [blue]-- node[scale=0.7,midway,left] {$x_9$} (P3);
    \draw[thick] (P3) [red] -- node[scale=0.7,midway,left] {$x_7$}(P4);
  \foreach \i in {1,...,\n} {
        \coordinate (Q\i) at ({\i*360/\n}:2);
    }
\draw[thick] (P1)--(Q1) ;
\draw[thick] (P2)--({(2-0.2)*360/6}:2);
\draw[thick] (P2)--({(2+0.2)*360/6}:2);
\draw[thick] (P3)--({(3-0.2)*360/6}:2);
\draw[thick] (P3)--({(3+0.2)*360/6}:2);
\draw[thick] (P4)--({(4-0.2)*360/6}:2);
\draw[thick] (P4)--({(4+0.2)*360/6}:2);
\draw[thick] (P5)--(Q5);
\draw[thick] (P6)--(Q6);
\node[font=\Large] at (0,0) {DCI};
\end{tikzpicture}};
\hfill
 \draw (B) node {\begin{tikzpicture}[scale=0.8,baseline]
    % Define the number of sides and the radius
    \def\n{5}
    \def\radius{1.5}
    \def\delta{10}
    \def\outer{2}

    % Draw the nonagon
    \foreach \i in {1,...,\n} {
        \coordinate (P\i) at ({\i*360/\n}:\radius);
    }
    \draw[thick] (P1)-- (P2)-- (P3) -- (P4)-- (P5)-- cycle ;
  \foreach \i in {1,...,\n} {
        \coordinate (Q\i) at ({\i*360/\n}:2);
    }
    \draw[thick] (P2) [red] -- (P3)node[scale=0.7,midway,left] {$x_7$};
\draw[thick] (P1)--(Q1);
\draw[thick] (P4)--(Q4);
\draw[thick] (P5)--(Q5);
\node[font=\Large] at (0,0) {LI};

 \foreach \shift in {- \delta, \delta} {
        \draw[thick] 
            ({2*360/\n}:\radius) -- ({2*360/\n+\shift}:\outer);
    }

\foreach \shift in {-3, 3} {
        \draw[thick] 
            ({2*360/\n}:\radius) -- ({2*360/\n+\shift}:\outer);
    }

 \foreach \shift in {- \delta, \delta} {
        \draw[thick] 
            ({3*360/\n}:\radius) -- ({3*360/\n+\shift}:\outer);
    }
\node[xshift=0.3cm,yshift=0.4cm] at (P2) {$p_{\hat{1}}$};
\node[xshift=0.4cm,yshift=0.5cm] at (P1) {$p_{2}$};
\node[xshift=0.4cm,yshift=0.5cm] at (P5) {$p_{3}$};
\node[xshift=0.4cm,yshift=-0.5cm] at (P4) {$p_{4}$};
\node[xshift=-0.4cm,yshift=-0.5cm] at (P3) {$p_{\hat{5}}$};
\end{tikzpicture}};

  \draw (C) node {\begin{tikzpicture}[scale=0.8,baseline]
    % Define the number of sides and the radius
    \def\n{5}
    \def\radius{1.5}
    \def\delta{10}
    \def\outer{2}

    % Draw the nonagon
    \foreach \i in {1,...,\n} {
        \coordinate (P\i) at ({\i*360/\n}:\radius);
    }
    \draw[thick] (P1)-- (P2)-- (P3) -- (P4)-- (P5)-- cycle ;
  \foreach \i in {1,...,\n} {
        \coordinate (Q\i) at ({\i*360/\n}:2);
    }
    \draw[thick] (P2) [blue]  -- (P3) node[scale=0.7,midway, left] {$x_9$};
\draw[thick] (P1) --(Q1);
\draw[thick] (P4)--(Q4);
\draw[thick] (P5)--(Q5);
\node[font=\Large] at (0,0) {LI};

 \foreach \shift in {- \delta, \delta} {
        \draw[thick] 
            ({2*360/\n}:\radius) -- ({2*360/\n+\shift}:\outer);
    }

 \foreach \shift in {- \delta, \delta} {
        \draw[thick] 
            ({3*360/\n}:\radius) -- ({3*360/\n+\shift}:\outer);
    }

     \foreach \shift in {-3, 3} {
        \draw[thick] 
            ({3*360/\n}:\radius) -- ({3*360/\n+\shift}:\outer);
    }
\node[xshift=0.3cm,yshift=0.4cm] at (P2) {$p_{\hat{1}}$};
\node[xshift=0.4cm,yshift=0.5cm] at (P1) {$p_{2}$};
\node[xshift=0.4cm,yshift=0.5cm] at (P5) {$p_{3}$};
\node[xshift=0.4cm,yshift=-0.5cm] at (P4) {$p_{4}$};
\node[xshift=-0.4cm,yshift=-0.5cm] at (P3) {$p_{\hat{5}}$};
\end{tikzpicture}};
  \draw[->] (1,0.5) --  (2.7,0.1)node[scale=0.7,red,rotate=-13.24,midway,above] {$x_7 \to \infty$};
   \draw[->] (-1,0.5) -- (-2.7,0.1) node[scale=0.7,blue,rotate=13.24,midway,above] {$x_9 \to \infty$};
\end{tikzpicture}
\caption{There exist two inequivalent ways to break the symmetry of the DCI subalphabet with six points and three adjacent masses, and we take the union of the resulting LI letters for 5-point 2-mass hard integrals (the easy configuration is treated straightforwardly as well). Also considering all relevant permutations of the external legs, we find a highly nontrivial partial overlap with the alphabet of~\cite{Abreu:2024yit}.}
\label{fig:6hto5}
\end{figure}
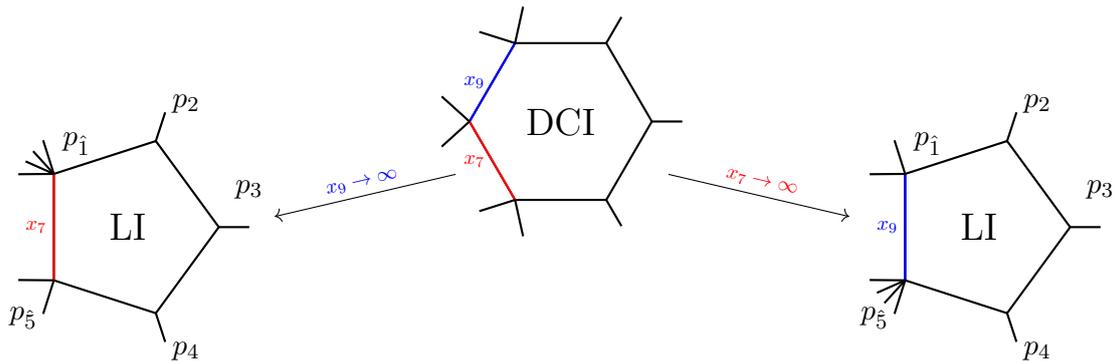

In addition, this general method of breaking dual conformal invariance applies not only to individual DCI integrals, but also to collections thereof: The alphabet of the collection will  simply be the (multiplicatively independent set in the)  union of the alphabet of each integral. In this work, we will apply the method in question to the cluster-algebraic alphabet believed to describe the finite part of the appropriately normalised nine-particle amplitude in planar $\cN=4$ SYM theory~\cite{Henke:2021ity}, which we review in more detail in subsection~\ref{subsec:Gr49AB_Review}. This will certainly be contained in the alphabet of the master integrals contributing to the amplitude, and one can also not exclude the existence of a good choice of basis where the two coincide for the finite parts of the integrals (i.e. this basis will have no extra letters that cancel out in the amplitude).

Finally, we will also make use of the well-known consequence of integration-by-parts identities~\cite{Chetyrkin:1981qh}, that bases of integrals contain all topologies obtained from the top sector by removing some of the propagators. By virtue of eq.~\eqref{eq:CDE_intro}, this implies that the alphabet of the basis will certainly contain the alphabet of the sectors where propagators between external legs have been removed. In fact, it is easy to identify the corresponding subalphabet by demanding its independence from some of the kinematic variables of the top topology, as we explain in subsection~\ref{subsec:massivemomenta}.

Based on this, here we will thus first reduce the candidate alphabet for massless 9-particle scattering in $\cN=4$ SYM theory to subalphabets describing the scattering of $(9-k)$ external particles, out of which $k$ are massive/offshell and the remaining massless. The complete list of these reductions is summarised in table~\ref{tab:results}, and we will be mostly focusing on the prominent cases illustrated in figures~\ref{fig:7to6pt} and~\ref{fig:6hto5}. That is, we will concentrate on seven- and six-point DCI processes with at least two external massive legs adjacent, which will precisely allow us to break the symmetry so as to obtain LI candidate letters for 6-point 1-mass (and massless, in the corresponding limit) as well as 5-point 2-mass processes in QCD. This is essentially in line with the procedure applied in~\cite{Chicherin:2020umh} to the proposed alphabet of the 8-particle amplitude in $\cN=4$ SYM theory~\cite{Drummond:2019cxm,Arkani-Hamed:2019rds,Henke:2019hve}, which gave candidate letters for 5-point 1-mass (and massless) integrals, for further applications see also~\cite{He:2021eec,He:2022ctv}. We now proceed to summarise our main results.

\subsection{Summary of results}\label{subsec:results_summary}
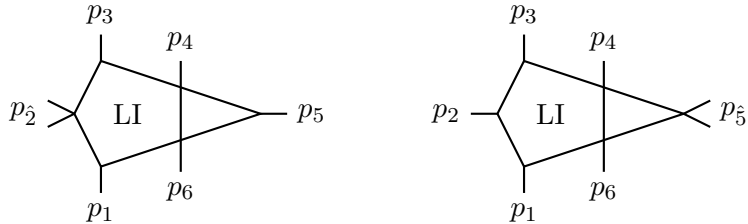
\begin{figure}
\begin{center}
\begin{tikzpicture}[scale=0.7,baseline]
\coordinate (y0) at (0,0);
\coordinate (y1) at (0,-0.5);
\coordinate (y2) at (0,0.5);
\coordinate (x1) at (-1.5,-1);
\coordinate (x2) at (-2,0);
\coordinate (x3) at (-1.5,1);
\coordinate (x5) at (1.5,0);
\coordinate (x6) at (0,-0.5);
\coordinate (p1) at (-1.5,-1.5) ;
\coordinate (p2A) at (-2.5,+0.25);
\coordinate (p2B) at (-2.5,-0.25);
\coordinate (p3) at (-1.5,1.5);
\coordinate (p4) at (0,1);
\coordinate (p5) at (2,0);
\coordinate (p6) at (0,-1.1);
\draw[thick] (y1)--(x1)--(x2)--(x3)--(y2)--(x5)--(x6)--(y1);
\draw[thick] (y1)--(y2);
\draw[thick] (p1) node[below]{$p_1$} -- (x1);

\node at (-1,0) {LI};

\node[left] at (-2.5,0) {$p_{\hat 2}$};
    \draw[thick] (p2A)  -- (x2);
    \draw[thick] (p2B)  -- (x2);
    \draw[thick] (p3) node[above]{$p_3$}-- (x3);
    \draw[thick] (p4) node[above]{$p_4$}-- (y2);
    \draw[thick] (p5) node[right]{$p_5$}-- (x5); 
    \draw[thick] (p6) node[below]{$p_6$}-- (x6);
\end{tikzpicture}\quad \qquad
\begin{tikzpicture}[scale=0.7,baseline]
\coordinate (y0) at (0,0);
\coordinate (y1) at (0,-0.5);
\coordinate (y2) at (0,0.5);
\coordinate (x1) at (-1.5,-1);
\coordinate (x2) at (-2,0);
\coordinate (x3) at (-1.5,1);
\coordinate (x5) at (1.5,0);
\coordinate (x6) at (0,-0.5);
\coordinate (p1) at (-1.5,-1.5) ;
\coordinate (p2) at (-2.5,0);
\coordinate (p3) at (-1.5,1.5);
\coordinate (p4) at (0,1);
\coordinate (p5A) at (2,+0.25);
\coordinate (p5B) at (2,-0.25);
\coordinate (p6) at (0,-1.1);
\draw[thick] (y1)--(x1)--(x2)--(x3)--(y2)--(x5)--(x6)--(y1);
\draw[thick] (y1)--(y2);
\draw[thick] (p1) node[below]{$p_1$} -- (x1);
\node[right] at (2,0) {$p_{\hat 5}$};
\node at (-1,0) {LI};
    \draw[thick] (p2) node[left]{$p_2$} -- (x2);
    \draw[thick] (p3) node[above]{$p_3$}-- (x3);
    \draw[thick] (p4) node[above]{$p_4$}-- (y2);
    \draw[thick] (p5A) -- (x5);
    \draw[thick] (p5B) -- (x5);
    \draw[thick] (p6) node[below]{$p_6$}-- (x6);
\end{tikzpicture}
\end{center}
\caption{Integral topologies expected to contain the nested square root letters of eq.~\eqref{eq:nested_sq_letters}.}
\label{fig:pent_triangles}
\end{figure}

{\bf 6-point 1-mass letters.} From the procedure sketched in the left half of figure~\ref{fig:7to6pt}, we obtain a total of 246 candidate letters. These are divided into 119 letters that are rational in the Mandelstam variables, 59 letters containing simple square roots that can be rationalised\footnote{We will henceforth denote these simple square roots as rationalisable, and similarly the letters containing them.} in the \emph{momentum twistor variables} discussed in subsection~\ref{subsec:Momentum twistors}, and 68 letters containing non-rationalisable square roots. As a first test, we check that these essentially contain the alphabet of the 6-point 1-loop integral, which is easily extracted as limit of the generic 6-point 1-loop integral from the results and ancillary file of~\cite{Dlapa:2023cvx}.

We may further eliminate letters that belong to subkinematics (5-point 2-mass etc), and are already known in the literature or will be discussed in what follows. In this manner, we end up with 29 letters with genuinely 6-point 1-mass kinematics, out of which 8 are rational in the Mandelstam variables, 9 have rationalisable and 12 have non-rationalisable square roots. Focusing on the latter, in the orientation where the massive leg is $p_{\hat 1}$ as shown in figure~\ref{fig:7to6pt}, these have the form
\begin{equation}\label{eq:nested_sq_letters}
l_{\pm}^{i} = \frac{A_{i}\pm B_{i}\epsilon_{2356}-C_{i}\sqrt{\Delta_{\pm}}}{{A_{i}\pm B_{i} \epsilon_{2356}+C_{i}\sqrt{\Delta_{\pm}}}},\,\quad \Delta_\pm= F\pm \epsilon_{2356} G\,,    \quad i=1,\ldots,6,
\end{equation}
where $A_i, B_i, C_i$ and $F, G$ are rational and polynomial expressions in the Mandelstam variables, respectively, and $\epsilon_{2356}$ is a rationalisable square root defined in eq.~\eqref{eq:epsilon_def} below.

Rather surprisingly, we observe that the letters in question have \emph{nested square roots}, namely the radicand of the square root also contains a square root in the Mandelstam variables. These have appeared quite recently in calculations of 5-point master integrals with massive and massless propagators that contribute to QCD amplitudes for $pp\to t\bar t+X$ production, where $X$ is a Higgs boson or a jet~\cite{FebresCordero:2023pww, Badger:2024fgb,Becchetti:2025oyb}. Whereas all these integral families are associated to elliptic geometries and integrate to more complicated function spaces, our results strongly suggest that \emph{nested square roots are ubiquitous even in quantum field theories with massless propagators, and for quantities purely expressible in terms of multiple polylogarithms}.

As we explain in section~\ref{sec:(6,1)letters}, where these as well as the remaining rational(isable) letters are analysed further, we expect that the nested square-root letters originate in 1-mass pentagon-triangle integral families such as the ones shown in figure~\ref{fig:pent_triangles}. We are confident that our results will aid the calculation of the 2-loop 6-point 1-mass master integrals, especially since the complicated nature of the letters~\eqref{eq:nested_sq_letters} poses a challenge for the existing, partial algorithms for determining them directly from the integrals~\cite{Jiang:2024eaj,Matijasic:2024too,Effortlessxxx,Correia:2025yao}. Finally, we perform 
another highly nontrivial check of our 6-point 1-mass predictions by taking the massless limit, as described next. 

\vspace{4pt}

\noindent {\bf 6-point massless letters.} In the $p^2_{\hat{1}}\to 0$ limit illustrated in figure~\ref{fig:7to6pt}, the 6-point 1-mass alphabet reduces to 95 rational, 45 rationalisable and 20 non-rationalisable square-root letters. Taking their closure under cyclic permutations, as needed so as to be able to describe 6-particle planar amplitudes, we produce 374 letters, further split into 225 that are rational, as well as 114 and 35 that have rationalisable and non-rationalisable square-roots, respectively. 

We then move on to compare our prediction to the recent computation of all integrals needed for describing the finite parts of planar two-loop six-point amplitudes in massless QCD~\cite{Abreu:2024fei,Henn:2025xrc}. As we detail in subsection~\ref{subsec:6p0m_results}, we excitingly find that again the alphabet of this computation is essentially contained in our prediction! By `essentially' here and throughout the text we mean that the only letters we do not reproduce are those which appear as easily identifiable components of our predicted letters, and thus can be added back in at the very end. We define these `trivially recoverable' letters and explain how to recover them at the end of subsection~\ref{subsec:pTr49AB}. We point out that we do predict letters appearing at $\mathcal{O}(\epsilon^{-1})$ and $\mathcal{O}(\epsilon^0)$ in the dimensionally regulated integrals, which were missed by the alternative cluster-algebraic approaches of~\cite{Pokraka:2025ali,Bossinger:2025rhf}.

We complete our highly encouraging check by additionally comparing with the letters appearing in the CDEs of six-point massless integrals not only in the aforementioned papers but also in the earlier, preliminary works~\cite{Henn:2021cyv,Henn:2022ydo,Henn:2024ngj}, as well as with more recent amplitude bootstrap calculations~\cite{Carrolo:2025agz,Carrolo:2026qpu}. The core information of this extensive comparison is summarised in table~\ref{tab:(6,0)comparison}. In this manner, we establish that our prediction contains another 102 rational and 60 rationalisable letters that have not appeared before. We anticipate that these will appear in higher-loop contributions to the same kinematics.

\vspace{4pt}

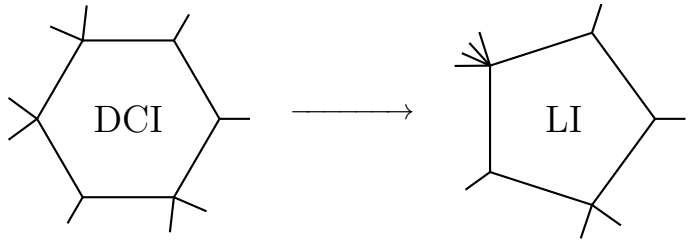
\begin{figure}
    \centering
    \begin{tikzpicture}[scale=0.8,baseline]
    % Define the number of sides and the radius
    \def\n{6}
    \def\radius{1.5}
    \def\delta{10}
    \def\outer{2}

    % Draw the nonagon
    \foreach \i in {1,...,\n} {
        \coordinate (P\i) at ({\i*360/\n}:\radius);
    }
    \draw[thick] (P1)-- (P2)-- (P3) -- (P4)-- (P5) -- (P6)-- cycle ;
  \foreach \i in {1,...,\n} {
        \coordinate (Q\i) at ({\i*360/\n}:2);
    }
\draw[thick] (P1)--(Q1) ;
\draw[thick] (P4)--(Q4) ;
\draw[thick] (P6)--(Q6);
\node[font=\Large] at (0,0) {DCI};;

% Special double line for side 2:
    % We take the angle of vertex 2 and offset it by ±delta
    \foreach \shift in {- \delta, \delta} {
        \draw[thick] 
            ({2*360/\n}:\radius) -- ({2*360/\n+\shift}:\outer);
    }

 % Special double line for side 3:
    % We take the angle of vertex 3 and offset it by ±delta
    \foreach \shift in {- \delta, \delta} {
        \draw[thick] 
            ({3*360/\n}:\radius) -- ({3*360/\n+\shift}:\outer);
    }
 % Special double line for side 5:
    % We take the angle of vertex 5 and offset it by ±delta
    \foreach \shift in {- \delta, \delta} {
        \draw[thick] 
            ({5*360/\n}:\radius) -- ({5*360/\n+\shift}:\outer);
    }

\end{tikzpicture} $\quad {\xrightarrow{\hspace*{40pt}}} \quad$ \begin{tikzpicture}[scale=0.8,baseline]
    % Define the number of sides and the radius
    \def\n{5}
    \def\radius{1.5}
    \def\delta{10}
    \def\outer{2}

    % Draw the nonagon
    \foreach \i in {1,...,\n} {
        \coordinate (P\i) at ({\i*360/\n}:\radius);
    }
    \draw[thick] (P1)-- (P2)-- (P3) -- (P4)-- (P5)-- cycle ;
  \foreach \i in {1,...,\n} {
        \coordinate (Q\i) at ({\i*360/\n}:2);
    }
\draw[thick] (P1)--(Q1);
\draw[thick] (P3)--(Q3);
\draw[thick] (P5)--(Q5);

% Special double line for side 2:
    % We take the angle of vertex 2 and offset it by ±delta
 \foreach \shift in {- \delta, \delta} {
        \draw[thick] 
            ({2*360/\n}:\radius) -- ({2*360/\n+\shift}:\outer);
    }

\foreach \shift in {-3, 3} {
        \draw[thick] 
            ({2*360/\n}:\radius) -- ({2*360/\n+\shift}:\outer);
    }
 
 % Special double line for side 4:
    % We take the angle of vertex 4 and offset it by ±delta
    \foreach \shift in {- \delta, \delta} {
        \draw[thick] 
            ({4*360/\n}:\radius) -- ({4*360/\n+\shift}:\outer);
    }

\node[font=\Large] at (0,0) {LI};
\end{tikzpicture}
\caption{The 5-point 2-mass LI alphabet is obtained from the 6-point 3-mass `medium' DCI subalphabet, in complete analogy with the derivation of the 6-point 1-mass alphabet illustrated in the left half of figure~\ref{fig:7to6pt}.}
    \label{fig:5pt2mE}
\end{figure}

\noindent {\bf 5-point 2-mass letters.} There exist two inequivalent `hard' and `easy' configurations for these planar kinematics, depending on whether the two massive legs are adjacent or not, respectively. Our prediction for the alphabet of the 5-point 2-mass easy configuration follows along the same lines with that of the six-point 1-mass configuration, and is depicted in figure~\ref{fig:5pt2mE}. Either before or after breaking DCI, which merely amounts to an identification of kinematic variables, it yields 40 rational, 12 rationalisable and 19 non-rationalisable square-root letters.

As briefly discussed in the previous subsection and also illustrated in figure~\ref{fig:6hto5}, for the 2-mass hard configuration a new qualitative feature appears: One may break the symmetry of the original 6-point 3-mass hard DCI subalphabet of 39 rational, 12 rationalisable and 23 square-root letters in two distinct ways. We thus take the union of these two mappings to LI 5-point 2-mass alphabets, which are related by a flip symmetry with respect to the external momenta. Hence the resulting alphabet naturally respects this flip symmetry, as expected, and after removing multiplicative dependencies it comprises of 55 rational, 17 rationalisable and 31 non-rationalisable square-root letters.

Next, we move on to compare our findings with the alphabet of all 2-loop planar 5-point 2-mass integrals, derived in~\cite{Abreu:2024yit}. Anticipating the extension of the latter calculation to the nonplanar case, in the latter reference a special set of permutations of the external legs has been considered, so we also apply an analogous permutation completion of our prediction as well. We find a total of 495 letters, arranged in 50 classes or orbits under special permutations.

The overlap in this case is significantly more nontrivial: One the one hand, a significant qualitative difference of the alphabet of the aforementioned reference is that it contains letters with two non-rationalisable square roots, which by construction are absent in our prediction. In addition, the latter does not reproduce one of the four different types of non-rationalisable square roots appearing in the former, and hence also neither of the letters containing it. Given the past success of cluster algebras in revealing letters of the integrals that do not appear in the final amplitude, it is natural to ask if something similar happens here as well. On the other hand, here too we find that our prediction contains 87 new letters, arranged in 8 orbits under special permutations, which thus serve as candidates at the three-loop order.

All the ancillary files accompanying this work may be found in the repository~\cite{code}. Their contents are described in the relevant parts of the main text, and also summarised in Appendix~\ref{appx}.

\section {Kinematics}\label{sec:Kinematics}
In this section we build our conventions for the kinematic variables we will employ in our calculations. We start with the usual Lorentz-invariant Mandelstam variables and certain useful functions thereof, and then move on to review momentum twistor variables that are ideally suited for the dual conformal invariant quantities that will be our point of departure. As these necessarily involve massless momenta, after that we describe how to combine them in pairs so as to describe massive particles. Finally, we translate the procedure for breaking dual conformal symmetry, described in the introductory section, from momenta to momentum twistors.

\subsection{Lorentz invariance (LI) and Mandelstam variables}\label{subsec_Kin_Mandelstam}

The kinematic dependence of scattering amplitudes involving $n$ external particles is commonly described by their momentum vectors $p^\mu_{1},\ldots,p^\mu_{n}$ with vector component index $\mu=0,\ldots d-1$ , subject to  momentum conservation,
\be\label{eq:mom_conserv}
\sum_{i=1}^n p^\mu_i=0\,,
\ee
as well as the on-shell condition, e.g. $p^2=\eta_{\mu\nu }p^\mu p^\nu=0$ for a massless particle. For definitiveness we take the mostly minus convention for the signature of our Minkowski metric $\eta_{\mu\nu}$.

Lorentz invariance implies that the dependence of scattering amplitudes on the momenta enters principally through their scalar products or Mandelstam variables, defined as
\begin{equation}
s_I:= \left (\sum_{i\in I} p_i\right)^2\,,
\end{equation}
for a subset of the particle labels $I$. In this work we will particularly be focusing on planar Mandelstam variables that are formed by consecutive particle labels,
\begin{equation}\label{eq:s_planar}
s_{i,i+1,\cdots,j}:=(p_{i}+p_{i+1}+\cdots+p_j)^2=x^2_{ij+1} \,,\;\; i<j\,,
\end{equation}
where we have also indicated the relation of these variables to distances of the dual space coordinates defined in~\eqref{eq:x_coords_def}.

When $m\le n$ of the external momenta are massive and the rest massless, if the dimension of these momenta is $d\ge n-1$ and $n\ge 3$ then it is not difficult to show that the number of independent, dimensionful Mandelstam invariants is\footnote{E.g. for $m=0$, the above formula follows by eliminating $p_n$ from momentum conservation and counting the number of nonzero planar variables~\eqref{eq:s_planar} formed by the remaining momenta, also recalling that $s_{1,\ldots,n-1}=p_n^2=0.$
}
\be\label{eq:count_vars_naive}
\frac{1}{2}(n-1)(n-2)+m-1\,.
\ee
However if $d<n-1$, then one also needs to take into account that no more than $d$ vectors can be linearly independent in $d$ dimensions. The correct number of independent Mandelstam invariants for any $d\ge 2$ and $n\ge  \lfloor d \rfloor+2$, where $\lfloor x \rfloor$ denotes the largest integer smaller than $x$, is instead given by
\be\label{eq:count_indep_vars}
N^{d}=nd-d-(n-m)-\frac{d(d-1)}{2}\,,
\ee
where to the total number of $nd$ of momentum components we have removed those eliminated by the $d$ momentum conservation equations, the $(n-m)$ on-shell conditions for the massless momenta and  the dimension $d(d-1)/2$ of the $SO(d-1,1)$ Lorentz group, as its generators leave scalar quantities invariant. 

The linear relations between the momenta in $d<n-1$ dimensions may be expressed in Lorentz-invariant form with the help of Gram determinants. For two subsets of momenta  $I$ and $J$ of the same cardinality, their Gram determinant is defined as,
\begin{equation} \label{eq:grammdef}
G(I;J):= \det p_i\cdot p_j\,,\quad G(I):=G(I;I) \; \quad \text { with } i\in I ,\; j\in J\;.
\end{equation}
Then it is a matter of simple linear algebra to show that if $G(I)=0$ the momenta in the set $I$ are linearly dependent. Restricting to our case of interest $d=4$, this implies that
\begin{equation}\label{eq:grammvanish}
G(I) =0 \;, \qquad \text{for } |I|\geq 5  \text { and } n> 5\;.
\end{equation}
Not all choices of $I$ represent independent relations between the momenta, a particular set that does may be chosen as
\be
G(1,\ldots,d,i)=0\,,\quad i=d+1,\ldots n-1\,.
\ee

In what follows, we will restrict to the case of interest $d=4$, $n\ge 4$, for which the number of independent Mandelstam invariants~\eqref{eq:count_indep_vars} reduces to
\be\label{eq:Mandelstam__indep_count}
N^{d=4}=3n-10+m\,.
\ee
Obviously, the number of independent dimensionless ratios we can form out of these Mandelstam variables will be $N^{d=4}-1$.

In order to fully specify the kinematics of physical particles with $p_0>0$, i.e. all their momentum 4-vectors, in addition to the Mandelstam invariants we also need to know the value of the quantities \cite{Eden:1966dnq}
\begin{equation}\label{eq:epsilon_def}
\epsilon_{ijkl}:= -4i \varepsilon_{\mu_{1}\mu_{2}\mu_{3}\mu_{4}}{p}^{\mu_{1}}_{i} {p}^{\mu_{2}}_{j} {p}^{\mu_{3}}_{k} {p}^{\mu_{4}}_{l}=-4i\det\begin{pmatrix}
p_i^0& p_i^1& p_i^2& p_i^3\\
p_j^0& p_j^1& p_j^2& p_j^3\\
p_k^0& p_k^1& p_k^2& p_k^3\\
p_l^0& p_l^1& p_l^2& p_;^3
\end{pmatrix}\,,
\end{equation}
where $\varepsilon_{\mu_{1}\mu_{2}\mu_{3}\mu_{4}}$ is the fully antisymmetric Levi-Civita tensor. From this definition, it is not difficult to show that the norm of these quantities as well as their relative signs are expressible in terms of Gram determinants of Mandelstam invariants,
\begin{equation}\label{eq:eps_to_Gram}
\epsilon_{I} \cdot \epsilon_{J} =16 G(I;J), \quad \epsilon^2_{I} = 16 G(I), \quad \text{with} \;\abs{I},\abs{J}=4.
\end{equation}
So in fact the sign of a single $\epsilon_{I}$, together with the Mandelstam invariants, is sufficient to fully specify the kinematics. This sign is not invariant under parity or spatial reflection transformations $P$,
\be\label{eq:parity}
P\cdot\left(x^0,\vec x\right)=(x^0,-\vec x)\,,
\ee
which is a symmetry of Minkowski space, though not necessarily of the particle interactions or of the scattering amplitude. 

Whereas Mandelstam invariants are Lorentz scalars, i.e. invariant under parity transformations, the $\epsilon_I$ are Lorentz pseudoscalars, in other words they flip sign,
\be\label{eq:parityOp}
P\cdot s_{i,\ldots,j}=s_{i,\ldots,j}\,,\quad P\cdot \epsilon_{ijkl}=-\epsilon_{ijkl}\,,
\ee

For our purposes, it will be helpful to define two additional pseudoscalar quantities which naturally arise in the leading singularities and letters of five-point and six-point integrals. These are, 
\begin{equation}
    \label{eq:delta5}
    \Delta_{5} = \Tr{\gamma^{5}\slashed{p}_1\slashed{p}_2\slashed{p}_3\slashed{p}_4}, \quad
    \Delta_{6} = \Tr{\gamma^{5}\slashed{p}_1\slashed{p}_2\slashed{p}_3\slashed{p}_4\slashed{p}_5 \slashed{p}_{6}},
\end{equation}
where $\slashed{p}:=\gamma_\mu p^\mu$, $\gamma^{5}:=i\gamma^{0}\gamma^{1}\gamma^{2}\gamma^{3}$ and $\gamma^\mu$ denotes the usual Dirac matrices. From the properties of the latter, it follows easily that $\Delta_5=\epsilon_{1234}$ and thus by virtue of~\eqref{eq:eps_to_Gram} its square will be proportional to a Gram determinant. Similarly, though more tediously, one may also show that
\be\label{eq:delta6toEps}
\Delta_6
=\sum_{1\le i<j\le6}(-1)^{i+j}(p_i\cdot p_j)
\epsilon_{1,\dots\widehat{i}\dots,\widehat{j}\dots6}\,,
\ee
where hatted indices are omitted, e.g. $\epsilon_{1234\widehat{5}\widehat{6}}=\epsilon_{1234}$.

Properties of Dirac matrices imply that $\Delta_{6}$ is fully antisymmetric under permutations of the $p_i$ in $d=4$, and in subsection~\ref{subsec:1loop6p1m} we will also show that its square equals a (special case of a) Cayley determinant,
\be
\Delta_6^2=-\det(x^2_{ij})\,,\quad i,j=1,\ldots,6\,,
\ee
where the dual coordinates have been defined in eq.~\eqref{eq:x_coords_def}. By virtue of \eqref{eq:eps_to_Gram}, \eqref{eq:delta6toEps} is also clear that a single of the quantities $\epsilon_I, \Delta_5, \Delta_6$ would suffice to express all the rest, and up to a sign this single quantity is the square root of a determinant of Mandelstam invariants.\footnote{Note that momentum conservation implies additional linear relations among the $\epsilon_I$. For an $n$-particle process, $\frac{(n-1)!}{(n-5)!4!}$ of them will be linearly independent (but still algebraically related).} As this replacement would lead to very large expressions, however, in what follows we will refrain from applying it, also keeping in mind that the pseudoscalar quantities are algebraically related.

\subsection{Dual conformal invariance (DCI) and momentum twistors}\label{subsec:Momentum twistors}

Dual-conformal invariant quantities such as planar $\cN=4$ SYM amplitudes are most naturally described in terms of \emph{momentum twistor}~\cite{Hodges:2009hk} variables.  Indeed, these enter in the starting point of our analysis, the proposed alphabet for the 9-particle SYM amplitude~\cite{Henke:2021ity}, so in this subsection we will review them briefly. For a more extensive introduction see e.g.~\cite{Arkani-Hamed:2010pyv,Papathanasiou:2022lan}.

What makes momentum twistors the variables of choice for describing massless, planar, DCI kinematics is that they:
\begin{enumerate}
    \item Trivialise momentum conservation, eq.~\eqref{eq:mom_conserv}.
    \item Trivialise the on-shell condition for the external massless particles,
    \be\label{eq:massless_momenta}
p_{i}^2=x^2_{ii+1}=0\,.
\ee
\item Rationalise the pseudoscalars $\epsilon_I,\Delta_5,\Delta_6$, eqs.~\eqref{eq:epsilon_def},\eqref{eq:delta5} in the previous subsection.
\end{enumerate}

The reader solely interested in the practical application of momentum twistors may jump directly to eq.~\eqref{eq:momtwistmat} and the paragraph containing it below, but before that let us devote a few paragraphs to motivate their origin and aforementioned properties. The main idea behind momentum twistors is that conformal transformations in four dimensions coincide with (hyperbolic) $SO(4,2)$ rotations in six dimensions~\cite{Dirac:1936fq}, also implying that four-dimensional vectors may be mapped to null projective six-dimensional vectors. We may apply this idea to the dual position variables $x_i$ of eq.~\eqref{eq:x_coords_def}, which by construction fulfil point 1 above. Denoting the resulting six-dimensional vectors as $X_i$, the aforementioned mapping translates
\be\label{eq:x_to_X}
x^2_{ij}\propto X_i\cdot X_j\,,
\ee
where the proportionality constant cancels out in conformally invariant quantities. Group theory then tells us that the six-dimensional vector representation of $SO(4,2)$ is locally equivalent to an antisymmetric combination of two fundamental $SU(2,2)$ (or, after complexifying, $SL(4,\mathbb{C})$) representations $Z_i, Z_i'$,
\be\label{eq:XtoZ}
X_i^{AB}=Z_i^A Z_i^{'B}-Z_i^B Z_i^{'A}\,,\qquad A,B=1,\ldots,4\,. 
\ee
Since they descend from a projective vector, they too are defined modulo rescalings $Z_i\sim \lambda Z_i$ for nonzero $\lambda$ (and similarly for primed), so they live in complex projective space $\mathbb{CP}^3$.

We call the $Z_i, Z_i'$ momentum twistors and $\mathbb{CP}^3$ momentum twistor space. By the logic of the previous paragraph, a point $x_i$ in dual coordinate space is mapped to a pair of points or a line $(Z_{i}, Z_{i}')$ in momentum twistor space. Similarly, it may be shown that if two dual coordinates are null separated, then their corresponding twistor lines must intersect, i.e. share a common twistor. This follows from the fact that $SO(4,2)$ scalar products translate to
\be\label{eq:X_to_Z}
X_i\cdot X_j=\varepsilon_{ABCD}X_i^{AB}X_j^{CD}=\varepsilon_{ABCD}Z_i^A Z_i^{'B} Z_j^C Z_j^{'D}\,,
\ee
where the presence of the Levi-Civita tensor allows us to drop the explicit antisymmetrisation in eq.~\eqref{eq:XtoZ}. Specialising the above formula to $j=i+1$, by virtue of eq.~\eqref{eq:x_to_X} we see that we solve the on-shell condition~\eqref{eq:massless_momenta} and thus satisfy point 2 above by simply letting $Z'_{i+1}=Z_i$. Finally, rationalisation of  the (DCI versions of the) pseudoscalar quantities of point 3 above is also related to eq.~\eqref{eq:x_to_X}, and more specificially to the fact that square distances $x_{ij}^2$ become linear in the $X_i$, as reviewed in e.g.~\cite{Bourjaily:2021lnz}.

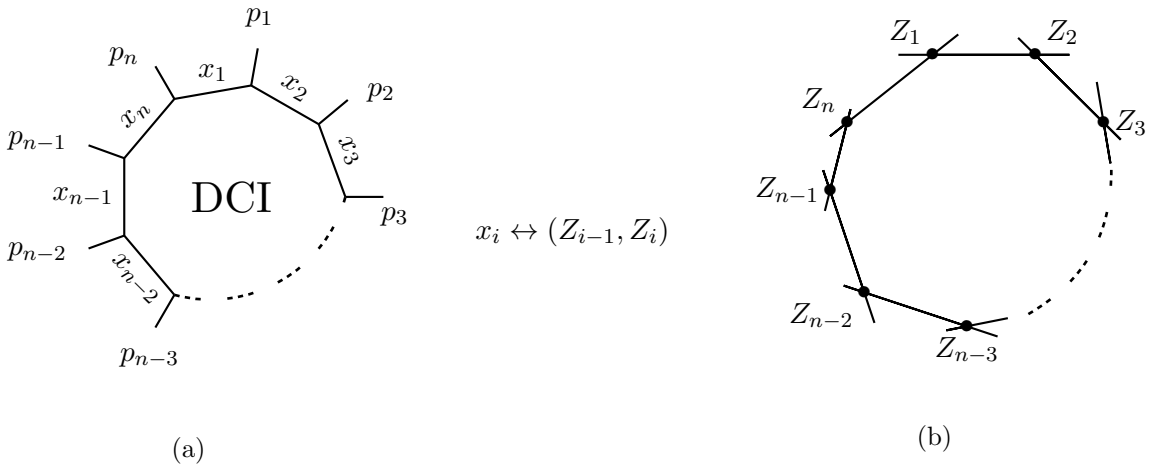
\begin{figure}
    \centering
    \begin{subfigure}[c]{0.35\textwidth}
     \begin{center}
     \begin{tikzpicture}
    % Define the number of sides and the radius
    \def\n{9}
    \def\radius{1.5}

    % Draw the nonagon
    \foreach \i in {1,...,\n} {
        \coordinate (P\i) at ({\i*360/\n}:\radius);
    }
    \draw[thick]  (P9) -- node[midway,above,rotate=-60] {$x_{3}$}  (P1) -- (P2)node[midway,above,rotate=-30] {$x_2$} -- (P3)node[midway,above] {$x_{1}$}  -- (P4)node[midway,above,rotate=45] {$x_{n}$} -- (P5)node[midway,left] {$x_{n-1}$} -- (P6)node[midway,below,rotate=-45] {$x_{n-2}$};
    
    \draw[line width=1pt, dash pattern=on 2pt off 2pt on 2pt off 2pt on 2pt off 10pt] (P6)  to [bend right=45] (P9);
  \foreach \i in {1,...,\n} {
        \coordinate (Q\i) at ({\i*360/\n}:2);
    }
\node[scale=1.5] at (0,0) {DCI};
\draw[thick] (P1)--(Q1) node[right, pos=1.3] {$p_2$};
\draw[thick] (P2)--(Q2) node[above, pos=1.3] {$p_1$};
\draw[thick] (P3)--(Q3) node[left, pos=1.3] {$p_{n}$};
\draw[thick] (P4)--(Q4) node[left, pos=1.3] {$p_{n-1}$};
\draw[thick] (P5)--(Q5) node[left, pos=1.3] {$p_{n-2}$};
\draw[thick] (P6)--(Q6) node[below, pos=1.3] {$p_{n-3}$};
\draw[thick] (P9)--(Q9) node[below, pos=1.3] {$p_{3}$};
\end{tikzpicture}
     \caption{\phantom{a}}
    \end{center}
    \end{subfigure} 
\hfill
    \begin{minipage}[c]{0.2\textwidth}
    \hspace{0.1cm}
    $ x_i \leftrightarrow (Z_{i-1}, Z_i)$
  \end{minipage}
\hfill
    \begin{subfigure}[c]{0.35\textwidth}
    \begin{center}
    \begin{tikzpicture}[scale=0.9]
\coordinate (P1) at (0,0);
\coordinate (P2) at (1.5,0);
\coordinate (P3) at (2.5,-1);
\coordinate (P4) at (2.75,-2.5);
\coordinate (P5) at (2,-3.7);
\coordinate (P6) at (0.5,-4);
\coordinate (P7) at (-1,-3.5);
\coordinate (P8) at (-1.5,-2);
\coordinate (P9) at (-1.25,-1);
\coordinate (Q3) at ({2.5+0.25*0.45},{-1-1.5*0.45});
\draw[thick] (P1) 
node[anchor=south east]{$Z_1$}
node {$\bullet$};
\draw[thick] (P2) 
node[anchor=south west]{$Z_2$}
node {$\bullet$};
\draw[thick] (P3) 
node[right]{$Z_3$}
node {$\bullet$};
\draw[thick] (P6) 
node[below]{$Z_{n-3}$}
node {$\bullet$};
\draw[thick] (P7) 
node[anchor=north east]{$Z_{n-2}$}
node {$\bullet$};
\draw[thick] (P8) 
node[left]{$Z_{n-1}$}
node {$\bullet$};
\draw[thick] (P9) 
node[anchor=south east]{$Z_{n}$}
node {$\bullet$};
\draw[thick] (P1) -- (P2) -- ++(0.5,0) -- ++(-2.5,0);
\draw[thick] (P2) -- (P3) -- ++(0.25,-0.25) -- ++(-1.5,1.5);
\draw[line width=1pt, dash pattern=on 2pt off 2pt on 2pt off 2pt on 2pt off 10pt] (P6)++({(0.5-2)*(-0.6)},{(-4+3.7)*(-0.6)}) to [bend right=30] (Q3);
\draw[thick] (P3) -- ++({0.25*0.4},{-1.5*0.4}) -- ++({-0.25*0.8},{1.5*0.8});
\draw[thick] (P6) -- ++({(0.5-2)*0.2},{(-4+3.7)*0.2})-- ++({(0.5-2)*(-0.6)},{(-4+3.7)*(-0.6)});
\draw[thick] (P6) -- (P7) -- ++({(-1-0.5)*0.2},{(-3.5+4)*0.2})-- ++({(-1-0.5)*(-1.5)},{(-3.5+4)*(-1.5)});
\draw[thick] (P7) -- (P8) -- ++({(-1.5+1)*0.2},{(-2+3.5)*0.2})-- ++({(-1.5+1)*(-1.5)},{(-2+3.5)*(-1.5)});
\draw[thick] (P8) -- (P9) -- ++({(1.5-1.25)*0.2},{(2-1)*0.2})-- ++({(1.5-1.25)*(-1.5)},{(2-1)*(-1.5)});
\draw[thick] (P8) -- (P9) -- ++({(-1.25)*0.2},{(-1)*0.2})-- ++({(-1.25)*(-1.5)},{(-1)*(-1.5)});
\end{tikzpicture}
    \end{center}
    \caption{\phantom{b}}
    \end{subfigure}
   \caption{Description of $n$-particle massless DCI kinematics in terms of (a) momenta/dual space variables and (b) momentum twistor, as well as their relation.}
    \label{fig:9ptkinematics}
\end{figure}

So after the dust settles, $n$-particle massless planar DCI kinematics are equivalently described by $n$ cyclically ordered momentum twistors $Z_i\in\mathbb{CP}^3$, as depicted in figure~\ref{fig:9ptkinematics}, and as encoded in a $4\times n$ matrix,
\begin{equation}
 \label{eq:momtwistmat}
    \mathbf{Z}:=
    \begin{pNiceArray}{ccc}
   Z_{1} & \ldots & Z_{n}
    \end{pNiceArray}\,,
\end{equation}
modulo 15 $SL(4,\mathbb{C})$ transformations and $n$ individual $Z_i\to \lambda Z_i$ rescalings.  After gauge-fixing these symmetries,\footnote{We provide more details on how to do this in practice in subsection~\ref{subsec:BreakDCI}.} one is thus left with a \emph{momentum twistor parametrisation} with $4n-15-n=3n-15$ unconstrained variables. As is evident from eqs~\eqref{eq:x_to_X},\eqref{eq:X_to_Z}, the $SL(4,\mathbb{C})$-invariant building blocks in momentum twistor space  are the \emph{4-brackets} or \emph{Pl\"ucker coordinates},
\be\label{eq:Plucker_def}
\boxed{\br{ijkl}:=\det{Z_{i}Z_{j}Z_{k}Z_{l}}=\varepsilon_{ABCD} Z_i^A Z_j^B Z_k^C Z_l^D}\,,
\ee
and DCI quantities are formed by combinations thereof that are invariant under rescalings $Z_i\to \lambda Z_i$; For example the four-mass box cross ratios~\eqref{eq:BoxCrossRatios} are reexpressed as
\be\label{eq:BoxCrossRatiosZ}
u=\frac{\br{8123}\br{4567}}{\br{8145}\br{2367}}\,,\quad v=\frac{\br{2345}\br{8167}}{\br{8145}\br{2367}}\,.
\ee
Since the number $\frac{n!}{(n-4)!4!}$ of distinct Pl\"ucker coordinates is greater than the number of components of the matrix $\mathbf{Z}$, not all of them will be independent. They obey Pl\"ucker relations,
\begin{equation} \label{eq:pluckrel}
    \br{i_1 i_2 i_3 j_a} \br{j_b j_c j_d j_e}\varepsilon^{abcde}=0\;,
\end{equation}
which, similarly to the Gram determinant constraints, encode the fact that no more that four momentum twistors $Z_i\in \mathbb{CP}^3$ can be linearly independent. Any momentum twistor parametrisation automatically satisfies the Pl\"ucker relations, and through the map of momenta to twistors also the Gram determinant constraints, owing to their inherently four-dimensional nature.

Finally, we note that the space of $n$-particle massless planar DCI kinematics in momentum twistor space is also equivalent to a particular quotient of the Grassmannian $Gr(4,n)$, namely the space of all 4-planes going through the origin of an $n$-dimensional space. The latter is invariant under $GL(4,\mathbb{C})$ transformations of the basis vectors on each plane, so combining one of the momentum twistor rescalings with $SL(4,\mathbb{C})$, it is evident that eq~\eqref{eq:momtwistmat} equivalently describes
\begin{equation}
\label{eq:conf}
    Gr(4,n)/(\mathbb{C}^{\star})^{n-1}\,,
\end{equation}
where the quotient is with respect to the remaining $n-1$ rescalings. This observation is crucial for endowing the space of kinematics with cluster-algebraic structure~\cite{Arkani-Hamed:2012zlh,Golden:2013xva}, and also elucidates how the parity transformation~\eqref{eq:parity} acts in momentum twistor language: Given that for the Grassmannian the latter is known to exchange $k$-planes with $(n-k)$-planes, we have that
\be\label{eq:parityZ}
P\cdot Z_i^A\propto \varepsilon_{ABCD}Z_{i-1}^B Z_{i}^C Z_{i+1}^D\,,
\ee
where again the proportionality factor cancels out in conformally invariant quantities.

\subsection{Massive momenta from sums of massless ones}
\label{subsec:massivemomenta}

\begin{figure}
        \centering
        \begin{subfigure}[c]{0.5\textwidth}
            \begin{tikzpicture}
    % Define the number of sides and the radius
    \def\n{8}
    \def\radius{1.5}
    \def\outer{2}
    \def\delta{10} % small angular offset in degrees
    % Draw the nonagon
    \foreach \i in {1,...,\n} {
        \coordinate (P\i) at ({\i*360/\n}:\radius);
    }
    \draw[thick]  (P8) -- node[midway,above,rotate=-60] {$x_{r+2}$}  (P1) -- (P2)node[midway,above,rotate=-30] {$x_{r+1}$} -- (P3)node[midway,above,rotate=30] {$x_{r-1}$}  -- (P4)node[midway,above,rotate=60] {$x_{r-2}$} -- (P5)node[midway,left,pos=0.7] {$x_{r-3}$};
    
    \draw[line width=1pt, dash pattern=on 2pt off 2pt on 2pt off 2pt on 2pt off 10pt] (P5)  to [bend right=45] (P8);
  \foreach \i in {1,...,\n} {
        \coordinate (Q\i) at ({\i*360/\n}:2);
    }
\node[scale=1.5] at (0,0) {DCI};
\draw[thick] (P1)--(Q1) node[right, pos=1.3] {$p_{r+1}$};
\draw[thick] (P3)--(Q3) node[left, pos=1.3] {$p_{r-2}$};
\draw[thick] (P4)--(Q4) node[left, pos=1.3] {$p_{r-3}$};
\draw[thick] (P5)--(Q5) node[left, pos=1.3] {$p_{r-4}$};
\draw[thick] (P8)--(Q8) node[below, pos=1.3] {$p_{r+2}$};

\draw[thick] 
            ({2*360/\n}:\radius) -- ({2*360/\n+\delta}:\outer) node[above,pos=1]{$k_{r}=p_r+p_{r-1}$};
\draw[thick] 
            ({2*360/\n}:\radius) -- ({2*360/\n-\delta}:\outer);
\end{tikzpicture}
        \end{subfigure}
        \hfill
        \begin{subfigure}[c]{0.4\textwidth}
            \begin{tikzpicture}[scale=0.9]
\coordinate (P1) at (0,0);
\coordinate (P2) at (1.5,0);
\coordinate (P3) at (2.5,-1);
\coordinate (P4) at (2.75,-2.5);
\coordinate (P5) at (2,-3.7);
\coordinate (P6) at (0.5,-4);
\coordinate (P7) at (-1,-3.5);
\coordinate (P8) at (-1.5,-2);
\coordinate (P9) at (-1.25,-1);
\coordinate (Q3) at ({2.5+0.25*0.45},{-1-1.5*0.45});
\draw[thick] (P1) 
node[anchor=south east]{$Z_r$}
node {$\bullet$} 
node[below, pos=1.3] {$\boldsymbol{\leftrightarrow}$};
\draw[thick] (P2) 
node[anchor=south west]{$Z_{r+1}$}
node {$\bullet$};
\draw[thick] (P3) 
node[right]{$Z_{r+2}$}
node {$\bullet$};
\draw[thick] (P6) 
node[below]{$Z_{r-4}$}
node {$\bullet$};
\draw[thick] (P7) 
node[anchor=north east]{$Z_{r-3}$}
node {$\bullet$};
\draw[thick] (P8) 
node[left]{$Z_{r-2}$}
node {$\bullet$};
\draw[thick] (P9) 
node[anchor=south east]{$Z_{r-1}$}
node {$\bullet$}
node[below,rotate=70] {$\boldsymbol{\leftrightarrow}$};
\draw[thick] (P1) -- (P2) -- ++(0.5,0) -- ++(-2.5,0);
\draw[thick] (P2) -- (P3) -- ++(0.25,-0.25) -- ++(-1.5,1.5);
\draw[line width=1pt, dash pattern=on 2pt off 2pt on 2pt off 2pt on 2pt off 10pt] (P6)++({(0.5-2)*(-0.6)},{(-4+3.7)*(-0.6)}) to [bend right=30] (Q3);
\draw[thick] (P3) -- ++({0.25*0.4},{-1.5*0.4}) -- ++({-0.25*0.8},{1.5*0.8});
\draw[thick] (P6) -- ++({(0.5-2)*0.2},{(-4+3.7)*0.2})-- ++({(0.5-2)*(-0.6)},{(-4+3.7)*(-0.6)});
\draw[thick] (P6) -- (P7) -- ++({(-1-0.5)*0.2},{(-3.5+4)*0.2})-- ++({(-1-0.5)*(-1.5)},{(-3.5+4)*(-1.5)});
\draw[thick] (P7) -- (P8) -- ++({(-1.5+1)*0.2},{(-2+3.5)*0.2})-- ++({(-1.5+1)*(-1.5)},{(-2+3.5)*(-1.5)});
\draw[thick] (P8) -- (P9) -- ++({(1.5-1.25)*0.2},{(2-1)*0.2})-- ++({(1.5-1.25)*(-1.5)},{(2-1)*(-1.5)});
\draw[thick] (P8) -- (P9);
\draw[dashed,blue] (P9) -- (P1);
\end{tikzpicture}
        \end{subfigure}
        \caption{Momenta/dual coordinates and momentum twistors for 1-mass DCI kinematics. Independence from $x_i$ on the LHS becomes independence from the dashed line on the RHS, or equivalently to translation invariance of $Z_{r-1}, Z_{r}$ along their lines, indicated by arrows.}
        \label{fig:8pt1}
    \end{figure}
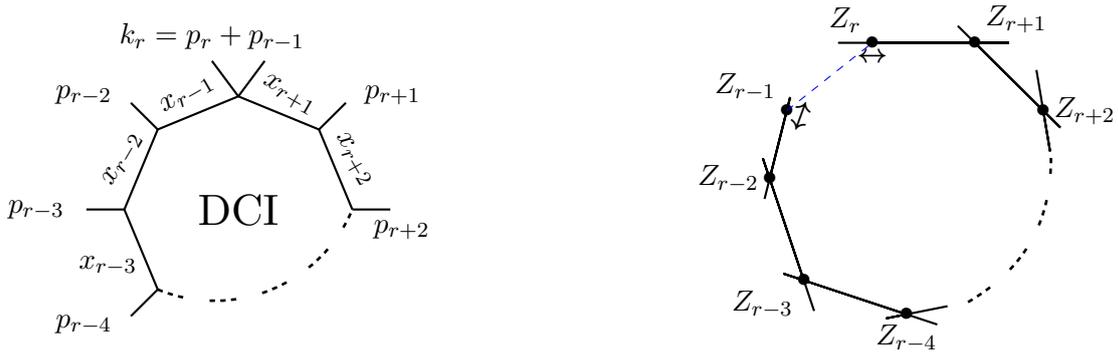

Momentum twistors are convenient kinematic variables not only when the external momenta are massless, as reviewed in the previous subsection. As we explain here, they apply to massive momenta as well, and in the next sections we will indeed employ them to obtain predictions for 6-point 1-mass and 5-point 2-mass LI integrals. 

The main idea, also discussed in the introduction, is that we can always represent massive legs as sums of massless ones. Let's assume that we form the 1-mass $(n-1)$-point configuration shown on the left-hand side (LHS) of figure~\ref{fig:8pt1}, such that
\be\label{eq:massive_momentum}
k_{r}:=p_{r-1}+p_{r}=x_{r+1}-x_{r-1}\,,
\ee
where all $p_i^2=0$ for $i=1,\ldots n$ including, specifically,
\be\label{eq:massive_momentum_constituents}
p_{r-1}^2=(x_r-x_{r-1})^2=0=(x_{r+1}-x_{r})^2=p_{r}^2
\ee
and both formulas were also expressed in the dual space variables of eq.~\eqref{eq:x_coords_def}.

Evidently, the kinematics will be independent of the point $x_r$. Naively this would give us a four-parameter freedom to shift it without any effect, but since we require its distance \eqref{eq:massive_momentum_constituents} from the neighboring points to remain lightlike, we in fact have a two-parameter redundancy.\footnote{For example, in the centre-of-mass frame of the momenta $p_{r-1},p_{r}$, where without loss of generality their spatial components are chosen to only lie on the $\mu=3$ axis, this redundancy pertains to shifts in the $\mu=1,2$ directions.} Eliminating this from the degrees of freedom of the lightlike constituents of $k_i$, we thus have $3+3-2=4$ degrees of freedom, as expected for a massive particle.

From the above discussion and figure~\ref{fig:8pt1}, it is clear that if we describe the parent massless $n$-point kinematics by the planar Mandelstam invariants $s_{i,i+1\ldots, j}$ of eq.~\eqref{eq:s_planar}, only those of them where the set of indices $\{i,i+1\ldots, j\}$ contains either both or none of the $\{r-1,r\}$ will be relevant for our 1-mass kinematics. Equivalently, the fact that 1-mass kinematics is independent of $x_r$ implies that Lorentz-invariant quantities $f(x_{ij}^2)$ do not depend on $x_{rk}^2=x_{kr}^2$ for any $k\ne r, r\pm1$. Stated in infinitesimal form, $f(x_{ij}^2)$ must be annihilated by the generators
\be
\frac{\partial}{\partial x^2_{rk}}\,\quad\,,
\ee
for any value of the $x^2_{rk}$, assuming that the spacetime dimension $d$ is large enough so that all Mandelstam invariants remain independent. Specifically for the case $n>d+1=5$ of interest, it is possible to show that only the generators with $k=r \pm 2$ are relevant, in line with the two-parameter symmetry we discussed in the previous paragraph.\footnote{This can be done e.g. by choosing the basis vectors as $p_{r-2},p_{r-1},p_{r}, p_{r+1}$, and encoding all Lorentz-invariant information in their $p_i\cdot p_j$ invariants, as well as the coefficients of the other vectors on this basis. Trading the  $p_i\cdot p_j$ products with the planar Mandelstam invariants~\eqref{eq:s_planar}, it then follows that dependence on $x_r$ only enters through $x^2_{r\; r\pm 2}$.}

Now let us move on describe the same kinematics in the language of momentum twistors, as shown on the right-hand side of figure~\ref{fig:8pt1}. The point $x_r$ corresponds to the line $(Z_{r-1},Z_{r})$, so shifts with respect to the former translate to changing the direction of the latter without changing the direction of the remaining twistor lines. This is equivalent to translating $Z_{r}$ and $Z_{i-1}$ along the lines they belong to,
\begin{equation}
\label{eq:momtwistmass}
    Z_{r-1} \rightarrow Z_{r-1}+\lambda Z_{r-2}\,, \quad Z_{r} \rightarrow Z_{r} +\mu Z_{r+1}\, ,
\end{equation}
where $\lambda, \mu \in \mathbb{C}$. Indeed, since $x^2_{ij}\propto\br{i-1,i,j-1,j}$, these transformations leave $x^2_{r\pm 1k}$ and thus also $x^2_{r\pm1r}$ invariant, while shifting all other $x^2_{rk}$, for different values of $k$. Therefore they reproduce the original  shifts of $x_i$ that preserve eq.~\eqref{eq:massive_momentum_constituents}.

Interestingly, each of the transformations~\eqref{eq:momtwistmass} is the momentum twistor space version of a Britto-Cachazo-Feng-Witten (BCFW) shift \cite{Elvang:2013cua}, namely the same deformation of momenta that is used in order to obtain BCFW recursion relations for tree-level amplitudes~\cite{Britto:2005fq}. We will make use of this terminology also in subsection~\ref{subsec:Subkinematic_alphabets}.

For a function of momentum twistors $f(Z_i)$, invariance under the infinitesimal versions of the transformations~\eqref{eq:momtwistmass} implies that it must be annihilated by the operators
\be\label{eq:ops8p1m}
\mathcal{O}_{r-2,r-1}\,,\quad \mathcal{O}_{r+1,r}\,,
\ee
which are defined in their most general form as~\cite{Drummond:2010cz}
\begin{equation}
    \label{eq:opdef}
    \mathcal{O}_{i,j} := Z_{i}^{I}\partial_{Z_{j}^{I}}\,,
\end{equation}
and which act on  Pl\"ucker coordinates as
\begin{equation}
    \label{eq:opact}
    \begin{aligned}
    &\mathcal{O}_{i,j} \br{jklm} = \br{iklm}\,.
    \end{aligned}
\end{equation}
That is, if the Pl\"ucker coordinate contains $Z_j$ then $\mathcal{O}_{i,j}$ replaces it with $Z_i$, otherwise it annihilates it.

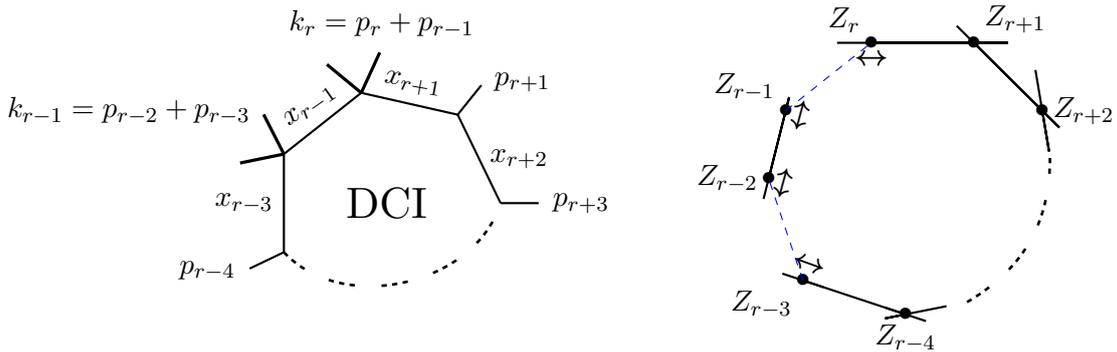
\begin{figure}
        \centering
        \begin{subfigure}[c]{0.5\textwidth}
            \begin{tikzpicture}
    % Define the number of sides and the radius
    \def\n{7}
    \def\radius{1.5}

    % Draw the nonagon
    \foreach \i in {1,...,\n} {
        \coordinate (P\i) at ({\i*360/\n}:\radius);
    }
    \draw[thick] (P7) --node[midway,right] {$x_{r+2}$} (P1) -- (P2)node[midway,above,rotate=-10] {$x_{r+1}$} -- (P3)node[pos=0.5,above,rotate=35] {$x_{r-1}$}  -- (P4)node[midway,left] {$x_{r-3}$} ;
  \foreach \i in {1,...,\n} {
        \coordinate (Q\i) at ({\i*360/\n}:2);
    }
    \draw[thick] (P1)--(Q1)node[midway,right,pos=1.1]{$p_{r+1}$};
\node[scale=1.5] at (0,0) {DCI};
\draw[very thick] (P2)--({(2-0.2)*360/7}:2) node[midway,above,pos=1.1]{$k_{r}=p_{r}+p_{r-1}$};
\draw[very thick] (P2)--({(2+0.2)*360/7}:2);
\draw[very thick] (P3)--({(3+0.2)*360/7}:2);
\draw[very thick] (P3)--({(3-0.2)*360/7}:2)node[midway,left,pos=1.1]{$k_{r-1}=p_{r-2}+p_{r-3}$};
\draw[thick] (P4)--(Q4)node[midway,left,pos=1.1]{$p_{r-4}$};
\draw[thick] (P7)--(Q7)node[midway,right,pos=1.1]{$p_{r+3}$};
\draw[line width=1pt, dash pattern=on 2pt off 2pt on 2pt off 2pt on 2pt off 10pt] (P4) to [bend right=60] (P7);
\end{tikzpicture}
        \end{subfigure}
        \hfill
        \begin{subfigure}[c]{0.4\textwidth}
            \begin{tikzpicture}[scale=0.9]
\coordinate (P1) at (0,0);
\coordinate (P2) at (1.5,0);
\coordinate (P3) at (2.5,-1);
\coordinate (P4) at (2.75,-2.5);
\coordinate (P5) at (2,-3.7);
\coordinate (P6) at (0.5,-4);
\coordinate (P7) at (-1,-3.5);
\coordinate (P8) at (-1.5,-2);
\coordinate (P9) at (-1.25,-1);
\coordinate (Q3) at ({2.5+0.25*0.45},{-1-1.5*0.45});
\draw[thick] (P1) 
node[anchor=south east]{$Z_{r}$}
node {$\bullet$}
node[below] {$\boldsymbol{\leftrightarrow}$};
\draw[thick] (P2) 
node[anchor=south west]{$Z_{r+1}$}
node {$\bullet$};
\draw[thick] (P3) 
node[right]{$Z_{r+2}$}
node {$\bullet$};
\draw[thick] (P6) 
node[below]{$Z_{r-4}$}
node {$\bullet$};
\draw[thick] (P7) 
node[anchor=north east]{$Z_{r-3}$}
node {$\bullet$}
node[above,rotate=-20] {$\boldsymbol{\leftrightarrow}$};
\draw[thick] (P8) 
node[left]{$Z_{r-2}$}
node {$\bullet$}
node[below,rotate=75] {$\boldsymbol{\leftrightarrow}$};
\draw[thick] (P9) 
node[anchor=south east]{$Z_{r-1}$}
node {$\bullet$}
node[below,rotate=75] {$\boldsymbol{\leftrightarrow}$};
\draw[thick] (P1) -- (P2) -- ++(0.5,0) -- ++(-2.5,0);
\draw[thick] (P2) -- (P3) -- ++(0.25,-0.25) -- ++(-1.5,1.5);
\draw[thick] (P3) -- ++({0.25*0.4},{-1.5*0.4}) -- ++({-0.25*0.8},{1.5*0.8});
\draw[thick] (P6) -- ++({(0.5-2)*0.2},{(-4+3.7)*0.2})-- ++({(0.5-2)*(-0.6)},{(-4+3.7)*(-0.6)});
\draw[thick] (P7) -- ++ ({(-1-0.5)*0.2},{(-3.5+4)*0.2});
\draw[thick] (P7) -- ++ ({(-1-0.5)*-0.2},{(-3.5+4)*-0.2});
\draw[thick] (P6) -- ++ ({(-1-0.5)*0.2},{(-3.5+4)*0.2});
\draw[thick] (P6) -- ++ ({(-1-0.5)*-0.2},{(-3.5+4)*-0.2});
\draw[thick] (P6) -- (P7);
\draw[thick] (P8) -- (P9) -- ++({(1.5-1.25)*0.2},{(2-1)*0.2})-- ++({(1.5-1.25)*(-1.5)},{(2-1)*(-1.5)});
\draw[line width=1pt, dash pattern=on 2pt off 2pt on 2pt off 2pt on 2pt off 10pt] (P6)++({(0.5-2)*(-0.6)},{(-4+3.7)*(-0.6)}) to [bend right=30] (Q3);
\draw[dashed,blue] (P7) -- (P8);
\draw[dashed,blue] (P9) -- (P1);
\end{tikzpicture}
        \end{subfigure}
        \caption{Momenta/dual coordinates and momentum twistors for 2-mass hard DCI kinematics, i.e. when the massive legs are adjacent.}
        \label{fig:7pt2m}
    \end{figure}

Our discussion generalises in a obvious manner when we form more massive legs by combining another two (or, for that matter, any number of) massless legs. For example, the two-mass hard case is depicted in figure~\ref{fig:7pt2m}, and Lorentz-invariant functions in these kinematics must be annihilated by
\be\label{eq:ops7p2m}
\mathcal{O}_{r-4,r-3}\,,\quad \mathcal{O}_{r-1,r-2}\,,\quad \mathcal{O}_{r-2,r-1}\,,\quad \mathcal{O}_{r+1,r}\,\,.
\ee

Having formulated how to describe massive external particles using momentum twistors, we are now ready to also formulate the breaking of dual conformal symmetry in terms of the latter.

\subsection{Breaking DCI and constructing parameterisations for momentum twistors}\label{subsec:BreakDCI}

As we discussed in subsection~\ref{subsec:MethodOutline} using the simple example of the four-mass box integral, if one considers the conformal group $SO(4,2)$ acting on $n$ dual coordinate points $x_i$, takes one of them to infinity and restricts to the subgroup of transformations that do not move this point, this breaks $SO(4,2)$ to the Poincar\'e group times scale transformations. At the level of the embedding coordinates and momentum twistors defined in subsection~\ref{subsec:Momentum twistors}, the point at infinity $x_\infty$ allows us to define square distances up to an overall scale as
\begin{equation}
\label{eq:mandtobr}
    x^2_{i,j} = \frac{X_i\cdot X_j}{(X_i\cdot X_\infty)(X_j\cdot X_\infty)}=\frac{\br{i-1ij-1j}}{\br{i-1iI_{\infty}}\br{j-1jX_{\infty}}}\,,
\end{equation}
where $X_\infty$ is the six-dimensional embedding vector avatar of $x_\infty$, equivalently a pair of twistors or a twistor line by virtue of eq.~\eqref{eq:XtoZ}.

Indeed, the above definition fixes the proportionality constant of eq.~\eqref{eq:x_to_X} such that all $x^2_{i\infty}=(X_\infty\cdot X_\infty)^{-1}\to \infty$ at the same rate, i.e. their ratios tend to one. The fact that the square distances~\eqref{eq:mandtobr} are defined modulo rescalings follows from the same property of $X_\infty$. Therefore, the variables that are invariant under the surviving symmetries are dimensionless ratios of $x^2_{ij}$, which as we saw in subsection~\ref{subsec:MethodOutline} are precisely what the cross ratios~\eqref{eq:CrossRatios} reduce to when one of their points is taken to be $x_\infty$. These ratios are also the natural arguments of the transcendental functions describing Feynman integrals, which are typically normalised to be dimensionless in canonical bases.

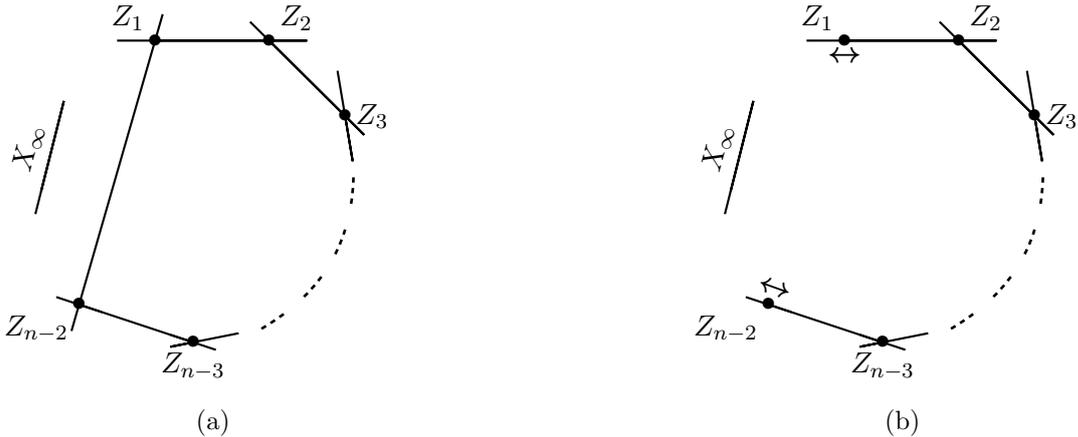
\begin{figure}
        \centering
        \begin{subfigure}[b]{0.4\textwidth}
            \begin{tikzpicture}
\coordinate (P1) at (0,0);
\coordinate (P2) at (1.5,0);
\coordinate (P3) at (2.5,-1);
\coordinate (P4) at (2.75,-2.5);
\coordinate (P5) at (2,-3.7);
\coordinate (P6) at (0.5,-4);
\coordinate (P7) at (-1,-3.5);
\coordinate (P8) at (-1.5,-2);
\coordinate (P9) at (-1.25,-1);
\coordinate (Q3) at ({2.5+0.25*0.45},{-1-1.5*0.45});
\draw[thick] (P1) 
node[anchor=south east]{$Z_1$}
node {$\bullet$};
\draw[thick] (P2) 
node[anchor=south west]{$Z_{2}$}
node {$\bullet$};
\draw[thick] (P3) 
node[right]{$Z_{3}$}
node {$\bullet$};
\draw[thick] (P6) 
node[below]{$Z_{n-3}$}
node {$\bullet$};
\draw[thick] (P7) 
node[anchor=north east]{$Z_{n-2}$}
node {$\bullet$};

\draw[thick] (P1) -- (P2) -- ++(0.5,0) -- ++(-2.5,0);
\draw[thick] (P2) -- (P3) -- ++(0.25,-0.25) -- ++(-1.5,1.5);
\draw[thick] (P3) -- ++({0.25*0.4},{-1.5*0.4}) -- ++({-0.25*0.8},{1.5*0.8});
\draw[thick] (P6) -- ++({(0.5-2)*0.2},{(-4+3.7)*0.2})-- ++({(0.5-2)*(-0.6)},{(-4+3.7)*(-0.6)});
\draw[thick] (P7) -- ++ ({(-1-0.5)*0.2},{(-3.5+4)*0.2});
\draw[thick] (P7) -- ++ ({(-1-0.5)*-0.2},{(-3.5+4)*-0.2});
\draw[thick] (P6) -- ++ ({(-1-0.5)*0.2},{(-3.5+4)*0.2});
\draw[thick] (P6) -- ++ ({(-1-0.5)*-0.2},{(-3.5+4)*-0.2});
\draw[thick] (P6) -- (P7);
\draw[thick] (-1.1,-3.85) -- (0.1,0.35);
\draw[thick] (P8) --node[above,rotate=75]{$X_{\infty}$} (P9) -- ++({(1.5-1.25)*0.2},{(2-1)*0.2})-- ++({(1.5-1.25)*(-1.5)},{(2-1)*(-1.5)});
\draw[line width=1pt, dash pattern=on 2pt off 2pt on 2pt off 2pt on 2pt off 10pt] (P6)++({(0.5-2)*(-0.6)},{(-4+3.7)*(-0.6)}) to [bend right=30] (Q3);
\end{tikzpicture}
            \caption{\phantom{a}}
        \end{subfigure}
        \hfill
        \begin{subfigure}[b]{0.4\textwidth}
            \begin{tikzpicture}
\coordinate (P1) at (0,0);
\coordinate (P2) at (1.5,0);
\coordinate (P3) at (2.5,-1);
\coordinate (P4) at (2.75,-2.5);
\coordinate (P5) at (2,-3.7);
\coordinate (P6) at (0.5,-4);
\coordinate (P7) at (-1,-3.5);
\coordinate (P8) at (-1.5,-2);
\coordinate (P9) at (-1.25,-1);
\coordinate (Q3) at ({2.5+0.25*0.45},{-1-1.5*0.45});
\draw[thick] (P1) 
node[anchor=south east]{$Z_1$}
node {$\bullet$}
node[below] {$\boldsymbol{\leftrightarrow}$};
\draw[thick] (P2) 
node[anchor=south west]{$Z_{2}$}
node {$\bullet$};
\draw[thick] (P3) 
node[right]{$Z_{3}$}
node {$\bullet$};
\draw[thick] (P6) 
node[below]{$Z_{n-3}$}
node {$\bullet$};
\draw[thick] (P7) 
node[anchor=north east]{$Z_{n-2}$}
node {$\bullet$}
node[above,rotate=-20] {$\boldsymbol{\leftrightarrow}$};

\draw[thick] (P1) -- (P2) -- ++(0.5,0) -- ++(-2.5,0);
\draw[thick] (P2) -- (P3) -- ++(0.25,-0.25) -- ++(-1.5,1.5);
\draw[thick] (P3) -- ++({0.25*0.4},{-1.5*0.4}) -- ++({-0.25*0.8},{1.5*0.8});
\draw[thick] (P6) -- ++({(0.5-2)*0.2},{(-4+3.7)*0.2})-- ++({(0.5-2)*(-0.6)},{(-4+3.7)*(-0.6)});
\draw[thick] (P7) -- ++ ({(-1-0.5)*0.2},{(-3.5+4)*0.2});
\draw[thick] (P7) -- ++ ({(-1-0.5)*-0.2},{(-3.5+4)*-0.2});
\draw[thick] (P6) -- ++ ({(-1-0.5)*0.2},{(-3.5+4)*0.2});
\draw[thick] (P6) -- ++ ({(-1-0.5)*-0.2},{(-3.5+4)*-0.2});
\draw[thick] (P6) -- (P7);
\draw[thick] (P8) --node[above,rotate=75]{$X_{\infty}$} (P9) -- ++({(1.5-1.25)*0.2},{(2-1)*0.2})-- ++({(1.5-1.25)*(-1.5)},{(2-1)*(-1.5)});
\draw[line width=1pt, dash pattern=on 2pt off 2pt on 2pt off 2pt on 2pt off 10pt] (P6)++({(0.5-2)*(-0.6)},{(-4+3.7)*(-0.6)}) to [bend right=30] (Q3);
\end{tikzpicture}
            \caption{\phantom{a}}
        \end{subfigure}
        \caption{When modded out by additional shift symmetries, $n$ momentum twistors including the two defining the infinity twistor line may also describe (a) $(n-2)$-point massless (b) $(n-3)$-point 1-mass, and more generally $(n-k-1)$-point $(k-1)$-mass LI kinematics.}
        \label{fig:6pt}
    \end{figure}

The take-home message of this discussion is that LI kinematics may be parameterised equally well in terms of momentum twistors, with the inclusion of an additional pair of twistors or equivalently twistor line $X_\infty$. A distinctive feature of this line is that it cannot intersect any of the remaining twistor lines, encoding the fact that $x_\infty$ cannot be lightlike separated from any other point $x_i$. Also combined with the procedure of subsection~\ref{subsec:massivemomenta} for representing massive legs with momentum twistors,  this leads to the visualisations of massless and one-mass LI kinematics that are depicted in the left- and right-had side of figure \ref{fig:6pt}, respectively.

Comparing the right-hand sides of figures \ref{fig:7pt2m} and \ref{fig:6pt}, we confirm that $(n-2)$-point 2-mass DCI kinematics reduces to $(n-3)$-point 1-mass LI kinematics also in terms of momentum twistors  (up to relabelling, $Z_r\to Z_1$ etc.), once we identify $(Z_{r-2},Z_{r-1}) \equiv X_{\infty}$. We have thus expressed the procedure of breaking DCI by choosing a point at infinity also in this language, and note that the shift symmetries  of $Z_{r-2},Z_{r-1}$ are precisely those that leave the corresponding twistor line invariant.

Summarising the lessons of the entire section, $n$ momentum twistors modded out by an additional, appropriately chosen $2k$ shift symmetries of the form
\be\label{eq:Z_shift}
\quad Z_{i} \rightarrow Z_{i} +\mu_{\pm i} Z_{i\pm 1}\, ,
\ee
may equally well describe
\begin{enumerate}
\item $(n-k)$-point $k$-mass DCI kinematics, $k\ge 0$\,,
\item $(n-k-1)$-point $(k-1)$-mass LI kinematics, $k\ge 1$.
\end{enumerate}
As a consistency check, we note that in all cases the number of independent, dimensionsless variables is
\be
3n-15-2k=3(n-k)-15+k=3(n-k-1)-11+(k-1)\,,
\ee
in agreement with the degree of freedom counts we did in the previous sections, see e.g. eq.~\eqref{eq:Mandelstam__indep_count} and below it.

We close this subsection by describing how one can gauge-fix these symmetries in practice, in order to construct non-redundant momentum twistor parametrisations for each kinematic configuration: One starts from a generic $4\times n$ momentum twistor matrix $\mathbf{Z}$ of eq.~\eqref{eq:momtwistmat} and applies 
\begin{enumerate}
    \item $GL(4)$ transformations, acting as invertible $4\times 4$ matrices $g$ from the left,
    \item $GL(1)$ rescalings of the momentum twistors, acting from the right as $D=\text{diag}(d_1,\ldots,d_n)$ invertible diagonal matrices with a single $d_i$ set to one,
    \item Shift symmetries~\eqref{eq:Z_shift}, also encoded by matrices $h$ acting on the right, which only differ from the identity by elements $h_{i\;i+1}$ or $h_{i+1\;i}$, with indices cyclically identified,
\end{enumerate}
consecutively, in order to set $n+15+2k$ components of $\mathbf{Z}$ to constant values. For example, $GL(4)$ transformations allow us to set any four columns of $\mathbf{Z}$ to the identity matrix, say the first four. $GL(1)$ rescalings combined with residual $GL(4)$ transformations $g=\text{diag}(d^{-1}_1,\ldots,d_4^{-1})$ that do not alter the identity matrix can then be used to set another $n-1$ components to nonzero constants such as $\pm 1$.  A particular choice, closely related to the web-parameterisation~\cite{Speyer2005} in terms of cluster variables, is
\be\label{eq:ZnWebVar}
\left(
\begin{array}{cccccccc}
 1 & 0 & 0 & 0 & -1 & Z_6^1&\ldots & Z_n^1\\
 0 & 1 & 0 & 0 & 1 & Z_6^2&\ldots & Z_n^2\\
 0 & 0 & 1 & 0 & -1 & Z_6^3&\ldots & Z_n^3 \\
 0 & 0 & 0 & 1 & 1 & 1 &\ldots& 1\\
\end{array}
\right)\,,
\ee
and of course any permutation of the columns is also an equally valid parameterisation.

In the absence of shift symmetries, i.e. if $k=0$ we are done, otherwise we continue with their application in order to set another $2k$ components of the matrix $\mathbf{Z}$ to constant values. We will indeed make use of this procedure in order to construct a 6-point 1-mass parameterisation in section~\ref{sec:(6,1)letters}.

\section{The nine-particle SYM alphabet and its reductions}
\label{sec:9ptred}

In this section, we will carry out the reduction of the proposed alphabet of the 9-particle (massless) amplitude in $\cN=4$ SYM theory to subalphabets describing $(9-k)$-point $k$-mass DCI scattering, mainly based on the relation between the respective kinematics, covered in subsection~\ref{subsec:massivemomenta}. By breaking DCI as discussed in the introduction and further elaborated in subsection~\ref{subsec:BreakDCI}, from them we will then obtain predictions for $6$- and $5$-particle LI scattering in QCD in the following section. For completeness, we begin by recalling how our initial $\cN=4$ SYM alphabet is obtained with the help of cluster algebras and their generalisations.

\subsection{Review of symbology from cluster algebras and generalisations}\label{subsec:Gr49AB_Review}

As we recalled in subsection~\ref{subsec:Momentum twistors}, the space of $n$-particle DCI kinematics is closely related to the Grassmannian $Gr(4,n)$,  which is known to have a \emph{cluster algebra} structure. Very briefly, rank-$d$ cluster algebras consist of \emph{cluster variables} $a_i$ that are grouped into overlapping sets $\{a_1,\ldots,a_d\}$, the \emph{clusters}, and they can be constructed from an initial cluster by an operation known as \emph{mutation}. All their essential features can be illustrated in simple rank-two examples, where the initial cluster is $\{a_1,a_2\}$, the remaining clusters are $\{a_{i},a_{i+1}\}$ for diffent values of $i\ne 1$, the mutation operation is given by
\begin{equation}\label{eq:rank2mutation}
    a_{m+1}=
\tfrac{1+a^p_m}{a_{m-1}}\,,
\end{equation}   
and each positive integer value of $p$ defines a different cluster algebra. For $p=1$ it is easy to show that $a_6=a_1$, yielding a finite cluster algebra, i.e one with a finite number of variables and clusters. On the contrary, for $p\ge 2$ the mutation never lands us back to where we started, and we end up with an infinite cluster algebra. In any case, by iterating eq.~\eqref{eq:rank2mutation} it is evident that all cluster variables will be rational functions of $a_1$ and $a_2$.

$Gr(4,n)$ cluster algebras may be similarly defined as rank-$(3n-15)$ cluster algebras whose initial cluster variables are certain Pl\"ucker coordinates~\eqref{eq:Plucker_def}, so that all remaining variables will also be rational functions thereof. Building on the relevance of these cluster algebras for the integrand of $\cN=4$ SYM amplitudes~\cite{Arkani-Hamed:2012zlh}, in~\cite{Golden:2013xva} a remarkable observation was made: The symbol alphabet of integrated $n$-particle $\cN=4$ SYM amplitudes, in the appropriate normalisation, precisely coincides with the collection of all variables of the finite $Gr(4,n)$ cluster algebras for $n=6,7$.

It is of course natural to assume that the same pattern continues also for $n\ge 8$. If true, this would provide us with a prediction for the alphabet of any $\cN=4$ SYM amplitude, also extending to QCD by the DCI breaking mechanism we explained in the previous two chapters. However this assumption fails for the following two reasons: $Gr(4,n)$ cluster algebras
\begin{enumerate}
\item become infinite for $n\ge 8$, thereby losing any predictive power, and
\item cannot account for non-rational letters also involving square roots such as~\eqref{eq:Delta_4mb},
\end{enumerate}
that have been observed in several cases of Feynman integrals and scattering amplitudes.

The long-standing question of how to overcome these limitations was finally answered in the works~\cite{Arkani-Hamed:2019rds,Drummond:2019cxm,Henke:2019hve}, that appeared simultaneously. In short, point 1 above was resolved by introducing a stopping criterion for mutations, based on the relation of Grassmannians with simpler geometric objects retaining their essential features, which are known as tropical Grassmannians. From the finite subset of all $Gr(4,n)$ clusters and variables thus obtained, point 2 above is then resolved by considering \emph{infinite mutation sequences}, whose essence may be illustrated in the so called $A_1^{(1)}$ affine, or loosely speaking barely infinite, rank-two cluster algebra of eq.~\eqref{eq:rank2mutation} with $p=2$: By noticing that the quantity
\be\label{eq:A2_invariant}
\mathcal{P}=\frac{1+a_1^2+a_2^2}{a_1 a_2}=\frac{1+a_2^2+a_3^2}{a_2 a_3}=\ldots=\frac{1+a_i^2+a_{i+1}^2}{a_i a_{i+1}}\,, \;\;\forall \;\; i\in \mathbb{Z}\,,
\ee
is invariant under mutations, we can replace the nonlinear mutation rule with the linear recurrence
\be
a_{m+1}=a_m \mathcal{P}-a_{m-1}\,,
\ee
as may be readily verified by replacing $\mathcal{P}$ from~\eqref{eq:A2_invariant} with $i=m-1$. This recurrence now has a sensible $m\to \infty$ infinite mutation limit, where it reduces to the second-order equation,
\be
\beta =\mathcal{P}-\frac{1}{\beta}\,,\quad \beta=\lim_{m\to \infty} \frac{a_{m+1}}{a_m}\,,
\ee
which is readily solved by
\be
\beta_{\pm}=\frac{\mathcal{P}\pm \sqrt{\mathcal{P}^2-4}}{2}\,.
\ee
Evidently, we have succeeded in producing square roots by this limiting procedure! We note that different motivations had already let to its study in the mathematics community~\cite{Canakci2018,Reading2018b}.

In refs~\cite{Arkani-Hamed:2019rds,Drummond:2019cxm,Henke:2019hve}, this idea was applied to the first infinite $Gr(4,8)$ cluster algebra, by essentially looking at its $A_1^{(1)}$ subalgebras. How the latter are embedded in the former matters, generically giving different values for the resulting square roots. Rather than the strict limit, ref.~\cite{Drummond:2019cxm} additionally took the direction of approach into account, which has the effect of producing more letters with the same square root, that differ in the rational terms accompanying them. Support for the correctness of this prescription was provided by comparing the resulting letters with those of explicit amplitude calculations, and this was further reinforced by a different approach for generalising infinite cluster algebras, based on scattering diagrams~\cite{Herderschee:2021dez}.

\subsection{The structure of the alphabet and its reformulation}\label{subsec:pTr49AB}

The procedure we reviewed in the previous subsection for taming the infinities of $Gr(4,n)$ cluster algebras was applied to the $n=9$ case in~\cite{Henke:2021ity}, see also~\cite{Ren:2021ztg} for the rational letters and radicands of square-root letters. Let us here describe the structure of the candidate finite alphabet for nine-particle DCI scattering it has produced, which will be the starting point of our analysis.

The alphabet in question has dihedral as well as parity symmetry, where the former is reviewed in eqs~\eqref{eq:parity} and~\eqref{eq:parityZ} and the latter pertains to symmetry under cyclic transformations $Z_i\to Z_{i+1}$, enlarged by a dihedral flip such as $Z_{i}\to Z_{n+1-i}$. First, it contains 3078 $Gr(4,9)$ cluster variables, which by definition are rational in the Pl\"ucker coordinates. We stress that not all of these will remain rational when reexpressed in terms of momenta or cross ratios, because momentum twistors additionally rationalise the simple square roots of the pseudoscalar quantities of eqs.~\eqref{eq:epsilon_def},\eqref{eq:delta5}. So to avoid confusion, we will \emph{denote the parity-even and parity-odd letters that are rational in the Pl\"ucker coordinates as rational and rationalisable, respectively}. The 1737 rational and 1341 rationalisable letters may be found in the~\texttt{Gr49RationalDCI.m} ancillary file.

In addition, we have 2349 square-root letters of the form
\be\label{eq:Gr49NRLetters}
\frac{A_{ij}-\sqrt{\Delta_j}}{A_{ij}+\sqrt{\Delta_j}}\,,\quad j=1,\ldots 324\,,
\ee
arranged in 36 cyclic classes, where $A_{ij}$ and $\Delta_j$ are again rational in the Pl\"ucker coordinates, but not necessarily in the Mandelstam variables. In the ancillary file~\texttt{algebraicLetters-\allowbreak Representatives.m} representative letters of each dihedral class, containing the same square root, are given. The full dihedral class and square-root alphabet is obtained by cyclically permutting the momentum twistor labels of the representatives nine times. The form presented here is simpler that the one of~\cite{Henke:2021ity}, which had the feature that the square-root letters could be immediately obtained from the rational(isable) ones, but at the cost extra proportionality factors in the numerator and denominator of~\eqref{eq:Gr49NRLetters}, also absorbed inside the square root. These have now been removed, and the form of the radicands has been chosen to line up with that of~\cite{Ren:2021ztg}.

Finally, we note that if square-root letters of the form \eqref{eq:Gr49NRLetters} appear in a Feynman integral, then usually $\Delta_j$ also appear as rational letters of the latter. The procedure of the previous subsection does not automatically also provide the $\Delta_j$, but these can be clearly identified and added back in once we have the more complicated letters containing them. The same holds true also for parity-odd letters that have the same general form~\eqref{eq:Gr49NRLetters} in Mandelstam variables, even if their square roots rationalise in momentum twistors. We will denote these radicands in either choice of kinematic variables as \emph{trivially recoverable} letters. As pointed out in~\cite{Henke:2021ity}, the only difference in the final results of the scattering diagrams approach~\cite{Herderschee:2021dez} for $Gr(4,8)$ as compared to~\cite{Drummond:2019cxm} is the production of the trivially recoverable letters; it would be very interesting to explore if the same pattern holds for higher $n$.

\subsection{Results: Subalphabets for $(9-k)$-point $k$-mass DCI kinematics}\label{subsec:Subkinematic_alphabets}

\begin{table}
\centering
\setlength{\tabcolsep}{8pt} % Default value: 6pt
\renewcommand{\arraystretch}{2} % Default value: 1
\begin{tabular}{|c|c|c|c|c|c|}
\hline
Kinematics & $\substack{\text{Dimensionless}\\\text{variables}}$
      & $\substack{\text{Rational}\\\text{letters}}$
      & $\substack{\text{Rationalisable}\\\text{letters}}$
      & $\substack{\text{Square-root}\\\text{letters}}$
      \\ \hline

\multicolumn{1}{|c|}{\raisebox{-2ex}{\begin{tikzpicture}[scale=0.3]
    % Define the number of sides and the radius
    \def\n{9}
    \def\radius{1.5}

    % Draw the nonagon
    \foreach \i in {1,...,\n} {
        \coordinate (P\i) at ({\i*360/\n}:\radius);
    }
    \draw[thick] (P1) -- (P2)-- (P3) -- (P4)-- (P5) -- (P6)-- (P7)-- (P8) -- (P9)-- cycle ;
  \foreach \i in {1,...,\n} {
        \coordinate (Q\i) at ({\i*360/\n}:2);
    }
\draw[thick] (P1)--(Q1) ;
\draw[thick] (P2)--(Q2);
\draw[thick] (P3)--(Q3) ;
\draw[thick] (P4)--(Q4) ;
\draw[thick] (P5)--(Q5);
\draw[thick] (P6)--(Q6);
\draw[thick] (P7)--(Q7);
\draw[thick] (P8)--(Q8);
\draw[thick] (P9)--(Q9);
\node[scale=0.9] at (0,0) {DCI};
\end{tikzpicture}}} 
      & 12 & 1737 & 1341 & 2349 \\ \hline

\multicolumn{1}{|c|}{\text{BCFW shift}}
      & 11 & 979 & 271 & 692 \\ \hline

\multicolumn{1}{|c|}{\raisebox{-2ex}{\begin{tikzpicture}[scale=0.3]
    % Define the number of sides and the radius
    \def\n{8}
    \def\radius{1.5}
    \def\radius{1.5}
    \def\outer{2}
    \def\delta{10} % small angular offset in degrees

    % Draw the nonagon
    \foreach \i in {1,...,\n} {
        \coordinate (P\i) at ({\i*360/\n}:\radius);
    }
    \draw[thick] (P1)-- (P2)-- (P3) -- (P4)-- (P5) -- (P6)-- (P7)-- (P8) -- cycle ;
  \foreach \i in {1,...,\n} {
        \coordinate (Q\i) at ({\i*360/\n}:2);
    }
\draw[thick] (P1)--(Q1) ;
\draw[thick] (P3)--(Q3) ;
\draw[thick] (P4)--(Q4) ;
\draw[thick] (P5)--(Q5);
\draw[thick] (P6)--(Q6);
\draw[thick] (P7)--(Q7);
\draw[thick] (P8)--(Q8);
\node[scale=0.9] at (0,0) {DCI};

 \foreach \shift in {- \delta, \delta} {
        \draw[thick] 
            ({2*360/\n}:\radius) -- ({2*360/\n+\shift}:\outer);
    }
\end{tikzpicture}}} 
      & 10 & 421 & 271 & 395 \\ \hline

\color{blue}{\raisebox{1ex}{\begin{tikzpicture}[scale=0.3,baseline]
    % Define the number of sides and the radius
    \def\n{7}
    \def\radius{1.5}
    \def\outer{2}
    \def\delta{10} % small angular offset in degrees
    % Draw the nonagon
    \foreach \i in {1,...,\n} {
        \coordinate (P\i) at ({\i*360/\n}:\radius);
    }
    \draw[thick] (P1)-- (P2)-- (P3) -- (P4)-- (P5) -- (P6)-- (P7)-- cycle ;
  \foreach \i in {1,...,\n} {
        \coordinate (Q\i) at ({\i*360/\n}:2);
    }
\draw[thick] (P1)--(Q1) ;
\draw[thick] (P4)--(Q4) ;
\draw[thick] (P5)--(Q5);
\draw[thick] (P6)--(Q6);
\draw[thick] (P7)--(Q7);

 % Special double line for side 2:
    % We take the angle of vertex 2 and offset it by ±delta
    \foreach \shift in {- \delta, \delta} {
        \draw[thick] 
            ({2*360/\n}:\radius) -- ({2*360/\n+\shift}:\outer);
    }

     % Special double line for side 3:
    % We take the angle of vertex 3 and offset it by ±delta
    \foreach \shift in {- \delta, \delta} {
        \draw[thick] 
            ({3*360/\n}:\radius) -- ({3*360/\n+\shift}:\outer);
    }
    \node[scale=0.9] at (0,0) {DCI};

\end{tikzpicture}}
\raisebox{1ex}{$\sim$}
\raisebox{1ex}{\begin{tikzpicture}[scale=0.3,baseline]
    % Define the number of sides and the radius
    \def\n{6}
    \def\radius{1.5}
    \def\outer{2}
    \def\delta{10} % small angular offset in degrees

    % Draw the hexagon
    \foreach \i in {1,...,\n} {
        \coordinate (P\i) at ({\i*360/\n}:\radius);
    }
    \draw[thick] (P1)--(P2)--(P3)--(P4)--(P5)--(P6)--cycle;

    % Outer points
    \foreach \i in {1,...,\n} {
        \coordinate (Q\i) at ({\i*360/\n}:\outer);
    }

    % Normal radial lines
    \draw[thick] (P1)--(Q1);
    \draw[thick] (P3)--(Q3);
    \draw[thick] (P4)--(Q4);
    \draw[thick] (P5)--(Q5);
    \draw[thick] (P6)--(Q6);

    % Special double line for side 2:
    % We take the angle of vertex 2 and offset it by ±delta
    \foreach \shift in {- \delta, \delta} {
        \draw[thick] 
            ({2*360/\n}:\radius) -- ({2*360/\n+\shift}:\outer);
    }
    \node at (0,0) {LI};
\end{tikzpicture}}}
      & 8 & 119 & 59 & 68 \\ \hline

\multicolumn{1}{|c|}{\raisebox{-2ex}{\begin{tikzpicture}[scale=0.3]
    % Define the number of sides and the radius
    \def\n{7}
    \def\radius{1.5}
    \def\outer{2}
    \def\delta{10} % small angular offset in degrees
    % Draw the nonagon

    % Draw the nonagon
    \foreach \i in {1,...,\n} {
        \coordinate (P\i) at ({\i*360/\n}:\radius);
    }
    \draw[thick] (P1)-- (P2)-- (P3) -- (P4)-- (P5) -- (P6)-- (P7)-- cycle ;
  \foreach \i in {1,...,\n} {
        \coordinate (Q\i) at ({\i*360/\n}:2);
    }
\draw[thick] (P1)--(Q1) ;
\draw[thick] (P3)--(Q3);
\draw[thick] (P5)--(Q5);
\draw[thick] (P6)--(Q6);
\draw[thick] (P7)--(Q7);
\node[scale=0.9] at (0,0) {DCI};

 % Special double line for side 2:
    % We take the angle of vertex 2 and offset it by ±delta
    \foreach \shift in {- \delta, \delta} {
        \draw[thick] 
            ({2*360/\n}:\radius) -- ({2*360/\n+\shift}:\outer);
    }

 % Special double line for side 4:
    % We take the angle of vertex 4 and offset it by ±delta
    \foreach \shift in {- \delta, \delta} {
        \draw[thick] 
            ({4*360/\n}:\radius) -- ({4*360/\n+\shift}:\outer);
    }
\end{tikzpicture}}} 
      & 8 & 113 & 53 & 46 \\ \hline

\multicolumn{1}{|c|}{\raisebox{-2ex}{\begin{tikzpicture}[scale=0.3]
    % Define the number of sides and the radius
    \def\n{7}
    \def\radius{1.5}
    \def\delta{10}
    \def\outer{2}

    % Draw the nonagon
    \foreach \i in {1,...,\n} {
        \coordinate (P\i) at ({\i*360/\n}:\radius);
    }
    \draw[thick] (P1)-- (P2)-- (P3) -- (P4)-- (P5) -- (P6)-- (P7)-- cycle ;
  \foreach \i in {1,...,\n} {
        \coordinate (Q\i) at ({\i*360/\n}:2);
    }
\draw[thick] (P1)--(Q1) ;
\draw[thick] (P3)--(Q3);
\draw[thick] (P4)--(Q4);
\draw[thick] (P6)--(Q6);
\draw[thick] (P7)--(Q7);
\node[scale=0.9] at (0,0) {DCI};

 % Special double line for side 2:
    % We take the angle of vertex 2 and offset it by ±delta
    \foreach \shift in {- \delta, \delta} {
        \draw[thick] 
            ({2*360/\n}:\radius) -- ({2*360/\n+\shift}:\outer);
    }

 % Special double line for side 5:
    % We take the angle of vertex 5 and offset it by ±delta
    \foreach \shift in {- \delta, \delta} {
        \draw[thick] 
            ({5*360/\n}:\radius) -- ({5*360/\n+\shift}:\outer);
    }

\end{tikzpicture}}} 
      & 8 & 128 & 64 & 81 \\ \hline

\color{blue}{\raisebox{1ex}{\begin{tikzpicture}[scale=0.3,baseline]
    % Define the number of sides and the radius
    \def\n{6}
    \def\radius{1.5}
    \def\delta{10}
    \def\outer{2}

    % Draw the nonagon
    \foreach \i in {1,...,\n} {
        \coordinate (P\i) at ({\i*360/\n}:\radius);
    }
    \draw[thick] (P1)-- (P2)-- (P3) -- (P4)-- (P5) -- (P6)-- cycle ;
  \foreach \i in {1,...,\n} {
        \coordinate (Q\i) at ({\i*360/\n}:2);
    }
\draw[thick] (P1)--(Q1) ;
\draw[thick] (P5)--(Q5);
\draw[thick] (P6)--(Q6);
\node[scale=0.9] at (0,0) {DCI};

% Special double line for side 2:
    % We take the angle of vertex 2 and offset it by ±delta
    \foreach \shift in {- \delta, \delta} {
        \draw[thick] 
            ({2*360/\n}:\radius) -- ({2*360/\n+\shift}:\outer);
    }

 % Special double line for side 3:
    % We take the angle of vertex 3 and offset it by ±delta
    \foreach \shift in {- \delta, \delta} {
        \draw[thick] 
            ({3*360/\n}:\radius) -- ({3*360/\n+\shift}:\outer);
    }
 % Special double line for side 4:
    % We take the angle of vertex 4 and offset it by ±delta
    \foreach \shift in {- \delta, \delta} {
        \draw[thick] 
            ({4*360/\n}:\radius) -- ({4*360/\n+\shift}:\outer);
    }

\end{tikzpicture}}
\raisebox{1ex}{$\sim$}
\raisebox{1ex}{\begin{tikzpicture}[scale=0.3,baseline]
    % Define the number of sides and the radius
    \def\n{5}
    \def\radius{1.5}
    \def\outer{2}
    \def\delta{10} 

    % Draw the nonagon
    \foreach \i in {1,...,\n} {
        \coordinate (P\i) at ({\i*360/\n}:\radius);
    }
    \draw[thick] (P1)-- (P2)-- (P3) -- (P4)-- (P5)-- cycle ;
  \foreach \i in {1,...,\n} {
        \coordinate (Q\i) at ({\i*360/\n}:2);
    }
\draw[thick] (P1)--(Q1);
\draw[thick] (P4)--(Q4);
\draw[thick] (P5)--(Q5);

\foreach \shift in {- \delta, \delta} {
        \draw[thick] 
            ({2*360/\n}:\radius) -- ({2*360/\n+\shift}:\outer);
    }
\foreach \shift in {- \delta, \delta} {
        \draw[thick] 
            ({3*360/\n}:\radius) -- ({3*360/\n+\shift}:\outer);
    }

\node at (0,0) {LI};
\end{tikzpicture}}}
      & 6 & 39 & 12 & 23 \\ \hline

\color{blue}{\raisebox{1ex}{\begin{tikzpicture}[scale=0.3,baseline]
    % Define the number of sides and the radius
    \def\n{6}
    \def\radius{1.5}
    \def\delta{10}
    \def\outer{2}

    % Draw the nonagon
    \foreach \i in {1,...,\n} {
        \coordinate (P\i) at ({\i*360/\n}:\radius);
    }
    \draw[thick] (P1)-- (P2)-- (P3) -- (P4)-- (P5) -- (P6)-- cycle ;
  \foreach \i in {1,...,\n} {
        \coordinate (Q\i) at ({\i*360/\n}:2);
    }
\draw[thick] (P1)--(Q1) ;
\draw[thick] (P4)--(Q4) ;
\draw[thick] (P6)--(Q6);
\node[scale=0.9] at (0,0) {DCI};

% Special double line for side 2:
    % We take the angle of vertex 2 and offset it by ±delta
    \foreach \shift in {- \delta, \delta} {
        \draw[thick] 
            ({2*360/\n}:\radius) -- ({2*360/\n+\shift}:\outer);
    }

 % Special double line for side 3:
    % We take the angle of vertex 3 and offset it by ±delta
    \foreach \shift in {- \delta, \delta} {
        \draw[thick] 
            ({3*360/\n}:\radius) -- ({3*360/\n+\shift}:\outer);
    }
 % Special double line for side 5:
    % We take the angle of vertex 5 and offset it by ±delta
    \foreach \shift in {- \delta, \delta} {
        \draw[thick] 
            ({5*360/\n}:\radius) -- ({5*360/\n+\shift}:\outer);
    }

\end{tikzpicture}}
\raisebox{1ex}{$\sim$}
\raisebox{1ex}{\begin{tikzpicture}[scale=0.3,baseline]
    % Define the number of sides and the radius
    \def\n{5}
    \def\radius{1.5}
    \def\delta{10}
    \def\outer{2}

    % Draw the nonagon
    \foreach \i in {1,...,\n} {
        \coordinate (P\i) at ({\i*360/\n}:\radius);
    }
    \draw[thick] (P1)-- (P2)-- (P3) -- (P4)-- (P5)-- cycle ;
  \foreach \i in {1,...,\n} {
        \coordinate (Q\i) at ({\i*360/\n}:2);
    }
\draw[thick] (P1)--(Q1);
\draw[thick] (P3)--(Q3);
\draw[thick] (P5)--(Q5);

% Special double line for side 2:
    % We take the angle of vertex 2 and offset it by ±delta
    \foreach \shift in {- \delta, \delta} {
        \draw[thick] 
            ({2*360/\n}:\radius) -- ({2*360/\n+\shift}:\outer);
    }

 % Special double line for side 4:
    % We take the angle of vertex 4 and offset it by ±delta
    \foreach \shift in {- \delta, \delta} {
        \draw[thick] 
            ({4*360/\n}:\radius) -- ({4*360/\n+\shift}:\outer);
    }

\node at (0,0) {LI};
\end{tikzpicture}}}
      & 6 & 40 & 12 & 19 \\ \hline

\multicolumn{1}{|c|}{\raisebox{-2ex}{\begin{tikzpicture}[scale=0.3]
    % Define the number of sides and the radius
    \def\n{6}
    \def\radius{1.5}
    \def\delta{10}
    \def\outer{2}

    % Draw the nonagon
    \foreach \i in {1,...,\n} {
        \coordinate (P\i) at ({\i*360/\n}:\radius);
    }
    \draw[thick] (P1)-- (P2)-- (P3) -- (P4)-- (P5) -- (P6)-- cycle ;
  \foreach \i in {1,...,\n} {
        \coordinate (Q\i) at ({\i*360/\n}:2);
    }
\draw[thick] (P1)--(Q1) ;
\draw[thick] (P3)--(Q3);
\draw[thick] (P5)--(Q5);

% Special double line for side 2:
    % We take the angle of vertex 2 and offset it by ±delta
    \foreach \shift in {- \delta, \delta} {
        \draw[thick] 
            ({2*360/\n}:\radius) -- ({2*360/\n+\shift}:\outer);
    }

 % Special double line for side 4:
    % We take the angle of vertex 4 and offset it by ±delta
    \foreach \shift in {- \delta, \delta} {
        \draw[thick] 
            ({4*360/\n}:\radius) -- ({4*360/\n+\shift}:\outer);
    }
 % Special double line for side 6:
    % We take the angle of vertex 6 and offset it by ±delta
    \foreach \shift in {- \delta, \delta} {
        \draw[thick] 
            ({6*360/\n}:\radius) -- ({6*360/\n+\shift}:\outer);
    }

\node[scale=0.9] at (0,0) {DCI};
\end{tikzpicture}}} 
      & 6 & 30 & 6 & 0 \\ \hline

\raisebox{1ex}{\begin{tikzpicture}[scale=0.3,baseline]
    % Define the number of sides and the radius
    \def\n{5}
    \def\radius{1.5}
    \def\delta{10}
    \def\outer{2}

    % Draw the nonagon
    \foreach \i in {1,...,\n} {
        \coordinate (P\i) at ({\i*360/\n}:\radius);
    }
    \draw[thick] (P1)-- (P2)-- (P3) -- (P4)-- (P5)-- cycle ;
  \foreach \i in {1,...,\n} {
        \coordinate (Q\i) at ({\i*360/\n}:2);
    }

\draw[thick] (P5)--(Q5);

% Special double line for side 1:
    % We take the angle of vertex 1 and offset it by ±delta
    \foreach \shift in {- \delta, \delta} {
        \draw[thick] 
            ({1*360/\n}:\radius) -- ({1*360/\n+\shift}:\outer);
    }
% Special double line for side 2:
    % We take the angle of vertex 2 and offset it by ±delta
    \foreach \shift in {- \delta, \delta} {
        \draw[thick] 
            ({2*360/\n}:\radius) -- ({2*360/\n+\shift}:\outer);
    }

 % Special double line for side 3:
    % We take the angle of vertex 3 and offset it by ±delta
    \foreach \shift in {- \delta, \delta} {
        \draw[thick] 
            ({3*360/\n}:\radius) -- ({3*360/\n+\shift}:\outer);
    }
 % Special double line for side 4:
    % We take the angle of vertex 4 and offset it by ±delta
    \foreach \shift in {- \delta, \delta} {
        \draw[thick] 
            ({4*360/\n}:\radius) -- ({4*360/\n+\shift}:\outer);
    }

\node[scale=0.9] at (0,0) {DCI};
\end{tikzpicture}}
\raisebox{1ex}{$\sim$}
\raisebox{1ex}{\begin{tikzpicture}[scale=0.3,baseline]
    % Define the number of sides and the radius
    \def\n{4}
    \def\radius{1.5}
    \def\delta{10}
    \def\outer{2}

    % Draw the nonagon
    \foreach \i in {1,...,\n} {
        \coordinate (P\i) at ({\i*360/\n-45}:\radius);
    }
    \draw[thick] (P1)-- (P2)-- (P3) -- (P4)-- cycle ;
  \foreach \i in {1,...,\n} {
        \coordinate (Q\i) at ({\i*360/\n-45}:2);
    }
\draw[thick] (P4)--(Q4);

% Special double line for side 1:
    % We take the angle of vertex 1 and offset it by ±delta
    \foreach \shift in {- \delta, \delta} {
        \draw[thick] 
            ({1*360/\n-45}:\radius) -- ({1*360/\n-45+\shift}:\outer);
    }

% Special double line for side 2:
    % We take the angle of vertex 2 and offset it by ±delta
    \foreach \shift in {- \delta, \delta} {
        \draw[thick] 
            ({2*360/\n-45}:\radius) -- ({2*360/\n-45+\shift}:\outer);
    }

% Special double line for side 3:
    % We take the angle of vertex 3 and offset it by ±delta
    \foreach \shift in {- \delta, \delta} {
        \draw[thick] 
            ({3*360/\n-45}:\radius) -- ({3*360/\n-45+\shift}:\outer);
    }

\node at (0,0) {LI};
\end{tikzpicture}}
      & 4 & 12 & 0 & 8 \\ \hline
\end{tabular}

\caption{Dimensions of $(9-k)$-point $k$-mass DCI alphabets, with $k\ge 1$ derived here from the known $k=0$ results~\cite{Henke:2021ity} of the first line, as subspaces invariant under $2k$ operators. `BCFW shift' refers to the subspace invariant under one operator, see subsection~\ref{subsec:massivemomenta}. By breaking DCI, the {\color{blue}{colour-coded cases}} will yield predictions for QCD in the next sections.}
\label{tab:results}
\end{table}

As we recalled in the introduction, the alphabet of a canonical basis of with $n$-point integrals in the top sector necessarily also contains that of lower-point integrals appearing in lower sectors. Otherwise said, the space of logarithms of the former will contain that of the latter as a subspace, and here we will apply the same logic to the $(9-k)$-point $k$-mass DCI alphabet with $k=0$, analysed in the previous subsection, to find its subalphabets for $k=1,\ldots,4$.

In subsection~\ref{subsec:massivemomenta} we explained that functions of these lower-point kinematics are independent of certain of the parent kinematic variables, and formulated this invariance in terms of operators that must annihilate them. So the alphabet subspace we are after will simply be the kernel or nullspace of these operators, which in momentum twistor language were defined in eqs.~\eqref{eq:opdef}-\eqref{eq:opact}. More concretely, if $\mathcal{O}$ denotes an operator of the latter form, and our initial alphabet is $\{\alpha_i\}$, then we are looking for the values of the constants $c_i$ that satisfy
    \begin{equation}\label{eq:oneOperatorProblem}
      \sum_k c_k \mathcal{O}(\log(\alpha_k))= \sum_k c_k\frac{\mathcal{O}(\alpha_k)}{\alpha_k}=0  \;.
    \end{equation}
In our case, the initial alphabet is the one related to the $Gr(4,9)$ cluster algebra and reviewed in the previous section, whereas the operators depend on the particular kinematic configuration considered. For example, the operators corresponding to the $(9-k)$-point $k$-mass configuration with $k=1$ in the orientation shown in figure~\ref{fig:8pt1} are given in eq~\eqref{eq:ops8p1m}, those in the hard $k=2$ configuration of figure~\ref{fig:7pt2m} are given in eq~\eqref{eq:ops7p2m} and so on.

For each operator $\mathcal{O}_{ij}$ dictated by the kinematics, we solve a different linear problem of the form \eqref{eq:oneOperatorProblem}, where the letters $\alpha_i$ are expressed in terms of Pl\"ucker coordinates, and the action of $\mathcal{O}=\mathcal{O}_{ij}$ on them follows from the definition of the latter. After this has been calculated, in order to solve the equations we may evaluate the Pl\"ucker coordinates in a momentum twistor parameterisation such as~\eqref{eq:ZnWebVar} for $n=9$. Since the equations must hold for any value of the kinematic parameters, by looking at the factors of the independent functions appearing, in principle they may be reduced to equations with numeric coefficients among the $c_i$. However these functions are complicated high-degree polynomials in many variables times square roots, so one may equally well aim to solve them with exact arithmetic, after evaluating them in a sufficient number of kinematic points. This too requires care, as it inevitably leads to very large exact numbers that may in practice render the calculation infeasible. If only rational numbers appear, then this issue can be sidestepped with finite field methods that have e.g. been implemented in the 	\texttt{FINITEFLOW} \cite{Peraro:2019svx} code and accompanying \texttt{Mathematica} interface.

Given that no out-of-the-box solutions sufficed for the calculation of our linear system, which has $\mathcal{O}(10^4)$ coefficients and also contains square-roots, let us thus briefly summarise the main insights that made it possible (the reader solely interested in the final results may skip to the next paragraph):
\begin{enumerate}
\item It suffices to calculate the nullspace of a single operator $\mathcal{O}_{12}$, as the ones of the remaining $\mathcal{O}_{i\;i\pm 1}$ operators may be obtained by dihedral transformations.
\item The nullspace of multiple operators is just the intersection of the nullspaces of each.
\item The linear system of point 1 may be broken down into separate subsystems for the rational letters and for each set of letters~\eqref{eq:Gr49NRLetters} with the same square root, because
\begin{equation}\label{eq:oneOperatorProblemAlgebraic}
       \mathcal{O} \log(\frac{A-\sqrt{\Delta}}{A+\sqrt{\Delta}})= \frac{1}{\sqrt{\Delta}}\frac{2 \Delta \mathcal{O}(A)-\mathcal{O}(\Delta) A}{A-\Delta^2} \,,
\end{equation}
and each radicand $\Delta=\Delta_j$ is linearly independent of all radicands and rational factors.
\item All subsystems of point 3 can be solved by finite field methods, as eq.~\eqref{eq:oneOperatorProblemAlgebraic} implies that any non-rational terms will appear at most as overall factors $\Delta^{-1/2}$ that can be scaled out. 
\end{enumerate}
This reduction procedure was implemented in the ancillary \texttt{Mathematica} file~\texttt{alphabet_\allowbreak reduction.nb}, which outputs any subalphabet in any orientation, taking as input the initial alphabet of subsection~\ref{subsec:pTr49AB}, as well as the file \texttt{nullSpaceData.m} containing the precomputed nullspace of point 1 above.\footnote{For the square-root letters, instead of applying a single operator on all of them, we have equivalently applied all operators on a representative letter of each cyclic class.}

In this manner, we have produced all DCI subalphabets illustrated in table~\ref{tab:results}. We could also continue with subalphabets where more than two massless legs are joined to form a massive leg, but these are contained in the alphabet related to the $Gr(4,8)$ cluster algebra, which has already been analysed in~\cite{Chicherin:2020umh}. The first line in the table corresponds to the known nine-point massless DCI alphabet, whereas the second line to the nullspace of a single operator as explained in point 1 above, whose action was interpreted in subsection~\ref{subsec:massivemomenta} as (the momentum twistor analogue of) an infinitesimal BCFW shift. We also remind the reader that by the conventions of the previous subsection, rational (rationalisable) letters are parity even (odd) rational expressions of the Pl\"ucker coordinates.

Next, we will focus on the colour-coded cases, where the DCI breaking mechanism analysed in the previous two sections will allow us to obtain predictions for LI 5- and 6-point letters relevant for QCD.

\section{Candidate letters for six-point one-mass processes} \label{sec:(6,1)letters}

In the previous section we saw how to obtain alphabets for $(9-k)$-point $k$-mass DCI kinematics with $k\ge 1$, as subspaces of the known $k=0$ case. Here we will focus on the $k=2$ alphabet, also appearing as the top color-coded blue entry in table~\ref{tab:results}, which by the DCI breaking procedure we explained in sections~\ref{sec:Intro} and~\ref{sec:Kinematics} equivalently describes 6-point 1-mass LI kinematics.

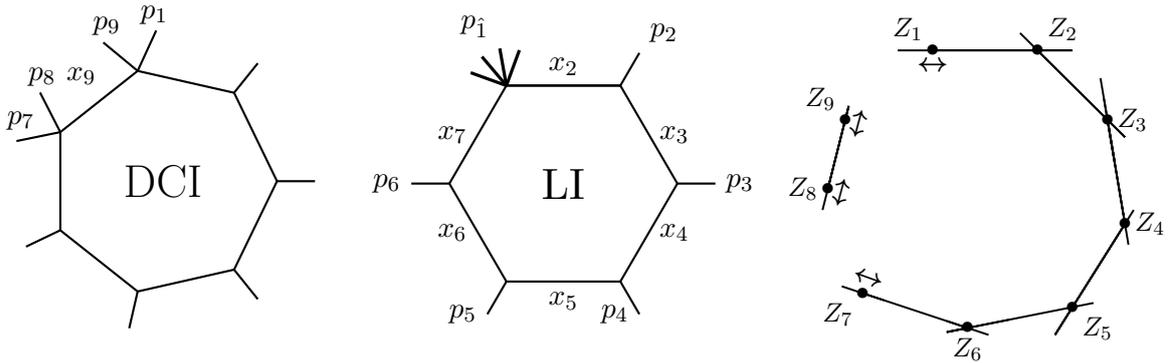
\begin{figure}
        \centering
        \begin{subfigure}[c]{0.31\textwidth}
       \raisebox{20pt}{\begin{tikzpicture}[scale=1]
    % Define the number of sides and the radius
    \def\n{7}
    \def\radius{1.5}
    \def\outer{2}
    \def\delta{10} % small angular offset in degrees
    % Draw the nonagon
    \foreach \i in {1,...,\n} {
        \coordinate (P\i) at ({\i*360/\n}:\radius);
    }
    \draw[thick] (P1)-- (P2)-- (P3) node[pos=0.4,above left] {$x_9$} -- (P4)-- (P5) -- (P6)-- (P7)-- cycle ;
  \foreach \i in {1,...,\n} {
        \coordinate (Q\i) at ({\i*360/\n}:2);
    }
\draw[thick] (P1)--(Q1) ;
\draw[thick] (P4)--(Q4) ;
\draw[thick] (P5)--(Q5);
\draw[thick] (P6)--(Q6);
\draw[thick] (P7)--(Q7);

\draw[thick] (P2)--({(2-0.2)*360/7}:2) node[midway,above,pos=0.9]{$p_1$};
\draw[thick] (P2)--({(2+0.2)*360/7}:2) node[midway,above,pos=0.9]{$p_9$};
\draw[thick] (P3)--({(3+0.2)*360/7}:2) node[midway,above,pos=0.9]{$p_7$};
\draw[thick] (P3)--({(3-0.2)*360/7}:2) node[midway,above,pos=0.9]{$p_8$};
    \node at (0,0) {\LARGE DCI};
\end{tikzpicture}}
\end{subfigure}
        \begin{subfigure}[c]{0.32\textwidth}
            \begin{tikzpicture}
    % Define the number of sides and the radius
    \def\n{6}
    \def\radius{1.5}

    % Draw the nonagon
    \foreach \i in {1,...,\n} {
        \coordinate (P\i) at ({\i*360/\n}:\radius);
    }
    \draw[thick] (P1) -- node[midway,above] {$x_{2}$} (P2) -- node[midway,left] {$x_{7}$} (P3);
    \draw[thick] (P6)-- node[midway,right] {$x_{3}$}  (P1)--(P2);
    \draw[thick] (P3) -- node[midway,left] {$x_{6}$}  (P4);
    \draw[thick] (P4) -- node[midway,below] {$x_{5}$} (P5);    
    \draw[thick] (P5) -- node[midway,right] {$x_{4}$}  (P6);

  \foreach \i in {1,...,\n} {
        \coordinate (Q\i) at ({\i*360/\n}:2);
    }
    \draw[thick] (P1) --(Q1) node[anchor=south west] {$p_{2}$};
\draw[very thick] (P2)  -- ++(70:0.5);
\draw[very thick] (P2)--++(100:0.5) node[anchor=south east] {$p_{\hat{1}}$};
\draw[very thick] (P2)--++(130:0.5);
\draw[very thick] (P2)--++(160:0.5);
\draw[thick] (P3) --(Q3)node[left] {$p_{6}$};
\draw[thick] (P4) --(Q4)node[left] {$p_{5}$};
\draw[thick] (P5) --(Q5)node[left] {$p_{4}$};
\draw[thick] (P6)--(Q6) node[right] {$p_{3}$} ;
\node[scale=1.5] at (0,0) {LI};
\end{tikzpicture}
        \end{subfigure}
        \hfill
        \begin{subfigure}[c]{0.32\textwidth}
            \scalebox{0.92}{\begin{tikzpicture}
\coordinate (P1) at (0,0);
\coordinate (P2) at (1.5,0);
\coordinate (P3) at (2.5,-1);
\coordinate (P4) at (2.75,-2.5);
\coordinate (P5) at (2,-3.7);
\coordinate (P6) at (0.5,-4);
\coordinate (P7) at (-1,-3.5);
\coordinate (P8) at (-1.5,-2);
\coordinate (P9) at (-1.25,-1);
\coordinate (Q3) at ({2.5+0.25*0.45},{-1-1.5*0.45});
\draw[thick] (P1) 
node[anchor=south east]{$Z_1$}
node {$\bullet$}
node[below] {$\boldsymbol{\leftrightarrow}$};
\draw[thick] (P2) 
node[anchor=south west]{$Z_{2}$}
node {$\bullet$};
\draw[thick] (P3) 
node[right]{$Z_{3}$}
node {$\bullet$};
\draw[thick] (P4) 
node[right]{$Z_{4}$}
node {$\bullet$};
\draw[thick] (P5) 
node[anchor=north west]{$Z_{5}$}
node {$\bullet$};
\draw[thick] (P6) 
node[below]{$Z_{6}$}
node {$\bullet$};
\draw[thick] (P7) 
node[anchor=north east]{$Z_{7}$}
node {$\bullet$}
node[above,rotate=-20] {$\boldsymbol{\leftrightarrow}$};
\draw[thick] (P8) 
node[left]{$Z_{8}$}
node {$\bullet$}
node[below,rotate=75] {$\boldsymbol{\leftrightarrow}$};
\draw[thick] (P9) 
node[anchor=south east]{$Z_{9}$}
node {$\bullet$}
node[below,rotate=75] {$\boldsymbol{\leftrightarrow}$};

\draw[thick] (P1) -- (P2) -- ++(0.5,0) -- ++(-2.5,0);
\draw[thick] (P2) -- (P3) -- ++(0.25,-0.25) -- ++(-1.5,1.5);
\draw[thick] (P3) -- ++({0.25*0.4},{-1.5*0.4}) -- ++({-0.25*0.8},{1.5*0.8});
\draw[thick] (P6) -- ++({(0.5-2)*0.2},{(-4+3.7)*0.2})-- ++({(0.5-2)*(-0.6)},{(-4+3.7)*(-0.6)});
\draw[thick] (P7) -- ++ ({(-1-0.5)*0.2},{(-3.5+4)*0.2});
\draw[thick] (P7) -- ++ ({(-1-0.5)*-0.2},{(-3.5+4)*-0.2});
\draw[thick] (P6) -- ++ ({(-1-0.5)*0.2},{(-3.5+4)*0.2});
\draw[thick] (P6) -- ++ ({(-1-0.5)*-0.2},{(-3.5+4)*-0.2});
\draw[thick] (P5) -- ++ ({(0.5-2)*-0.2},{(-4+3.7)*-0.2});
\draw[thick] (P6) -- (P7);
\draw[thick] (P8) --(P9) -- ++({(1.5-1.25)*0.2},{(2-1)*0.2})-- ++({(1.5-1.25)*(-1.5)},{(2-1)*(-1.5)});
\draw[thick] (P3) -- (P4)++(0.05,-0.3) -- ++(-0.35,2.1);;
\draw[thick] (P4) -- (P5) -- ++(-0.25,-0.4)-- ++({0.75*1.5},{1.2*1.5});
\draw[thick] (P5) -- (P6);

\end{tikzpicture}}
        \end{subfigure}
        \caption{Orientation of momenta/dual coordinates and momentum twistors for the 7-point 2-mass DCI or equivalently 6-point 1-mass LI alphabet, for $x_9\to \infty$, of section~\ref{sec:(6,1)letters}. $p_{\hat{1}}:=p_7+p_8+p_9+p_1$ and otherwise the labels of the left and center graphs coincide.}
        \label{fig:6ptonemass}
    \end{figure}

We choose the orientation of figure~\ref{fig:7pt2m} with $n=9$ and $r=1$, which for concreteness we also display in figure~\ref{fig:6ptonemass}. The relevant 6-point 1-mass Mandelstam variables are,
\begin{equation} \label{eq:sixpointonemassMand}
   \vec v_{6,1} =\{s_{\hat{1}2},s_{23},s_{34},s_{45},s_{56},s_{6\hat{1}},s_{\hat{1}23},s_{234},s_{345},p_{\hat{1}}^2\},
\end{equation}
where we have also adopted the convention
\be
p_{\hat{1}}:=p_7+p_8+p_9+p_1\,,
\ee
generalising to the other Mandelstam variables in an obvious manner, so as to line up the LI and DCI variables as shown in the picture. As we reviewed in subsection~\ref{subsec_Kin_Mandelstam}, not all the variables \eqref{eq:sixpointonemassMand} will be independent as in $d=4$ dimensions they obey the Gram determinant constraint $G({\hat{1}},2,3,4,5)=0$.

With the help of figure~\ref{fig:7pt2m}, it is not difficult to infer that the specialisation of the map between Mandelstam and momentum twistor variables~\eqref{eq:mandtobr} to our case reads,
\begin{align}
    & s_{{\hat{1}}2}  =\frac{\br{6723}}{\br{6789}\br{2389}} \quad && s_{23}  =\frac{\br{1234}}{\br{1289}\br{3489}} \quad
    &&& s_{34}  =\frac{\br{2345}}{\br{2389}\br{4589}} \nonumber \\
    & s_{45}  =\frac{\br{3456}}{\br{3489}\br{5689}}  \quad && s_{56}  =\frac{\br{4567}}{\br{4589}\br{6789}} \quad &&& s_{6{\hat{1}}} =\frac{\br{5612}}{\br{5689}\br{1289}}\label{eq:6p1mMand2br}\\ 
   & s_{{\hat{1}}23} =\frac{\br{6734}}{\br{6789}\br{3489}} \quad
    && s_{234}  =\frac{\br{1245}}{\br{1289}\br{4589}} 
    \quad &&& s_{345}=\frac{\br{2356}}{\br{2389}\br{5689}} \nonumber 
    \\ 
    & p_{{\hat{1}}}^2 = \frac{\br{6712}}{\br{6789}\br{1289}} \,,\nonumber
\end{align}
with the momentum twistor line $(Z_8,Z_9)$ corresponding to the point at infinity $X_\infty$.

By the general procedure described in subsection~\ref{subsec:BreakDCI}, we may construct the following momentum twistor parameterisation suitable for these kinematics,
\begin{equation}\label{eq:Z6p1m}
  \mathbf{Z}_{6,1} = \left(
    \begin{array}{ccccccccc}
     1 & 0 & 1 & 1 & 1 & 1 & 0 & 0 & 0 \\
     0 & 1 & 0 & 1 & x_1 & x_2 & 1 & 0 & 0 \\
     0 & 0 & 1 & 0 & x_3 & x_4 & x_5 & 1 & 0 \\
     0 & 0 & 0 & 1 & x_6 & x_7 & x_8 & 0 & 1 \\
    \end{array}
    \right)
\end{equation}
that evidently has the right number of 8 independent, dimensionless kinematic variables,\footnote{This is a representation where all Mandelstam invariants have been divided by $s_{23}$, or equivalently the latter has been sent to one.} and automatically satisfies the Gram determinant constraint.

As 7-point 2-mass DCI and 6-point 1-mass kinematics are described by the same momentum twistor configuration, the corresponding alphabets will look identical in this language. In the chosen orientation, they may be found in the \texttt{AB6p1mZ.m} ancillary file, where the non-rationalisable square-root letters have been organised into subsets according to which of the 11 square roots involved they contain. In order to reexpress the alphabet in terms of the more familiar Mandelstam variables~\eqref{eq:6p1mMand2br}, in principle we could pick a subset of 8 ratios thereof, evaluate them in terms of the parameterisation~\eqref{eq:Z6p1m}, solve these equations for the $x_i$ variables of the latter, and replace these in the same parameterisation of the entire alphabet. However this leads to overly complicated expressions, and in subsection~\ref{subsec:ZtoMand} we will provide an alternative method that we found more efficient for arriving at the desired result.

In any case, the parameterisation~\eqref{eq:6p1mMand2br}-\eqref{eq:Z6p1m} is perfectly sufficient for describing 6-point 1-mass LI kinematics, and it has the added benefit that it rationalises the square roots~\eqref{eq:epsilon_def}, \eqref{eq:delta5} of the Mandelstam variables, thus also all parity-odd letters containing them. Indeed, in the next subsection we will employ it to perform a sanity check of our candidate 6-point 1-mass alphabet.

\subsection{Comparison with the 1-loop alphabet}\label{subsec:1loop6p1m}

As a first test of our cluster-algebraic 6-point 1-mass prediction, we compare it with the alphabet of the 1-loop 1-mass hexagon integral, which may be straightforwadly derived as we show next. The reader solely interested in the comparison may directly skip to the last paragraph of this subsection.

When the external momenta are four-dimensional, the 1-loop hexagon is a sum of its pentagon subgraphs~\cite{Binoth:1999sp}, and the same will also hold for their respective alphabets, which are known~\cite{Abreu:2023rco,Abreu:2024yit}. Here we will instead be more general and obtain the alphabet for external momenta in $d>4$ dimensions, as a limit of respective result for the generic hexagon integral, computed and contained in the ancillary file of ref.~\cite{Dlapa:2023cvx}. Generic here means that all external legs and propagators are massive and different from each other, which gives rise to an alphabet of 63 rational and 391 square-root letters involving 11 square roots. The letters are naturally organised in classes associated to the graph or any of its subgraphs where a number of propagators has been contracted, so let us here display the former that are new. These have the form
\begin{align}\label{eq: letter one-contracted}
    W_{1,\ldots,(i-1),\ldots,{6}}&=
    \dfrac{\cY\begin{bmatrix}
        i\\1
    \end{bmatrix}-\sqrt{\cY\begin{bmatrix}
        i\\i
    \end{bmatrix}\cY\begin{bmatrix}
        1\\1
    \end{bmatrix}}}{\cY\begin{bmatrix}
        i\\1
    \end{bmatrix}+\sqrt{\cY\begin{bmatrix}
        i\\i
    \end{bmatrix}\cY\begin{bmatrix}
        1\\1
    \end{bmatrix}}},\quad i=2,\ldots 7,\\
    W_{1,\ldots,(i-1),\ldots,(j-1),\ldots,{6}}&=
    \dfrac{\cY\begin{bmatrix}
        1&j\\1&i
    \end{bmatrix}-\sqrt{-\cY\begin{bmatrix}
        1\\1
    \end{bmatrix}\cY\begin{bmatrix}
        1&i&j\\1&i&j
    \end{bmatrix}}}{\cY\begin{bmatrix}
        1&j\\1&i
    \end{bmatrix}+\sqrt{-\cY\begin{bmatrix}
        1\\1
    \end{bmatrix}\cY\begin{bmatrix}
        1&i&j\\1&i&j
    \end{bmatrix}}}\,,\quad 2\le i<j\le 7, \label{eq:letter_two_contracted}
\end{align}
and
\be
  \label{eq: rational letter}
        W_{1,2,\ldots,{6}}=\dfrac{\cY \begin{bmatrix}
        \cdot\\ \cdot
        \end{bmatrix}}{\cY\begin{bmatrix}
            1\\1
        \end{bmatrix}}\,,
\ee
where
\begin{equation}\label{eq:mCayley}
    \mathcal{Y}:=\begin{pmatrix}
        0&1&1&\cdots &1\\
        1&Y_{11}&Y_{12}&\cdots &Y_{1{6}}\\
        1&Y_{12} & Y_{22}&\cdots &Y_{2{6}}\\
        \vdots&\vdots&\vdots &&\vdots \\
        1&Y_{1{n}}&Y_{2{n}} &\cdots&Y_{{6}{6}}
    \end{pmatrix}\,,\qquad Y_{ij}=m_i^2+m_j^2-s_{i,i+1,\ldots ,j-1}\,,\quad \text{with } s_{i,i-1}:=0\,,
\end{equation}
is the  $n=6$ modified Cayley matrix, and we have also defined its minors where rows $i_1,\ldots i_k$ and columns $j_1,\ldots j_k$ have been removed as
\begin{equation}\label{eq:MinorNotation}
     \mathcal{Y}\begin{bmatrix}
        i_1&i_2&\cdots&i_k\\
        j_1&j_2&\cdots&j_k
    \end{bmatrix},\qquad 1\le i_1<i_2<\cdots i_k\le {7},\ 1\le j_1<j_2<\cdots j_k\le {7}\,,
\end{equation}
including the special case
\be\mathcal{Y}\begin{bmatrix}
    \cdot\\ \cdot
\end{bmatrix}:=\det\mathcal{Y}=-2^{5}\,G(1,\ldots,5)\,,
\ee
with the Gram determinant $G$ defined in eq.~\eqref{eq:grammdef}. Evidently, the 6+21 letters of eqs~\eqref{eq: letter one-contracted}-\eqref{eq:letter_two_contracted} contain square roots.

 Next, we take the limit where all internal masses and all but one external  momenta-squared vanish $m_i, p_{i\ne 1}, s_{12345}=p_6^2\to 0$,  which reduces the alphabet to 54 rational and 72 square-root letters containing 11 square roots. For concreteness, let us quote their radicands: First, we have four K\"allen polynomials related to massive triangle subgraphs,
 \begin{equation}\label{eq:triangeDs}
D_1 = \lambda(s_{23}, s_{45},s_{61})\,,\;D_2 = \lambda(s_{12},s_{34},s_{56}),  D_3 = \lambda(p_1^2,s_{234},s_{56})\,,\; D_4= \lambda(p_1^2, s_{23}, s_{456})\,, 
\end{equation}
\label{eq:kallen}
where the K\"allen polynomial is defined as
\begin{equation}
\lambda(a,b,c)=a^2+b^2+c^2 -2 a b - 2 a c - 2 b c \;.
\end{equation}
Then, we have six Gram determinants related to the pentagon subgraphs,
\be\label{eq:pentagonDs}
D_{i+4}=-\cY\begin{bmatrix}
        8-i\\8-i
    \end{bmatrix}=2^4 G(8-i,9-i,10-i,11-i)= \epsilon^2_{8-i,9-i,10-i,11-i}\,,\,\, i=1,\ldots,6\,,
\ee
where all indices of the Gram determinants are taken modulo 6, and their square roots are defined in eq.~\eqref{eq:epsilon_def}. Finally, we have the radicand that does not appear in any subgraphs
\be\label{eq:hexagonD}
D_{11}=-\cY\begin{bmatrix}
        1\\1
    \end{bmatrix}=\Delta_6^2,\ee
where we have also related it to the pseudoscalar defined in eq.~\eqref{eq:delta5}. Note that all of the above radicands also appear as rational letters of the alphabet. We additionally remark that the minors of eq.~\eqref{eq:letter_two_contracted} with three rows and columns removed become perfect squares in the 1-mass limit. As a consequence, only 4 out of the 15 genuinely 6-point letters of the latter equation remain multiplicatively independent.

Finally, we restrict our external momenta to four dimensions, which implies the vanishing of the five-particle Gram determinant of eq.~\eqref{eq:grammvanish}. The latter appears in the numerator of eq.~\eqref{eq: rational letter}, hence the letter in question also vanishes. Additionally, all letters of eq.~\eqref{eq: letter one-contracted} vanish or diverge, because for vanishing Gram determinant the radicand becomes the perfect square of the rational part of the letter. The same phenomenon has also been observed in the fully massless hexagon integral~\cite{Henn:2022ydo}. To arrive at these conclusions, and also to eliminate additional multiplicative dependencies, we have employed the inherently four-dimensional momentum twistor parametrisation of eq.~\eqref{eq:Z6p1m}. In this manner, we confirm that the finite letters~\eqref{eq:letter_two_contracted} are indeed no longer multiplicatively independent from the letters of the pentagon subgraphs, in agreement with~\cite{Binoth:1999sp}. Finding and eliminating multiplicative relations is simple if the letters are rational, and even in the presence of roots it has been implemented in the \texttt{Mathematica} package \texttt{SymBuild}~\cite{Mitev:2018kie}. For our calculations we also have developed a homemade implementation.

All in all, the alphabet of the four-dimensional 1-mass hexagon integral we have thus obtained consists of 53 rational and 47 square-root letters. These are included in the accompanying ancillary file~\texttt{AB6p1m1LDD.m}, together with the remaining 1+25 letters that are relevant beyond four dimensions. Using the kinematic parameterisation~\eqref{eq:6p1mMand2br}-\eqref{eq:Z6p1m}, we are now ready to compare the four-dimensional 1-loop letters with our cluster-algebraic prediction of 244 6-point 1-mass letters: We find that these contain all of the 1-loop letters, except for the rational letters consisting of the radicands~\eqref{eq:triangeDs}-\eqref{eq:pentagonDs}. Given that the square-root letters containing these radicands are contained in our prediction, as pointed out in section~\ref{subsec:pTr49AB} it is possible to simply add these rational letters in by hand. It is thus in this sense that the 1-loop alphabet is essentially contained in our prediction.

\subsection{Reexpressing the alphabet to Mandelstam variables}\label{subsec:ZtoMand}
In the preamble of this section we provided the momentum twistor parameterisation of our 6-point 1-mass alphabet, and in the previous subsection we applied it to successfully perform a 1-loop check. Given the difficulty in inverting the aforementioned parameterisation, let us here summarise an alternative method we employed in order to reexpress our alphabet to the more familiar Mandelstam variables.

Let us assume that we have a set of letters that are irreducible polynomials in the kinematic variables $x_1,\ldots, x_r$. For the purposes of this section these will be the variables of our momentum twistor parameterisation \eqref{eq:6p1mMand2br}-\eqref{eq:Z6p1m}, however we keep the discussion slightly more general since we also applied the method in the following sections. As long as the number of the letters we are considering is the same with the number of irreducible polynomials appearing in them, redefining our alphabet to consist of the latter is simply a change of basis and thus we have no loss of generality by requiring our letters to be irreducible polynomials.

Then, for every irreducible polynomial letter $q(x_1,\ldots,x_{8})$ we make an ansatz for it in terms of a homogeneous polynomial $p^{(n)}(s)$ of order $n$ in the Mandelstam variables, augmented by pseudoscalars $\epsilon_{I}$ \eqref{eq:epsilon_def}\footnote{As we mentioned in subsection~\ref{subsec_Kin_Mandelstam}, the $\epsilon_{I}$ are related and it is principle sufficient to consider one of them. However using an overcomplete set leads to simpler expressions, of possibly lower polynomial degree, justifying its selection.} multiplied by similar homogeneous polynomials of order of $n-2$,
\begin{equation} \label{eq:ansatz}
p^{(n)}(s)+\sum_{I} c^{(n-2)}_{I}(s) \epsilon_{I},
\end{equation}
with undetermined coefficients for each monomial term contained in them, and with each Mandelstam variable evaluated back in the $x_i$ variables by (relations analogous to) eqs.\eqref{eq:6p1mMand2br}-\eqref{eq:Z6p1m}.

Instead of directly matching $q(x_1,\ldots,x_{r})$ with its ansatz, we instead first find a kinematic subspace where $q=0$ by solving this equation in terms of the variable with the smallest order. If $q$ is linear in the variable in question then the solution will be rational, and if it is quadratic we may use the \texttt{RationalizeRoots} package \cite{Besier:2019kco} to find a rational reparameterisation. These two orders turned out sufficient for our purposes, and after finding a rational solution we plugged it back to the ansatz and demanded that it vanishes. The resulting equations for the free coefficients may be solved by finite field methods, and we end up with an expression~\eqref{eq:ansatz} that is proportional to the letter $q$.

Of course, this approach is not fully algorithmic since one still has to find the proportionality constant between $q(x_1,\ldots,x_{r})$ and its tentative expression in terms of Mandelstam variables. Furthermore, the fact that not all Mandelstam variables are independent implies that the solution will not be unique. However any particular choice suffices, and by evaluating the output for all letters in the set back in the $x_i$ variables, we may check whether they span the same space as the original set of letters, i.e. whether they provide an equivalent basis.

Applying this to the 119 rational and 59 rationalisable letters of our 6-point 1-mass LI alphabet, we have indeed succeeded in finding an equivalent basis in terms of Mandelstam variables, which is contained in the \texttt{AB6p1m.m} ancillary file. For the 68 non-rationalisable letters which have the form~\eqref{eq:Gr49NRLetters} we may follow the same procedure separately for their radicands $\Delta_{j}$ and rational(isable) terms $A_{ij}$. As we'll see in the next section, however, we can in fact first make our life simpler by restricting to the subset of letters with genuinely 6-point 1-mass kinematics.

\subsection{Eliminating subalphabets and identifying nested roots} 
\label{subsec:6p1m_purify}

Instead of deriving and presenting expressions in terms of Mandelstam variables for all 244 letters of our 6-point 1-mass LI or equivalently 7-point 2-mass DCI alphabet, we make the following observation: The nested manner with which this is part of the 9-point massless DCI alphabet implies that also 6-point DCI alphabets will be part of the latter. 

Referring to table~\ref{tab:results}, where the 7-point 2-mass DCI alphabet appears  as its first colour-coded blue entry, it is easy to figure out which subalphabets it will contain by deleting an edge of its kinematic visualisation in all possible ways. This yields subalphabets corresponding to two (different orientations of) 3-mass hard, two 3-mass easy, two 2-mass hard and one 1-mass 6-point DCI kinematics. There is no need to consider lower-point reductions separately, as these will already be included in the aforementioned subalphabets.

As we explained in section~\ref{sec:Kinematics}, the visualisation relevant for a given alphabet encodes which kinematic variables the latter is independent of, and this independence is tantamount to the annihilation by certain operators which take the form~\eqref{eq:opdef}-\eqref{eq:opact} in momentum twistor variables. So the 6-point DCI subalphabets will correspond to subspaces of the 7-point 2-mass DCI alphabet which are annihilated by additional operators. These subspaces may be easily computed with the commands implemented in the ancillary \texttt{Alphabet_reduction.nb} file, and we find that their union consists of 111 rational, 50 rationalisable and 46 non-rationalisable square-root letters.\footnote{We notice that in fact the 6-point 1-mass subalphabet is not needed to span this space.}

Examining the non-rationalisable letters that appear in this space of subalphabets, with the method of the previous subsection we find that the radicands $t_i$ appearing in the aforementioned subset contain those of the 1-loop triangle discussed in subsection~\ref{subsec:1loop6p1m}, $t_i\propto D_i$, $i=1,\ldots,4$, as well as
\begin{equation}
\begin{aligned}
t_{5} & =  p_1^4 \left(s_{45}-s_{345}\right){}^2-2 p_1^2 \left(s_{45}-s_{345}\right)+
\left(s_{12} s_{45}-\left(s_{23}+s_{123}\right) s_{345}\right)+\\
&+ \left(s_{12} s_{45}+\left(s_{23}-s_{123}\right) s_{345}\right){}^2, \\
t_{6} & =  p_1^4 \left(s_{23}-s_{61}\right){}^2-2 p_1^2 s_{61} \left(2 s_{23} s_{45}+\left(s_{61}-s_{23}\right) s_{123}\right)+s_{61}^2 s_{123}^2,\\
t_{7} & = \left(-p_1^2+s_{234}+s_{123}\right){}^2-4 s_{23} s_{56},\\
t_{8} & =p_1^4 s_{56}^2-2 p_1^2 s_{\hat{1}2} s_{56} \left(p_1^2-s_{234}+2 s_{34}\right)+s_{\hat{1}2}^2 \left(p_1^2-s_{234}\right){}^2 \\
t_{9} & = p_1^4 \left(s_{34}-s_{345}\right){}^2-2 p_1^2 \left(s_{34}-s_{345}\right) \left(s_{34} s_{61}-\left(s_{56}+s_{234}\right) s_{345}\right)+\\
&+ \left(s_{34} s_{61}+\left(s_{56}-s_{234}\right) s_{345}\right){}^2.
 \end{aligned}
\end{equation}
These may be indeed related to radicands of known 5-point 2-mass LI  letters~\cite{Abreu:2024yit}, upon appropriate identification of the kinematic variables.

Therefore it makes sense to restrict our attention to the \emph{complement of the space of lower-point subalphabets}. We denote the letters that span it as \emph{genuinely} 7-point 2-mass DCI or equivalently 6-point 1-mass LI, and by definition they are only defined modulo multiplicative combinations of the 111+50+46 subalphabet letters.

Continuing with the study of non-rationalisable letters, we now move on to identify the genuinely 6-point 1-mass LI ones and find that their radicands in momentum twistor variables have the form
\begin{align}
\Delta_{+}
& \propto
\Big(
  \br{1289}\br{3567}
  -\br{1267}\br{3589}
  +\br{1235}\br{6789}
\Big)^2
-4\,\br{1235}\br{1289}\br{3567}\br{6789}
 \,,\nonumber\\[8pt]
\Delta_{-}
& \propto
\Big(
  \br{1289}\br{2346}\br{4567}
  +\br{1267}\big(
      \br{2349}\br{4568}
      -\br{2348}\br{4569}
    \big)
  +\br{1234}\br{2456}\br{6789}
\Big)^2\nonumber
\\
&\quad
-4\,\br{1234}\br{1289}\br{2346}\br{2456}\br{4567}\br{6789}\,. \label{eq:6p1mRootsZ}
\end{align}
These may be rendered invariant under $Z_i\to \lambda_i Z_i$ in a canonical fashion by multiplying them with appropriate powers of Pl\"ucker coordinates of the form $\br{i\;i+1\;i+2\;i+3}$, owing to the fact that when the total number of twistors (including the infinity twistor line) is not a multiple of four, there exists a unique product picking up only one of the scaling factors $\lambda_i$. In our case, the combination which only picks a $\lambda_1$ scaling is specifically
\be
\left(\frac{\br{1234} \br{1239} \br{1289} \br{1789} \br{3456} \br{4567}\br{5678}}{\br{2345}^3 \br{6789}^3}\right)^{\frac{1}{4}}\,,
\ee
and the rest are obtained by cyclic permutations. After rendering $\Delta_+, \Delta_-$ projectively invariant in this fashion, we observe that these are mapped into one another under the parity transformation~\eqref{eq:parityZ}!

The fact that the $\Delta_\pm$ are not parity invariant implies that they themselves contain parity-odd rationalisable square roots, i.e. $\Delta_\pm$ are \emph{nested square roots} when expressed in terms of Mandelstam variables. As pointed out in subsection~\ref{subsec:results_summary}, these have been identified quite recently in the literature~\cite{FebresCordero:2023pww, Badger:2024fgb,Becchetti:2025oyb}, and only for integral families with massive internal particles that involve elliptic geometries and thus evaluate beyond the class of multiple polylogarithms considered here. Our work for the first time provides strong evidence that these nested roots appear also in purely polylogarithmic integrals with massless propagators, and even in the simple $\cN=4$ SYM theory.

As reviewed in subsection~\ref{subsec_Kin_Mandelstam}, all parity-odd quantities~$\epsilon_I, \Delta_i$ expressible as square roots in the Mandelstam variables are related. We may thus choose one of them when aiming to express $\Delta_\pm$ and the entire non-rationalisable square-root letter in Mandelstam variables, and here we have chosen $\epsilon_{2356}$ since it is the parity-odd quantity with lowest mass dimension that respects the surviving $p_2\leftrightarrow p_6, \; p_3\leftrightarrow p_5$ flip symmetry of our alphabet, as may be readily verified in figure~\ref{fig:7to6pt}. If a quantity can be written as
\be\label{eq:parity_decomposition}
H=F+\epsilon_{2356} G\,,
\ee
where $F, G$ are even under the action of the parity operator~\eqref{eq:parityOp}, $P\cdot F=F$ and similarly for $G$, then obviously
\be\label{eq:parity_components}
F=\frac{H+P\cdot H}{2}\,,\quad G=\frac{H-P\cdot H}{2\,\epsilon_{2356}}\,.
\ee
After finding a parity eigenbasis for the genuinely 7-point 2-mass DCI letters in terms of momentum twistors, we can make use of this decomposition with $H$ being their radicands $\Delta_\pm$ or their rational parts (analogous to $A_{ij}$ of eq.~\eqref{eq:Gr49NRLetters}) as follows: We evaluate the right-hand sides of the equalities~\eqref{eq:parity_components} with the help of the (parity conjugate of the) parameterisation~\eqref{eq:Z6p1m}, and check that these quantities are indeed parity even. This implies that the radicands and rational parts of the letters in question indeed have a decomposition of the form~\eqref{eq:parity_decomposition}. Then either the components $F, G$ separately, or if desired the entire letters, may be found with an appropriately tailored ansatz as described in subsection~\ref{subsec:ZtoMand}.

\subsection{New genuinely 6-point 1-mass letters}\label{subsec:6p1m_results}
So after the dust settles, we are ready to provide the genuinely 6-point 1-mass LI letters of our predicted alphabet, which do not belong to any 5-point 2-mass LI subalphabets. We have 9 parity even rational letters,
{\sloppy
\setlength{\jot}{4pt}   % spacing between aligned rows
\begin{align}
\alpha_{1} & = 
s_{23} s_{56}+s_{34} \left(s_{123}-s_{45}\right)+\left(s_{45}-s_{123}\right) s_{234},\nonumber
\\
\alpha_{2} & =
s_{34} \left(s_{45}-s_{123}\right) \left(p_{\hat{1}}^2-s_{61}\right)-\left(s_{23} s_{56}+\left(s_{45}-s_{123}\right) s_{234}\right) \left(s_{12}-s_{345}\right),\nonumber
\\
\alpha_{3} & = 
s_{234} \left(s_{45} \left(s_{12}-p_{\hat{1}}^2\right)+s_{123} \left(s_{61}-s_{345}\right)\right)+s_{34} \left(s_{45} \left(p_{\hat{1}}^2-s_{12}\right)+s_{123} \left(s_{345}-s_{61}\right)\right)+\nonumber
\\
&+s_{23} s_{56} \left(s_{345}-s_{61}\right),\nonumber
\\
\alpha_{4}& = 
s_{34} \left(p_{\hat{1}}^2-s_{61}\right) \left(p_{\hat{1}}^2+s_{45}-s_{61}-s_{123}\right)+\left(p_{\hat{1}}^2-s_{56}-s_{61}\right) (s_{45} \left(p_{\hat{1}}^2-s_{12}-s_{61}\right)+\nonumber
\\
&+s_{345} (-p_{\hat{1}}^2+s_{61}+s_{123})),\nonumber
\\
\alpha_{5} & = 
s_{45} s_{61} \left(s_{61} \left(p_{\hat{1}}^2 \left(p_{\hat{1}}^2+s_{34}-s_{56}\right)+s_{12} \left(s_{56}-p_{\hat{1}}^2\right)\right)+p_{\hat{1}}^2 s_{234} \left(s_{12}-p_{\hat{1}}^2\right)\right)+\nonumber
\\
&+s_{345} \left(p_{\hat{1}}^2 s_{23}+s_{61} \left(s_{123}-p_{\hat{1}}^2\right)\right) \left(s_{61} \left(p_{\hat{1}}^2-s_{56}\right)-p_{\hat{1}}^2 s_{234}\right),
\\
\alpha_{6} & = s_{34} \left(s_{12} \left(-2 p_{\hat{1}}^2+s_{23}-s_{45}+s_{61}\right)+p_{\hat{1}}^2 \left(p_{\hat{1}}^2+s_{45}-s_{61}\right)+s_{23} \left(s_{61}-p_{\hat{1}}^2\right)+s_{12}^2\right)+\nonumber
\\
&+\left(p_{\hat{1}}^2-s_{12}-s_{234}\right) \left(s_{45} \left(p_{\hat{1}}^2-s_{12}\right)+s_{345} \left(-p_{\hat{1}}^2+s_{12}+s_{23}\right)\right),\nonumber
\\
\alpha_{7} & =
p_{\hat{1}}^2 s_{56} s_{123} s_{345}-p_{\hat{1}}^2 s_{12} \left(s_{34} s_{123} \left(s_{61}-p_{\hat{1}}^2\right)+s_{345} \left(s_{56} \left(p_{\hat{1}}^2-s_{23}\right)+s_{123} \left(p_{\hat{1}}^2-s_{234}\right)\right)\right)-\nonumber
\\
&-s_{12}^2 \left(s_{34} \left(p_{\hat{1}}^2 \left(p_{\hat{1}}^2+s_{45}-s_{61}\right)+s_{23} \left(s_{61}-p_{\hat{1}}^2\right)\right)-s_{345} \left(p_{\hat{1}}^2-s_{23}\right) \left(p_{\hat{1}}^2-s_{234}\right)\right),\nonumber
\\
\alpha_{8} & = s_{45}(s_{56}(-p_{\hat{1}}^2 \left(p_{\hat{1}}^2-s_{61}+s_{123}\right)+s_{12} \left(p_{\hat{1}}^2-s_{23}-s_{61}+s_{123}+s_{234}\right)+\nonumber
\\
&+s_{23} \left(p_{\hat{1}}^2-s_{61}+s_{345}\right))+s_{56}^2 \left(p_{\hat{1}}^2-s_{12}\right)+s_{123} \left(p_{\hat{1}}^2-s_{61}\right) \left(p_{\hat{1}}^2-s_{12}-s_{234}\right))-\nonumber
\\
&-s_{34}(s_{56} \left(p_{\hat{1}}^2 s_{45}+s_{23} \left(s_{61}-p_{\hat{1}}^2\right)+\left(p_{\hat{1}}^2-s_{61}\right) \left(p_{\hat{1}}^2-s_{123}\right)\right)-\nonumber
\\
&-s_{123} \left(p_{\hat{1}}^2-s_{61}\right) \left(p_{\hat{1}}^2+s_{45}-s_{123}-s_{234}\right))+\nonumber
\\
&+(p_{\hat{1}}^2-s_{56}-s_{61})(s_{56} \left(p_{\hat{1}}^2-s_{23}-s_{123}\right)+\nonumber
\\
&+s_{123} \left(-p_{\hat{1}}^2+s_{123}+s_{234}\right))s_{345},\nonumber
\end{align}
}
{\sloppy
\setlength{\jot}{4pt}   % spacing between aligned rows
\begin{flalign*}
\begin{split}
\alpha_{9} & =s_{45}(s_{56}^2 \left(p_{\hat{1}}^2-s_{12}\right)+s_{123} \left(p_{\hat{1}}^2-s_{61}\right) \left(p_{\hat{1}}^2-s_{12}-s_{234}\right)+s_{56}(-p_{\hat{1}}^2 \left(p_{\hat{1}}^2-s_{61}+s_{123}\right)+
\\
&+s_{12} \left(p_{\hat{1}}^2-s_{23}-s_{61}+s_{123}+s_{234}\right)+s_{23} \left(p_{\hat{1}}^2-s_{61}+s_{345}\right)))-
\\
&-s_{34}(s_{56} (p_{\hat{1}}^2 s_{45}+s_{23} \left(s_{61}-p_{\hat{1}}^2\right)+\left(p_{\hat{1}}^2-s_{61}\right) \left(p_{\hat{1}}^2-s_{123}\right)-
\\
&-(p_{\hat{1}}^2-s_{61})s_{123}p_{\hat{1}}^2+s_{45}-s_{123}-s_{234})+p_{\hat{1}}^2-s_{56}-s_{61})\cdot
\\
& \cdot s_{345} \left(s_{56} \left(p_{\hat{1}}^2-s_{23}-s_{123}\right)+s_{123} \left(-p_{\hat{1}}^2+s_{123}+s_{234}\right)\right)\,,
\end{split}
\end{flalign*}
}
which evidently have degrees between 2 and 5 in the Mandelstam variables.

Next, we have 8 parity-odd rationalisable letters, expressible as ratios where the numerator and denominator products of between four and seven factors, each of which is of order 2, 3 or 4 in the Mandelstam variables. We found it simpler to express them in terms of six $\epsilon_I$ as well as $\Delta_6$, but since they remain quite complicated we have chosen to include them in the ancillary file \texttt{AB6p1mGenuine.m} together with the rational letters we quoted above. 

Finally, we have 6+6 nonrationalisable letters in parity conjugate pairs,
\begin{equation} \label{eq:nested_letters2}
l_{\pm}^{i} = \frac{A_{i}\pm B_{i}\epsilon_{2356}+C_{i}\sqrt{\Delta_{\pm}}}{{A_{i}\pm B_{i} \epsilon_{2356}-C_{i}\sqrt{\Delta_{\pm}}}},\quad i=1,\ldots,6, \quad P\cdot  l_{\pm}^{i}=l_{\mp}^{i}\,,
\end{equation}
containing nested square roots also related by parity,
\be
\Delta_{\pm }  = F(s) \pm G(s) \epsilon_{2356},\quad P\cdot \Delta_{\pm }=\Delta_{\mp }
\ee
where the quantities $A_i, B_i, C_i, F, G$ are parity even, rational expressions of the Mandelstam variables, and we recall that the parity-odd quantities $\epsilon_{ijkl}$ have been defined in eqs.~\eqref{eq:epsilon_def}-\eqref{eq:eps_to_Gram}. Momentum twistor expressions for them (and in fact for the entire alphabet) may be found in the ancillary file~\texttt{AB6p1mZ.m},\footnote{In particular, these appear as the last two lists of letters in the \texttt{AB6p1mAlgZ} variable, defined in the file in question.} which may be readily evaluated in the parameterisation~\eqref{eq:6p1mMand2br}-\eqref{eq:Z6p1m} of the 6-point 1-mass Mandelstam variables. We remind the reader that Pl\"ucker coordinates $\br{ijkl}$ simply correspond to the minors of the parameterisation matrix $\mathbf{Z}_{6,1}$ by virtue of~\eqref{eq:Plucker_def}, and that $\Delta_{\pm }$ have already been expressed in terms of them in eq.~\eqref{eq:6p1mRootsZ}, up to simple factors. Explicit, though lengthy expressions in terms of Mandelstam variables have also been obtained for the radicands $\Delta_\pm$ as well as for a full letter and its parity conjugate. These may be found in the accompanying ancillary file~\texttt{nestLetter6p1m.m}.

We are confident that our alphabet predictions will aid the calculation of the full planar master integrals for 6-point 1-mass QCD processes. Particularly for the nested square-root letters~\eqref{eq:nested_letters2}, we expect that these originate from pentagon-triangle topologies. This is supported by the fact that in the massless limit of our alphabet $\Delta_+/\Delta_-$ becomes proportional to  letter $\alpha_{190}$ of the two-loop massless six-point alphabet~\cite{Henn:2024ngj}, which appears in the massless pentagon-triangle sector as well. Further evidence in this direction comes from~\cite{Badger:2024fgb}, where nested roots appear also appear in pentagon-triangle topologies, this time with five external points where two of them, as well as certain propagators connecting them, have the same nonzero mass. More concretely, it is plausible that the pentagon-triangle topologies yielding the nested square-root letters are those illustrated in figure~\ref{fig:pent_triangles}, since the roots in question have the same up-down flip symmetry.

Finally, the strong indication that our results provide for the presence of nested square-root letters in polylogarithmic integrals in general, has broader implications for the existing approaches for determining the alphabet directly from the integrals~\cite{Jiang:2024eaj,Matijasic:2024too,Effortlessxxx,Correia:2025yao}: It is our understanding that these approaches only work for or assume letters that are either rational or of the form~\eqref{eq:Gr49NRLetters} in Mandelstam variables, implying that nested square root letters~\eqref{eq:nested_letters2} are out of reach for them. A possibility for overcoming this limitation is extending them to momentum twistor variables, where as we have seen the nested square root letters do take the simpler form~\eqref{eq:Gr49NRLetters}.

\section{Candidate letters for six-point massless processes}\label{sec:6p0m}

Having obtained a cluster-algebraic prediction for the alphabet of 6-point 1-mass integrals in the previous section, it is of course natural to consider its massless limit. In all cases examined, this limit indeed yields the correct alphabet of the corresponding massless integrals, see e.g.~\cite{Dlapa:2023cvx}. There is also evidence that this limit makes sense even at the level of the master integrals, since it belongs to the kinematic subspace of their Landau equations, where it is known that their system of differential equations drops rank~\cite{Bitoun:2017nre,Fevola:2023kaw}.

Here we will thus analyse this limit and obtain candidate 6-point massless letters, which will allow us to further check our approach by comparing with existing results in the literature~\cite{Abreu:2024fei,Henn:2025xrc}, as well as explore the possibility of providing new predictions beyond them. In subsection~\ref{subsec:masslesslim} we describe the procedure of taking the massless limit using momentum twistor variables, and the corresponding expressions it leads to. Then in subsection~\ref{subsec:6p0m_cyclic} we carry out the cyclic completion of the resulting alphabet and transcribe it to Mandelstam variables, which will be necessary for comparing with the literature. Finally we address the latter task and present the novel letters we find in subsection~\ref{subsec:6p0m_results}.

 \subsection{The massless limit with momentum twistors} \label{subsec:masslesslim}

By eq.~\eqref{eq:6p1mMand2br} we have that $p_{\hat{1}}^2\propto \br{6712}$, so the limit $p^2_{\hat{1}}\to 0$ amounts to $Z_1,Z_2,Z_6,Z_7$ belonging to the same plane, or equivalently to lines $(Z_1,Z_2)$ and $(Z_6,Z_7)$ intersecting. This is also in agreement with the momentum twistor geometry for massless and 1-mass kinematics illustrated in figure~\ref{fig:6pt}, where it is evident that if before the limit the kinematics are invariant under shifts of $Z_1$ and $Z_7=Z_{n-2}$ along the aforementioned lines, without loss of generality we can choose these two twistors to coincide in the strict limit. We may parameterise this in a continuous fashion as    
\begin{equation}
        \label{eq:masslesslim}
    Z_{7}=Z_1 +\textcolor{blue}{\delta}  Z_{*} \;,\quad \text{for any}\; Z_{*} \; \text{such that} \; \br{126*}\neq 0\;,
    \end{equation}
    with the strict limit corresponding to $\delta\to 0$.    We may choose $Z_{*}=c_3 Z_3+c_4 Z_4+c_5 Z_5$ for arbitrary $c_i$, and make sure that in the end the result does not depend on them, after we also trade $\delta$ for the dimensionless parameter
    \be
    \delta'=\frac{\br{*289}}{\br{1289}}\delta\;.
    \ee
    
We take the limit~\eqref{eq:masslesslim} directly in the expressions of the 244 6-point 1-mass letters in terms of Pl\"ucker coordinates, and keep only the leading term in the series expansion in $\delta$. The only potential issue that might arise is that certain terms might secretely vanish because of the Pl\"ucker relations~\eqref{eq:pluckrel}. The few cases where this happens can be identified and manually set to zero by testing whether keeping the leading term in $\delta$ commutes with evaluating the eight surviving twistors at a random kinematic point of any $Gr(4,8)$ parameterisation. As some radicands become perfect squares in the strict $\delta\to 0$ limit, whether the numerator or denominator of the non-rationalisable square-root letters of the form~\eqref{eq:Gr49NRLetters} vanishes there will depend on the choice of branch. We ensure we pick the same branch for all kinematic points by using a parameterisation in terms of so called cluster $x$-coordinates, which for positive values always yield a positive quantity inside the square.

Applying this procedure to the 119+59 rational and rationalisable 1-mass letters, we find that for only one we need to expand to $\mathcal{O}(\delta)$, otherwise the rest remain finite in the limit. Eliminating relations among them, the produce 134 rational(isable) 6-point massless letters. Similarly, in the limit the 68 non-rationalisable square-root letters reduce to 6 rational(isable) and 20 non-rationalisable, multiplicatively independent letters. Further distinguishing between rational and rationalisable letters on the basis of their parity properties, all in all we end up with 95 rational, 45 rationalisable and 20 non-rationalisable 6-point massless letters, where the latter involve 3 square roots. These may be found in the \texttt{AB6p0mZ.m} file accompanying this submission.

\subsection{Translation to Mandelstam variables and cyclic completion}\label{subsec:6p0m_cyclic}

To describe 6-point massless kinematics we choose the following set of planar Mandelstam variables,
\begin{equation}\label{eq:6p0mMand}
  \vec v_{6,0}=  \{s_{12},s_{23},s_{34},s_{45},s_{56},s_{61},s_{123},s_{234},s_{345}\}\;.
\end{equation}
which, as in the 1-mass case, obey a Gram determinant constraint among them. Their relations to momentum twistor variables follow simply from the massless limit~\eqref{eq:masslesslim} of the respective 1-mass relations~\eqref{eq:6p1mMand2br},
\begin{align}\label{eq:6p0mMand2br}
    & s_{12}  =\frac{\br{6123}}{\br{6178}\br{2378}} \quad && s_{23}  =\frac{\br{1234}}{\br{1278}\br{3478}} \quad
    &&& s_{34}  =\frac{\br{2345}}{\br{2378}\br{4578}} \nonumber \\
    & s_{45}  =\frac{\br{3456}}{\br{3478}\br{5678}}  \quad && s_{56}  =\frac{\br{4561}}{\br{4578}\br{6178}} \quad &&& s_{61}  =\frac{\br{5612}}{\br{5678}\br{1278}}\\ 
    & s_{123}  =\frac{\br{6134}}{\br{6178}\br{3478}} \quad
    && s_{234}  =\frac{\br{1245}}{\br{1278}\br{4578}} 
    \quad &&& s_{345}=\frac{\br{2356}}{\br{2378}\br{5678}}\,, \nonumber
\end{align}
where we have dropped the hatted index on the now massless $p_1$ for simplicity, as well as relabelled the twistors $Z_8\to Z_7$ and $Z_9\to Z_8$ of the infinity twistor line.

The momentum twistor parameterisation of 6-point massless kinematics \cite{Henn:2021cyv},
\begin{equation}\label{eq:Z6p0m}
   \mathbf{Z}_{6,0} = \left(
    \begin{array}{cccccccc}
     1 & 0 & x_1 & x_1 x_2 & x_1 x_3 & x_1 x_6 & 0 & 0 \\
    0 & 1 & 1 & x_8 & 1 & 1 & 0 & 0 \\
    0 & 0 & 0 & 1 & x_4 & 1 & 0 & 1 \\
    0 & 0 & 1 & 0 & x_5 & x_7 & -1 & 0 \\
    \end{array}
    \right)\,,
\end{equation}
is perfectly sufficient for describing the corresponding alphabet prediction we obtained in the previous subsection. An equivalent basis in terms of the more global Mandelstam variables~\eqref{eq:6p0mMand} may be obtained by the method of subsection~\ref{subsec:ZtoMand}.

Finally, we perform the cyclic completion of our prediction. The latter is expected to apply to planar integrals and amplitudes, owing to the origin of our method in planar $\cN=4$ SYM theory. Planar amplitudes of the identical particles, e.g. gluons, have this symmetry under cyclic permutations of the external legs. So it is natural to consider the closure of our predicted 6-point massless alphabet, which will also allow us to more directly compare with the literature.

The symmetry in question amounts to invariance under a cyclic permutation $p_i\to p_{i+1}$ with $p_7\equiv p_1$ in terms of momenta. From the momentum twistor geometry for massless kinematics illustrated on the left of figure~\ref{fig:6pt}, as well as relations~\eqref{eq:6p0mMand2br}, it is evident that in momentum twistors it is equivalent to keeping  $Z_7, Z_8$ of the line at infinity unchanged while shifting the rest, $Z_i\to Z_{i+1}$, $i=1,\ldots,6$ with $Z_7\equiv Z_1$. The cyclic completion results from the application of this shift six times, which we may carry out in either set of variables. In the end it is always simpler to employ the momentum twistor parameterisation (\eqref{eq:6p0mMand2br} and~\eqref{eq:Z6p0m}), as it contains fewer square roots and facilitates the elimination of the multiplicative dependencies among letters. The thus resulting cyclic completion of our 6-point massless alphabet consists of 225 rational, 114 rationalisable and 35 non-rationalisable square-root letters which may be found in terms of Mandelstam variables in the accompanying ancillary file~\texttt{AB6p0m.m}. In addition to the rationalisable roots $\epsilon_I, \Delta_6$ defined in \eqref{eq:epsilon_def},\eqref{eq:delta5}, these also contain the non-rationalisable roots,
\be
\begin{aligned}
    r_{1}  &= \sqrt{\lambda(s_{12},s_{34},s_{56})}\,, \quad
    r_{2}  =  \sqrt{\lambda(s_{23},s_{45},s_{61})}\,, \\
    r_{3}  &= \sqrt{\lambda(s_{12},s_{12}+s_{45}-s_{123}-s_{345},s_{45})},\quad r_{4}  = T r_{3}, \quad  r_{5}  = T r_{4},
\end{aligned}
\ee
where $T p_{i}= p_{i+1}$ with indices cyclically identified is the operator of the aforementioned cyclic permutations.

\subsection{
  Comparison with~\cite{Abreu:2024fei,Henn:2025xrc} and new 6-point massless letters}\label{subsec:6p0m_results}

We now check our cluster-algebraic prediction against existing results in the literature. Canonical differential equations (CDE) for 6-point 2-loop massless planar integrals contributing to generic gauge theories have been derived in \cite{Abreu:2024fei,Henn:2025xrc} in the limit of four-dimensional external and internal momenta. The latter restriction is sufficient for computing the finite parts of respective 2-loop amplitudes, and significantly simplifies the calculation since the hardest integrals contributing can be shown to essentially start at $\mathcal{O}(\epsilon)$ in the dimensional regulator, thus decoupling in the strict four-dimensional limit where $\epsilon\to 0$.

The canonical differential equations in $D=d=4$ obtained in this manner contain 245 letters, however by constructing the space of functions containing the finite part of the amplitude in fact reveals that only 167 of them contribute. For completeness we will carry out the comparison with the 245, henceforth denoted `CDE' letters, also monitoring the subset of 167 `amplitude' letters.

We have checked that the alphabets \cite{Abreu:2024fei} and \cite{Henn:2025xrc} are equivalent, and for concreteness we will carry our comparison with the latter reference. This comparison is most simply carried out in the momentum twistor parameterisation~\eqref{eq:6p0mMand2br}-\eqref{eq:Z6p0m}, since many square roots rationalise. We take the union of the logarithms of the two alphabets and, as already mentioned in subsection~\ref{subsec:1loop6p1m}, find linear relations between them e.g. with (a homemade version of) the approach~\cite{Mitev:2018kie}.

Our findings are summarised in table~\ref{tab:(6,0)comparison} for the different types of letters contained in the two alphabets. In addition to rational, rationalisable and square-root letters as defined in subsection~\ref{subsec:pTr49AB}, the alphabet~\cite{Henn:2025xrc} also contains overall roots $\sqrt{\Delta}$ as letters, which we list in different categories depending on whether they rationalise in momentum twistors or not. Since $2\log\sqrt \Delta=\log \Delta$ we can trade them for their rational radicands.

\begin{table}
    \centering
    \begin{tabular}{|l|c|c|c|c|}
    \hline
     \# 6-pt massless letters in:    & Amplitude &  CDE &  This work  &
         New letters  \\ \hline
         Rational &  99 \phantom{(}\checkmark\phantom{(} & 117\phantom{(}\checkmark\phantom{(} &   225 & 108   \\ \hline
         Rationalisable (RB) &  36 \phantom{(}\checkmark\phantom{(} &  72  & 114 & 60 \\ \hline
      RB overall root \{$\epsilon_{ijkl}, \Delta_6$\}  & 7 \textcolor{gray}{(\checkmark)}& 16 \textcolor{gray}{(\checkmark)}&  -  & -     \\
       \hline
       Non-rationalisable (NR) & 23 \phantom{(}\checkmark\phantom{)}  & 35 \checkmark  &  35  & -    \\
       \hline
       NR overall root $\{r_i\}$ & 2 \textcolor{gray}{(\checkmark)} & 5 \textcolor{gray}{(\checkmark)}&  -  & -    \\
       \hline
       Total&  167& 245 & 374 & 168\\
       \hline
    \end{tabular}
    \caption{Comparison of different types of letters from amplitude function space/CDE of~\cite{Abreu:2024fei,Henn:2025xrc} and this work. Matching and trivially recoverable letters, as defined in subsection~\ref{subsec:pTr49AB}, are denoted by $\checkmark$ and \textcolor{gray}{(\checkmark)}, respectively. The last column contains our new predictions. }
    \label{tab:(6,0)comparison}
\end{table}

With respect to the amplitude function space, we obtain all letters (denoted with a tick mark) except for these overall square roots, which however by the logic of the aforementioned section are trivially recoverable. Namely since we obtain letters of the form~\eqref{eq:Gr49NRLetters} containing the same square roots, we can easily add them back in. It is in this sense that our prediction essentially contains the letters of the amplitude function space! We denote trivially recoverable letters with a gray tick mark in parentheses.

It is interesting to point out that while the rationalisable trivially recoverable letters are part of the amplitude function space, they may in fact drop out from the final amplitude: In the 5-particle case, these have a single analogous letter $\epsilon_{1234}$ which has been indeed observed to have this property. This phenomenon has a cluster-algebraic interpretation~\cite{Chicherin:2020umh}, since cluster algebras and their generalisations can only produce quantities that are always positive in a subregion of the Euclidean kinematic region.  So there is mounting evidence that even in general gauge theories, amplitudes might share some of the positivity properties first discovered in $\cN=4$ SYM theory~\cite{Arkani-Hamed:2012zlh,Arkani-Hamed:2013jha}. Further support in our case comes from more recent calculations of 6-point massless quantities both in $\cN=4$ SYM theory~\cite{Carrolo:2025pue} and in QCD~\cite{Carrolo:2025agz,Carrolo:2026qpu}, which suggests that a subset of 137 out of the 167 amplitude letters is sufficient for the MHV sector.  Excitingly, this subset indeed does not contain the 9 trivially recoverable letters appearing in the relevant column of table~\ref{tab:(6,0)comparison}!

Moving to the comparison with the superset of CDE letters, we see that we in fact also obtain all of its rational and non-rationalisable letters. The only letters beyond trivially recoverable ones we don't predict are 18 rationalisable letters of the form\footnote{These (or more precisely their 6-point 1-mass analogues) also appear to change sign in the positive region.}
\be\label{eq:6p0m_missed}
\sum_{I,J}s_I \epsilon_J \in W_{[139,156]}\,,
\ee
for some planar Mandelstam invariants $s_I$ and parity-odd cross-products of momenta $\epsilon_J$. In the above equation we have also provided them in the conventions of~\cite{Henn:2025xrc}, where $[i,j]$ denotes all letters $W_k$ with $k=i,\ldots j$.  For completeness, we also note that in the latter conventions $\{\epsilon_{ijkl}\}=W_{[123,137]}$, $\Delta_6=W_{138}$ and $\{r_i\}=W_{[118,122]}$.

The fact that we reproduce 206 CDE letters of~\cite{Abreu:2024fei,Henn:2025xrc} (and can recover another 21 from them) provides very strong support for the validity of our approach. We remark that there also exists an alternative approach aiming to predict alphabets for $n$-particle massless LI scattering by identifying a so-called partial flag variety inside the $Gr(n-2, 2n-4)$ cluster algebra~\cite{Bossinger:2022eiy,2638233, Bossinger:2024apm}, and it has been recently applied to $n=6$~\cite{Pokraka:2025ali,Bossinger:2025rhf}. An advantage in this case is its direct relation to $Gr(4,8)$, rather than the limit of the much more intricate $Gr(4,9)$ considered here. A disadvantage is that it lacks a transparent stopping criterion in agreement with the data for the mutations, so at this stage it can only offer postdictions. In any case, it is also interesting to compare the latter with our findings. We observe what the two approaches have in common the 18+16 rationalisable letters they don't predict, i.e. those of eq.~\eqref{eq:6p0m_missed} as well as the overall roots~\{$\epsilon_{ijkl}, \Delta_6$\}. However \cite{Pokraka:2025ali,Bossinger:2025rhf} additionally miss another 36 $\mathcal{O}(\epsilon^0)$ and $\mathcal{O}(\epsilon^{-1})$ letters which our work here does predict. In the opposite direction, the letters of the latter reference which we do not directly reproduce are the 5 non-rationalisable roots $\{r_i\}$, but as already mentioned these may be trivially recovered from the rest.

Finally, as shown in the rightmost column of table~\ref{tab:(6,0)comparison} we also find 168 letters beyond those of the 2-loop master integral CDE we compared with. They are arranged in 28 cyclic orbits of length 6, so we will simply quote a single letter for each. The first orbit is generated by
\be\label{eq:erroneous_letter}
\beta_0=s_{124}\,,
\ee
and interestingly it has appeared in the earlier computation of 6-point 2-loop planar master integrals on the maximal cut~\cite{Henn:2021cyv}, though erroneously: In the list of letters quoted in this reference, those at positions 49-66 and 88-100 actually do not appear in the differential equations.\footnote{We thank Antonela Matijašić, Julian Miczajka and  Yang Zhang for clarifications on this point.} The orbit generated by~\eqref{eq:erroneous_letter} is in particular at positions 49-54.

The letters of the remaining 27 orbits are genuinely new. We note that a similar generation 5-point massless predictions from the $Gr(4,8)$ cluster algebra~\cite{Chicherin:2020umh} did not produce any overcomplete output that was not contained in the alphabet of the 2- and 3-loop master integrals~\cite{Chicherin:2025mvc}. This makes us optimistic that our new letters will appear in six-point massless integrals at higher loops. The first 17 cyclic classes are consist of polynomial letters of degrees 2,3 and 4,
\begin{align*}
    \beta_{1}& = s_{12} s_{45}-\left(s_{45}+s_{56}\right) s_{345},
    \\
    \beta_{2}& = s_{12} s_{45}-\left(s_{34}+s_{45}\right) s_{123},
    \\
     \beta_{3}& =  s_{12} \left(s_{34}+s_{45}\right)-\left(s_{34}+s_{123}\right) s_{345},
     \\
      \beta_{4}& = s_{12} \left(s_{45}+s_{56}\right)-s_{123} \left(s_{56}+s_{345}\right),
     \\
     \beta_{5}& =s_{12} s_{45}-s_{345} \left(-s_{45}+s_{123}+s_{345}\right),
     \\
     \beta_{6}& =s_{12} s_{45}-s_{123} \left(-s_{45}+s_{123}+s_{345}\right),
     \\
     \beta_{7}& =s_{12} s_{45}-s_{23} \left(s_{34}-s_{345}\right)-s_{123} \left(s_{345}-s_{34}\right),
     \\
     \beta_{8} & = s_{12} s_{45} \left(s_{56}+s_{345}\right)+\left(s_{34}-s_{56}-s_{345}\right) s_{345} \left(-s_{45}+s_{123}+s_{345}\right),
     \\
      \beta_{9} & = s_{12} s_{45} \left(s_{34}+s_{123}\right)-s_{123} \left(s_{34}-s_{56}+s_{123}\right) \left(-s_{45}+s_{123}+s_{345}\right),
      \\
      \beta_{10} & = s_{12} \left(-s_{61} s_{34}+\left(s_{234}-s_{56}\right) s_{34}+s_{56} s_{345}\right)+\left(s_{34}-s_{56}\right) \left(s_{56}-s_{234}\right) s_{345},
      \\
       \beta_{11} & = s_{12} s_{61} s_{45}+s_{61} \left(s_{23} \left(s_{45}-s_{345}\right)-s_{45} s_{123}\right)+\left(s_{23}-s_{45}\right) \left(s_{23}-s_{123}\right) s_{345},
      \\
     \beta_{12} & = s_{12} \left(s_{61} s_{34}+\left(s_{56}-s_{34}\right) s_{234}\right)+\left(s_{34}-s_{56}\right) \left(s_{34}-s_{234}\right) \left(s_{34}-s_{56}-s_{345}\right),
     \\
      \beta_{13} & = s_{12} s_{45} \left(-s_{23}+s_{45}+s_{234}\right)+\left(s_{34}+s_{45}\right) \left(s_{23} \left(s_{45}+s_{56}\right)-s_{123} \left(s_{45}+s_{234}\right)\right),
      \\
      \beta_{14} & = s_{12} s_{45} \left(-s_{61}+s_{45}+s_{234}\right)+\left(s_{45}+s_{56}\right) \left(s_{61} \left(s_{34}+s_{45}\right)-\left(s_{45}+s_{234}\right) s_{345}\right),
      \\
      \beta_{15} & = s_{12} s_{61} s_{45}+\left(s_{23}-s_{45}\right) \left(-s_{45}^2+\left(s_{123}+s_{345}\right) s_{45}+s_{23} \left(s_{45}-s_{123}\right)+s_{123} \left(s_{61}-s_{345}\right)\right), 
      \\
      \beta_{16} & = s_{12} \left(s_{45} s_{23}^2+\left(s_{34} \left(s_{45}+s_{56}\right)+s_{45} \left(s_{56}-s_{123}-s_{234}\right)-s_{56} s_{345}\right) s_{23}+s_{34} s_{123} \left(s_{61}-s_{45}-s_{234}\right)\right)-\\
      & -\left(s_{23} \left(s_{56}+s_{123}\right)-s_{123} \left(s_{123}+s_{234}\right)\right) \left(\left(s_{23}+s_{34}\right) s_{345}-s_{61} s_{34}\right),
      \\
      \beta_{17} & = s_{12} \left(s_{45} s_{61}^2+\left(s_{34} \left(s_{45}+s_{56}-s_{123}\right)+s_{45} \left(s_{56}-s_{234}-s_{345}\right)\right) s_{61}+s_{56} \left(s_{23}-s_{45}-s_{234}\right) s_{345}\right) + \\
      & +\left(s_{23} s_{56}-\left(s_{61}+s_{56}\right) s_{123}\right) \left(s_{61} \left(s_{34}+s_{345}\right)-s_{345} \left(s_{234}+s_{345}\right)\right).
\end{align*}

We also have 10 cyclic orbits of rationalisable letters. The letters generating the first 8 are
\begin{equation}
\beta_{17+i} =\frac{F_{i}-\epsilon_{1234}}{F_{i}+\epsilon_{1234}}\,, \quad i=1,\ldots,8\,,
\end{equation}
where the polynomials involved are defined as
\begin{align}
F_{1} & = s_{12} \left(s_{23}-s_{234}\right)+s_{123} \left(s_{234}-s_{34}\right)+s_{23} \left(s_{56}-s_{34}\right),
\\
F_{2} & = s_{12} \left(s_{23}+s_{234}\right)-s_{23} \left(s_{34}+s_{56}\right)+s_{123} \left(s_{34}-s_{234}\right),
\\
F_{3} & = s_{12} \left(s_{234}-s_{61}\right)-\left(s_{56}-s_{234}\right) \left(2 s_{23}+s_{345}\right)+s_{61} \left(-2 s_{23}-s_{34}+s_{56}\right),
\\
F_{4} & = s_{12} \left(s_{234}-s_{61}\right)-\left(s_{56}-s_{234}\right) \left(2 s_{234}+s_{345}\right)+s_{61} \left(-s_{34}+s_{56}-2 s_{234}\right),
\\
F_{5} & = s_{12} \left(s_{23}-s_{234}\right)+s_{23} \left(-s_{34}+s_{56}+2 s_{234}\right)+\left(s_{34}-s_{234}\right) \left(s_{123}+2 s_{234}\right),
\\
F_{6} & = s_{12} \left(s_{61}-2 s_{45}-s_{234}\right)+\left(2 s_{45}+s_{56}+s_{234}\right) s_{345} -s_{61} \left(s_{34}+2 s_{45}+s_{56}\right),
\\
F_{7} & = 2 s_{12}^2+\left(-2 s_{34}+s_{45}-2 \left(s_{123}+s_{345}\right)\right) s_{12}+s_{45} s_{56}-s_{56} s_{345}+s_{123} s_{345}+\\
&+ s_{34} \left(-s_{45}+s_{123}+2 s_{345}\right),
\\
F_{8} & = 2 s_{12}^2-s_{45} s_{56}+s_{34} \left(s_{45}-s_{123}\right)+2 s_{56} s_{123}+s_{56} s_{345}+s_{123} s_{345} +
\\
& +s_{12}\left(s_{45}-2\left(s_{56}+s_{123}+s_{345}\right)\right),
\end{align}
The remaining two cyclic orbits of rationalisable letters are more complicated, involving also $\Delta_6$. They can be generated by
\begin{equation}\label{eq:masslessdeltaletters}
\beta_{13} =\frac{Q_{0}-Q_{1}\Delta_{6}-Q_{2}\epsilon_{1234}}{Q_{0}+Q_{1}\Delta_{6}+Q_{2}\epsilon_{1234}}, \quad \beta_{14} =\frac{Q_{3}-Q_{4}\Delta_{6}-Q_{5}\epsilon_{1234}-Q_{6}\epsilon_{1236}}{Q_{3}+Q_{4}\Delta_{6}+Q_{5}\epsilon_{1234}+Q_{6}\epsilon_{1236}}
\end{equation}
where
\begin{align*}
Q_{1} & =s_{23}+s_{34}-s_{234},
\\
Q_{2} & =s_{23} s_{56}-s_{123} s_{234},
\\
Q_{0} & =s_{12} (-\left(s_{45}+s_{123}\right) s_{234}^2+\left(s_{23} \left(s_{45}+s_{56}+s_{123}\right)+s_{34} \left(s_{45}+2 s_{123}\right)\right) s_{234}-
\\
& \quad - s_{23} \left(s_{23}+2 s_{34}\right) s_{56}) +s_{56} \left(-s_{34}+s_{56}+s_{345}\right) s_{23}^2+(s_{123} (s_{34} \left(s_{56}-s_{61}\right)+
\\
& \quad +\left(s_{34}-2 s_{56}\right) s_{234})+\left(s_{34} s_{56}-\left(s_{56}+s_{123}\right) s_{234}\right) s_{345}) s_{23}-\\
& \quad -s_{123} (s_{34}-s_{234}) \left(s_{61} s_{34}+s_{234} \left(s_{123}+s_{345}\right)\right),
\\
Q_{4} & =s_{23}-s_{45}-s_{234}+s_{345},
\\
Q_{5} & =s_{61} s_{45}-s_{61} s_{345}+s_{234} s_{345},
\\
Q_{6} & = -s_{61} s_{34}-s_{56} s_{34}+s_{234} s_{34}-s_{23} s_{56}+s_{45} s_{234}+s_{56} s_{234},
\\
Q_{3} & =s_{61}\left(s_{123} \left(s_{34} \left(s_{345}-s_{56}\right)+s_{234} \left(s_{56}+s_{345}\right)\right)-s_{23} \left(\left(s_{34}-s_{56}\right) \left(s_{45}-s_{123}\right)+s_{56} s_{345}\right)\right)-
\\
& \quad -s_{61}^2 s_{34} s_{123}+s_{12} (s_{34} s_{61}^2-(s_{34} \left(s_{45}-s_{56}+s_{234}\right)+s_{23} \left(s_{34}+s_{45}-s_{56}-s_{345}\right)+
\\
& \quad +s_{234} \left(s_{56}+s_{345}\right)) s_{61}-s_{23}^2 s_{56}+s_{45} \left(s_{34}-s_{234}\right) \left(s_{234}-s_{56}\right)+s_{23} (s_{34} \left(s_{234}-s_{56}\right)-
\\
& \quad -s_{45} \left(s_{56}-2 s_{234}\right)+s_{234} \left(s_{56}-s_{345}\right))+s_{234} \left(s_{45}+s_{234}\right) s_{345})+
\\ 
&\quad +s_{345} (s_{23} \left(-s_{34} s_{56}+s_{123} s_{56}+s_{234} s_{56}+s_{345} s_{56}-s_{123} s_{234}+s_{45} \left(s_{234}-s_{56}\right)\right)+
\\
& \quad +s_{123} \left(s_{34} s_{56}-s_{234} \left(s_{56}+s_{345}\right)\right)),
\\
\end{align*}

While some of these letters appear to have complicated expressions, it is intriguing to note that in certain cases they can be expressed much more elegantly in terms of momentum twistors. For example, the numerator of \eqref{eq:masslessdeltaletters} essentially becomes 
\begin{equation}
\br{ 1236}  \br{1278}  \br{ 1345}  \br{ 3578}  -\br{345 (378)\cap (12)} \br{ 1235}  \br{ 1678}\;.
\end{equation}
We derived this expression by first identifying all the shift symmetries $Z_i\partial_{Z_j}$  of the letter and then constructing an ansatz that makes these manifest.

\section{Candidate letters for five-point two-mass processes}
\label{sec:(5,2) letters)}

In this subsection, we study more closely the alphabets appearing as the lower two colour-coded blue cases in table \ref{tab:results}, which are suitable for 6-point 3-mass hard and medium DCI kinematics. By the DCI breaking mechanism explained in the first two sections, these apply equally well to 5-point 2-mass hard and easy LI kinematics, respectively, and we also compare our results in detail with existing calculations of the corresponding 2-loop planar integrals. This comparison can be more easily carried out after considering a particular symmetrisation of the letters with respect to the external particle labels, which we will also describe.

\subsection{The two inequivalent configurations}

We will consider our 6-point 3-mass DCI alphabets as further reductions of the 7-point 2-mass DCI alphabet in the orientation shown in the left of figure~\ref{fig:6ptonemass}, where two of the massive legs will always be $p_7+p_8$ and the $p_9+p_1$. For the hard case, the third massive leg will be $p_5+ p_6$, whereas for the medium case $p_4+p_5$. This introduces additional operators of the form~\eqref{eq:opdef} annihilating the respective subalphabets, giving rise to the counts seen on the lower two colour-coded blue entries of table~\ref{tab:results}.

%the leg $p_{\hat{ 1}}$ is always massive. For the hard configuration we will specifically also make leg  $p_{\hat{5}}$ massive, whereas in the easy one this will be $p_{\hat{4}}$. 

In complete analogy with the 7-point 2-mass hard DCI case, the alphabet of 6-point 3-mass medium DCI kinematics equivalently also describes 5-point 2-mass easy LI kinematics, up to properly relating the variables of the two. We will come back to this momentarily, but in any case we  obtain 40 rational, 12 rationalisable and 19 non-rationalisable 5-point 2-mass easy letters.

For the 6-point 3-mass hard DCI kinematics however, as we already mentioned in the introduction and illustrated in figure~\ref{fig:6hto5}, we may now choose either point $x_9$ (as we did so far) or point $x_7$ as the point at infinity. This amounts to two different maps between the DCI alphabet in terms of momentum twistors, but the two are related by the transformation $x_7\to x_9$ or $Z_9\to Z_6$ and $Z_8\to Z_7$ in terms of momentum twistors. Instead of two different maps from the same DCI alphabet, it is thus equivalent but simpler to complete the latter under the aforementioned transformation and just use one of the two maps.\footnote{As the DCI alphabet is invariant under the flip symmetry $Z_i\to Z_{6-i}$, it can be shown that the transformation exchanging the two possible choices of the point at infinity is equivalent to the flip symmetry of the 5-point 2-mass LI configuration with respect to the momenta.} After eliminating multiplicative dependencies between the 6-point 3-mass hard DCI alphabet and its image under the $Z_9\to Z_6$ and $Z_8\to Z_7$ transformation, we end up with a 5-point 2-mass hard LI alphabet of 55 rational, 17 rationalizable and 31 non-rationalisable letters. Both 5-point 2-mass hard and easy letters obtained this way may be found in terms of Pl\"ucker coordinates in the \texttt{AB6p2mZ.m} ancillary file.

Finally, we come to spell out the map between Pl\"ucker coordinates and Mandelstam invariants. To this end, we will first uniformise our conventions such that in both easy and hard configurations the relevant such invariants become
\begin{equation}\label{eq:5p2m_final_Mandelstams}
  \vec v_
  {5,2}=  \left\{s_{12},s_{23},s_{34},s_{45},s_{51},p_4^2,p_5^2\right\},
\end{equation}
in line with the existing literature. For the hard case, we go from the new variables~\eqref{eq:5p2m_final_Mandelstams} to the previous ones by the cyclic shift $p^{\text{new}}_{i}= p^{\text{old}}_{(i \bmod 5) + 1}, \; i=1,\ldots,5$ of the LI momenta. For the easy case, we first apply $p_{3} \leftrightarrow p_{4}$ to \eqref{eq:5p2m_final_Mandelstams} and then the same transformation as for the hard case.

Leaving the momentum twistor labels unchanged with respect to what we had before, their relation to the uniform Mandelstam invariants we just defined are
\begin{equation}
\label{eq:sToBrE}
\begin{aligned}
s_{12}= & \frac{\br{1234}}{\br{1289}\br{3489}}
\qquad
s_{34}=\frac{\br{3467}}{\br{3489}\br{6789}}
\qquad
s_{51}=\frac{\br{2367}}{\br{2389}\br{6789}}
\\[0.8em]
s_{23}= & -\frac{\br{2356}}{\br{2389}\br{5689}}
+\frac{\br{3456}}{\br{3489}\br{5689}}
+\frac{\br{2367}}{\br{2389}\br{6789}}
-\frac{\br{3467}}{\br{3489}\br{6789}}
\\[0.4em]
s_{45}= & \frac{\br{1234}}{\br{1289}\br{3489}}
-\frac{\br{1256}}{\br{1289}\br{5689}}
+\frac{\br{3456}}{\br{3489}\br{5689}}
+\frac{\br{1267}}{\br{1289}\br{6789}}-\frac{\br{3467}}{\br{3489}\br{6789}}
\\[0.8em]
p_4^2= &\frac{\br{3456}}{\br{3489}\br{5689}}
\qquad
p_5^2=\frac{\br{1267}}{\br{1289}\br{6789}}\,.
\end{aligned}
\end{equation}
for the easy configuration, which we may also evaluate in the appropriate momentum twistor parameterisation we have derived,
\be
\mathbf{Z}_{5,2e} =   \left(
    \begin{array}{ccccccccc}
    1 & 0 & 1 & 1 & 1 & 1 & 0 & 0 & 0 \\
    0 & 1 & 0 & 1 & 1 & 1 & 1 & 0 & 0 \\
    0 & 0 & 1 & 0 & z_1 & z_3 & z_4 & 1 & 0 \\
    0 & 0 & 0 & 1 & z_2 & z_5 & z_6 & 0 & 1 \\
    \end{array}\right)\,.
\ee
Similarly, for the hard configuration we have
\begin{align}
\label{eq:sToBrH}
   & s_{12}  =\frac{\br{1234}}{\br{1289}\br{3489}} \quad && s_{23}  =\frac{\br{2345}}{\br{2389}\br{4589}} \quad
    &&& s_{34}  =\frac{\br{3467}}{\br{3489}\br{6789}} \nonumber \\
    & s_{45}  =\frac{\br{1245}}{\br{1289}\br{4589}}  \quad && s_{51}  =\frac{\br{2367}}{\br{2389}\br{6789}} \quad &&& p_{4}^2  =\frac{\br{4567}}{\br{4589}\br{6789}} \\
    \quad &&& p_{5}^2 = \frac{\br{1267}}{\br{1289}\br{6789}}\,,
    \nonumber
\end{align}
 which may be evaluated with the help of
\begin{equation}
\label{eq:Z5p2m}
 \mathbf{Z}_{5,2h} =  \left(
    \begin{array}{ccccccccc}
    1 & 0 & 1 & 1 & 1 & 1 & 0 & 0 & 0 \\
    0 & 1 & 0 & 1 & 0 & 0 & 1 & 0 & 0 \\
    0 & 0 & 1 & 0 & y_1 & y_2 & y_3 & 1 & 0 \\
    0 & 0 & 0 & 1 & y_4 & y_5 & y_6 & 0 & 1 \\
    \end{array}\right). \quad 
\end{equation}
These parameterisations will also be useful when comparing with the existing literature, a task which we turn to next.

\subsection{\texorpdfstring{Completion under permutations and comparison with \cite{Abreu:2024yit}}{Comparison with Abreu 2024}}

All 2-loop 5-point 2-mass planar integrals have been calculated in~\cite{Abreu:2024yit}. Anticipating the extension to the nonplanar integrals, the set of letters thus obtained has also been enlarged by the their images under certain permutations, as we now explain. First, one needs to embed  the 5-point 2-mass kinematics into 7-point massless planar ones. In the orientation of eq.~\eqref{eq:5p2m_final_Mandelstams}, this amounts to replacing the massive legs 4 and 5 by the pairs $\{4,5\}, \{6,7\}$ of massless legs, e.g. $p_4^2\to \tilde s_{45}$, $s_{51}\to \tilde s_{671}$. Then,  all permutations of the seven massless legs are considered, and the resulting Mandelstam variables $\tilde s_{i\ldots j}$ are reexpressed in terms of those with adjacent legs with the help of identities. By virtue of eq.~\eqref{eq:count_vars_naive} there will be 14 distinct Mandelstam variables for these kinematics with $n=7$ and $m=0$. We keep the permutations where only a subset of the 7 Mandelstam variables corresponding to 5-point 2-mass kinematics appear, i.e. those where massless legs 4 and 5 (respectively 6 and 7) always appear together. This is not equivalent to directly considering the permutations of only the two massive and the three massless legs. Note that the set of valid permutations will depend on the form of the letter, i.e. on which Mandelstam variables appear on it. Evidently, these permutations mix easy and hard kinematics.

\begin{table}
    \centering
    \begin{tabular}{|c|c|c|c|c|c|c|c|c|}
    \hline
         Letter type: & Rational &  $\Delta_{5}$& $\Delta_{3}$ &
         $r_{1}$ & $r_{2}$ & $r_{3}$ & Double-root & Overall root\\ \hline
        Literature & 18 &  9 &  5 & 5 & 4 & 4 & 12 & 5\\ \hline
        Matched &  17 &  7 & 5 & 4 & 0 & 4  & 5 & 0\\ \hline
%       Mismatches & 1 & 2 &  0  & 1  & 4 & 0 & 7 & 5\\
%       \hline
       New & 5 & 6 &  0  & 0  & 0 & 0 & 0 &  0\\
       \hline
    \end{tabular}
    \caption{Counts of permutation orbits of 5-point 2-mass letters, separately for each of the roots \eqref{eq:5p2m_roots} appearing, if any. The first line refers to existing results, the second line to their subset matched by our calculation, and the third line to the new letters of the latter.}
    \label{tab:(5,2)comparison}
\end{table}

This procedure yields an alphabet of 570 letters, denoted as $W_i$. Since the orbits, or classes of letters  that are closed under the aforementioned set of permutations, is not given in the original reference, we have rederived them in the ancillary file \texttt{orbits5p2mACSZ.m}. We will denote them as $\mathcal{S}_j$, $j=1,\ldots,62$, where $j$ refers to the position they are listed in the file in question. Out of these orbits, 18 contain rational letters,  9 and 30 are of the general form~\eqref{eq:Gr49NRLetters} with rationalisable and non-rationalisable square roots, respectively, and finally another 5 orbits contain overall root letters  with their radicands. The radicands of all square roots appearing are specifically generated by
\begin{align}
    \begin{split}\label{eq:5p2m_roots}
        \Delta_{5} &= 16G(1,2,3,4), \\
        \Delta_{3}^{(1)} &= \lambda(p_{4}^2,p_{5}^2,s_{45}), 
        \\
        r_{1}^{(1)} & = p_4^4 s_{23}^2
        - 2 p_4^2 s_{23}(2p_5^2 - s_{15} + s_{23}) s_{45}
        + (s_{15} - s_{23})^2 s_{45}^2 , 
         \\
        r_{2}^{(1)} & =p_4^4 s_{12}^2
        + 2 p_4^2 s_{12}\big(p_5^2 s_{23} + (s_{15} - s_{34}) s_{45}\big)
        + \big(p_5^2 s_{23} + (s_{34} - s_{15}) s_{45}\big)^2, \\
       r_{3}^{(1)} & = 4 p_4^4 s_{12}(p_5^2 - s_{15}) s_{15}
        + \big(p_5^2(s_{23} + s_{34}) - s_{15}(s_{34} + s_{45})\big)^2,
    \end{split}
\end{align}
where $G(1,2,3,4)$ is a 4-point Gram determinant~\eqref{eq:grammdef} and $\lambda$ is the K\"allen function~\eqref{eq:kallen}, i.e $\Delta_{3}$ is related to massive triangle integrals. The roots $r_{1},r_{2},r_{3}$ originate from box-triangle topologies with different arrangements of the massive legs. The only square root that rationalizes in the momentum twistor parameterisation is $\Delta_5$. Note that these also appear as products in letters of the form
\be
\frac{a+b\sqrt{\Delta_{1}\Delta_{2}}}{a-b\sqrt{\Delta_{1}\Delta_{2}}}\,,
\ee
 with $(\Delta_{1},\Delta_{2})=\{(\Delta_{5} , r_{i}), (\Delta_{5}, \Delta_3),(\Delta_{3} ,r_{i})\}$.

Next, we check which of our 5-point 2-mass hard and easy letters are (not) contained in this previously computed alphabet by evaluating everything in the momentum twistor parameterisations provided in the previous subsection (both easy and hard parameterisations are equivalent at the level of the permutation-closed alphabet). We find a highly non-trivial overlap which we summarise in table \ref{tab:(5,2)comparison}, and also describe below.

%
%\begin{equation}
%W_i\;\;\text{ with } i\in \mathcal{S}_j\,,\, j=1,\ldots 62.    \end{equation}
%

Starting with the rational orbits, we deduce that our results contain all but the following rational orbit the permutation-closed alphabet of~\cite{Abreu:2024yit},
\begin{equation}
\mathcal{S}_{17}=\{154,\ldots,159\}\,,
\end{equation}
where here and below we will only denote the letter number, $W_i\to i$ for simplicity. We also reproduce most of the orbits containing the rationalisable $\Delta_5$,  except for
\be
\mathcal{S}_{43}=\{412\}\,, \quad
\mathcal{S}_{45}=\{434,\ldots,439\}.
\ee

Continuing with the orbits containing a single non-rationalisable root, we observe complete agreement for those with radicands $\Delta_3$ or $r_3$. For the ones with $r_1$, however, it appears that
\begin{equation}\label{eq:missed_orbit_r1}
   \mathcal{S}_{25}=\{246,253,259,267,277,281\}\;,
\end{equation}
involving in particular the permutations\footnote{The index refers to position in the respective orbit of overall root letters, to appear later on in the section.} $r_1^{(i)}$ with $i=1,3,5,7,10,11$, is not contained in our results.  This is to say that orbits might contain only a subset of all possible permutations of a square root, because the remaining permutations are not compatible with the rational terms of the letters. Perhaps more importantly, we find that $r_2$ is not contained in the square roots of our result. This implies that all orbits involving it, 
\begin{align}
\begin{split}
& \mathcal{S}_{29} = \{311, 317, 321, 325, 329, 333\}, \\
& \mathcal{S}_{30}= \{312, 315, 320, 323\},
   \\
& \mathcal{S}_{31} = \{313, 316, 319, 324, 327, 328, 331, 332\}\\
& \mathcal{S}_{32} = \{314, 318, 322, 326,330, 334\}\,,
  \end{split}
\end{align}
will not be part of our results either.

Next, we consider the orbits of letters involving products of two square roots. These are still of the general form~\eqref{eq:Gr49NRLetters} in terms of momentum twistor variables that we encounter  in our results, provided one of the two radicands is the rationalisable $\Delta_5$. So it turns out that we do reproduce 2+2+1 orbits where this comes as a product with $\Delta_3, r_1$ and $r_3$, respectively. But the same observations we made on letters with a single square root also imply the absence of the products with $r_2$,
\be
\mathcal{S}_{56} = \{553,\ldots,558\},
\ee
as well as with the permutations of $r_1$ that occur in the orbit~\eqref{eq:missed_orbit_r1},
\be
\mathcal{S}_{53} = \{523, 526, 529, 532, 537, 538\},
\ee
from our alphabet. Finally, by construction the latter cannot match products of two non-rationalisable roots, 
\begin{align*}
&\mathcal{S}_{46} = \{452,\ldots, 457, 478, 486, 508\},\\
&\mathcal{S}_{51} = \{483, 484, 491, 492, 498, 499, 505, 506, 513, 514, 520, 521\},\\   
&\mathcal{S}_{47} = \{458, 460, 462, 463, 465, 466, 481, 482, 489, 490, 495, 496, 502, 503, 509, 511, 518, 519\},\\
&\mathcal{S}_{48} = \{459, 461, 464, 467, 468, 469, 479, 480, 487, 488, 494, 497, 501, 504, 510, 512, 516, 517\},\\
&\mathcal{S}_{49} = \{470,\ldots,475\},\\
\end{align*}
as well as overall square roots $\sqrt{\Delta}, \Delta=\{\Delta_5,\Delta_3,r_1,r_2,r_3\}$, in this particular order,
\begin{align*}
 	& S_{62} = \{215\}\,,    \\
    & S_{58} =  \{172,\ldots,178\}\,, \\
    & S_{59} = \{179,\ldots,196\}\,, \\
    & S_{60} = \{197,\ldots,202\}\,,\\
    & S_{61}=\{206,\ldots, 214\}\,.
\end{align*}
Note that the orbits of the latter equation could be recovered from the previous ones, in the sense explained in subsection~\ref{subsec:pTr49AB}.

Last but not least, here too our calculation produces letters that have not appeared before. We transcribe these from momentum twistor to Mandelstam variables with the help of the method outlined in subsection~\ref{subsec:ZtoMand}, and generate their orbits as outlined at the beginning of this subsection. In this manner, we obtain 4 rational orbits generated by
\begin{align}
\gamma_{1} &= p_4^2 p_5^2 (s_{23}-s_{45})
           - (p_5^2-s_{45})(p_5^2 s_{23}-s_{45}s_{51}) \nonumber \\
\gamma_{2} &= p_5^2 s_{23}s_{34}
            + s_{51}\Big(s_{34}(s_{23}-s_{45}-s_{51})
            + p_4^2(s_{12}+s_{51})\Big) \nonumber \\
\gamma_{3} &= s_{34}\Big(
              p_5^2 s_{12}s_{23}
              + p_5^2 (p_5^2-s_{34})(p_5^2-s_{34}-s_{45})
              - (p_5^2-s_{34})(p_5^2+s_{12}-s_{34}-s_{45}) s_{51}
            \Big) \nonumber \\
          &\quad
            - p_4^2 (p_5^2-s_{34})\Big(
              p_5^4 + s_{34}s_{51}
              - p_5^2(s_{12}+s_{34}+s_{51})
            \Big) \\
\gamma_{4} &=  p_4^2 (p_5^2-s_{45})\Big(
              p_5^2 s_{23}(p_5^2-s_{12}+s_{34})
              - p_5^2 s_{34}s_{45}
              + (s_{12}-p_5^2)s_{45}s_{51}
            \Big) \nonumber \\
          &\quad
            - p_4^4 p_5^2 \Big(p_5^2(s_{23}-s_{45}) + s_{12}s_{45}\Big)
            - s_{34}(p_5^2-s_{45})^2 (p_5^2 s_{23}-s_{45}s_{51})\,,\nonumber
\end{align}
and in total contain 54 letters. In addition, we obtain another 4 orbits that jointly contain 33 rationalisable letters of the form
\be
\frac{A+B\sqrt{\Delta_5}}{A-B\sqrt{\Delta_5}}\,,
\ee
where $A$ is a polynomial of degree 3 or 4 in the Mandelstam variables.\footnote{We notice that in this case closure under special permutations introduces relations between letters of different orbits, and we have made a choice on how to eliminate them. This phenomenon could be avoided at the expense of increasing the polynomial degree of $A$, and hence the complexity of the letters.} Their precise form can be found in the ancillary file \texttt{AB5p2mNewOrbits.m}, which in fact contains both the rational and rationalised orbits as a common list, in this order.

\section{Conclusions and Outlook}

In this work, we obtained for the first time cluster-algebraic predictions for the alphabet of master integrals relevant for QCD phenomenology, that make contact with and go beyond the current state of the art. We started with the alphabet for 9-particle massless scattering in $\cN=4$ SYM theory, related to the $Gr(4,9)$ cluster algebra~\cite{Henke:2021ity}, and first reduced it to subalphabets with fewer, massive legs as summarised in table~\ref{tab:results}. By breaking their dual conformal invariance and, if needed, considering their massless limits, we have specifically obtained candidate letters for planar 6-point 1-mass and massless, as well as 5-point 2-mass QCD processes.

Our 6-point 1-mass results provide input for the future calculation of the relevant 2-loop integrals. Of the 29 letters we find with genuinely these kinematics, those with nested roots~\eqref{eq:nested_letters2} also have broader significance: Their discovery in quantities that are believed to be purely polylogarithmic is a new feature that requires incorporation in the existing methods of determining the alphabet directly from the integrals. Taking the limit where the external massive leg becomes massless also produces a 6-point massless candidate alphabet, that enables a very strong check against existing results in the literature~\cite{Abreu:2024fei,Henn:2025xrc}. As illustrated in table~\ref{tab:(6,0)comparison}, our prediction essentially contains the letters expected to contribute to the finite part of the 2-loop amplitude, and also yields new letters that might appear at higher loops! In a similar vein, we also obtain a 5-point 2-mass candidate alphabet and compare with that of the corresponding 2-loop planar master integrals~\cite{Abreu:2024yit}, as shown in table~\ref{tab:(5,2)comparison}. On the one hand, the overlap here is not as complete as for the 6-point massless case. On the other hand, it is reasonable to expect that some of the master integral letters will drop out of the finite amplitude alphabet, and that  our results may provide cluster-algebraic guidance for identifying these, as was previously the case for 5-point integrals with fewer masses~\cite{Chicherin:2020umh}. Beyond these, here too we provide new letters that may aid future higher-loop calculations.

Our work also opens exciting avenues for future work. The procedure we have provided applies completely generally to alphabets related to the $Gr(4,n)$ cluster algebra, and transforms them to corresponding predictions for $(n-k-1)$-point QCD processes with up to $k-1$ massive legs. The next natural step would thus be the construction of the  $Gr(4,10)$ alphabet along the lines of~\cite{Arkani-Hamed:2019rds,Drummond:2019cxm,Henke:2019hve}, and the generation of 6-point 2-mass and 5-point 3-mass predictions for it.  Integration of our work with the alternative cluster-algebraic approach of~\cite{Bossinger:2024apm,Pokraka:2025ali,Bossinger:2025rhf} also has the potential of combining the strengths of both, namely a lighter determination of massless alphabets with a criterion for which subset of the infinite cluster variables to keep. The 6-point 1-mass alphabet and all future predictions produced along these lines serve as input for QCD amplitude bootstrap methods, in the same manner this has already been initiated for the 6-point massless case~\cite{Carrolo:2025agz,Carrolo:2026qpu}. To this end, on top of the determination of the letters also the study and discovery of cluster adjacency relations will be important~\cite{Drummond:2017ssj}.

Finally, any departures of the cluster-algebraic predictions from the existing literature offer fertile ground for investigations on both sides. As noted already, an alternative scattering diagrams approach based~\cite{Herderschee:2021dez} for obtaining finite alphabets from infinite cluster algebras has been observed to also produce overall root letters for $Gr(4,8)$, namely letters we have coined here as trivially recoverable. It would be worthwhile to explore if the same phenomenon persists at higher $n$, thereby providing sounder theoretical justification for the inclusion of these letters. On the side of direct evaluation of integrals, it would be very interesting to systematise the additional simplifications that occur at the level of the finite amplitude, and explore the possibility that they stem from positivity properties of the latter~\cite{Arkani-Hamed:2012zlh,Arkani-Hamed:2013jha}.

\section*{Acknowledgements}
We thank Dima Chicherin, Johannes Henn and Simone Zoia for collaboration at the initial stages of this work, as well as Samuel Abreu, Christoph Dlapa,  Antonela Matijašić, Julian Miczajka and  Yang Zhang for illuminating discussions and correspondence. We are grateful to the organisers of the 2024 MIAPbP programme \emph{Special Functions: From Geometry to Fundamental Interactions}, of the \emph{Mathematical Structures in Scattering Amplitudes} 2024 workshop at GGI Florence, and of the 2026 ESI Vienna programme \emph{Amplitudes and Algebraic Geometry} for their hospitality and support during the completion of this work. We acknowledge support from the Deutsche Forschungsgemeinschaft (DFG) under Germany’s
Excellence Strategy – EXC 2121 “Quantum Universe” – 390833306. GP’s work was supported in part by the UK Science and Technology Facilities Council grant ST/Z001021/1
and by the Munich Institute for Astro-, Particle and BioPhysics (MIAPbP) which is funded
by the DFG under Germany’s Excellence Strategy – EXC-2094 – 390783311.

\appendix
\section{Description of ancillary files}\label{appx}
The ancillary files accompanying this article~\cite{code} are divided into four directories: \texttt{Alphabet Reduction/} contains the initial data and \texttt{Mathematica} notebook implementing the reduction of the 9-point massless DCI alphabet to $(9-k)$-point $k$-mass subalphabets, as described in section~\ref{sec:9ptred} and summarised in table~\ref{tab:results}. \texttt{6pt1M/} includes our results particularly for the 6-point 1-mass LI or equivalently 7-point 2-mass DCI alphabet, that are presented in section~\ref{sec:(6,1)letters}. Then \texttt{6pt0M/} contains our prediction for the 6-point massless alphabet, obtained from the aforementioned 1-mass alphabet as discussed in section~\ref{sec:6p0m}. Finally, \texttt{5pt2M/} includes 5-point 2-mass LI or equivalently 6-point 3-mass DCI alphabets, before and after the closure under special permutations as discussed in section~\ref{sec:(5,2) letters)}.

\subsection*{Contents of \texttt{Alphabet Reduction/}}
\begin{itemize}
    \item \texttt{Alphabet_reduction.nb}: Mathematica notebook performing the alphabet reductions illustrated in table~\ref{tab:results}. Its has the dependencies listed below.
    \item \texttt{Gr49RationalDCI.m}: Rational part, in momentum twistors, of the cluster-algebraic 9-particle massless DCI alphabet of~\cite{Henke:2021ity}.
    \item \texttt{algebraicLettersRepresentatives.m}: Seeds of simplified square-root letters of the alphabet~\cite{Henke:2021ity}, grouped in subsets involving the same square root. Each subset contains one representative letter of each cyclic class, so the full square-root alphabet is generated by cyclically permuting their momentum twistor labels nine times.
    \item \texttt{nullSpaceData.m}: Seeds for computing subspaces of the aforementioned alphabet that are annihilated by any subset of the operators \eqref{eq:opdef}, i.e. nullspaces, in \texttt{Mathematica SparseArray} format.
\end{itemize}

\subsection*{Contents of \texttt{6pt1M/}}
\begin{itemize}
    \item \texttt{AB6p1m1LDD.m}: Alphabet of the 1-loop 6-point 1-mass integral with $D$-dimensional external momenta, derived in subsection~\ref{subsec:1loop6p1m} compared against our prediction below.
    \item \texttt{AB6p1mZ.m}: 6-point 1-mass LI or equivalently 7-point 2-mass DCI alphabet in terms of momentum twistors as obtained by our reduction procedure, along with the corresponding momentum twistor parameterisation. 
    \item \texttt{AB6p1m.m}: Explicit Mandelstam variable expressions for the subspace of rational letters of the said 6-point 1-mass LI alphabet, along with rules for converting them back to momentum twistors. 
    \item \texttt{AB6p1mGenuine.m}: The part of the aforementioned alphabet space that is genuinely 6-point 1-mass, as defined in subsection~\ref{subsec:6p1m_purify}.
    \item \texttt{nestLetter6p1m.m}: Explicit expression in terms of Mandelstam variables for one of the 6-point 1-mass nested root letters.
\end{itemize}

\subsection*{Contents of \texttt{6pt0M/}}
\begin{itemize}
    \item \texttt{AB6p0mZ.m}: 6-point massless LI alphabet in terms of momentum twistors, obtained as the massless limit of the aforementioned 6-point 1-mass alphabet, along with the corresponding momentum twistor parameterisation.
    \item \texttt{AB6p0m.m}: Explicit Mandelstam variable expressions for the cyclic closure of the alphabet of the previous bullet, along with rules for converting them back to momentum twistors. The positions of new letters are also provided. 
\end{itemize}

\subsection*{Contents of \texttt{5pt2M}}
\begin{itemize}
    \item \texttt{AB6p0mZ.m}: 5-point 2-mass easy and hard alphabets arising from our reduction procedure expressed in terms of momentum twistors, along with their corresponding parameterisations.
    \item \texttt{orbits5p2mACSZ.m}: The known 2-loop 5-point 2-mass planar alphabet of~\cite{Abreu:2024yit}, arranged into orbits of the special permutations under which it is closed.
    \item \texttt{AB5p2mNewOrbits.m}: New letters of the special permutation closure of our 5-point 2-mass easy and hard alphabets, i.e. those that do not belong to the list of the previous bullet, in terms of Mandelstam variables. Maps back to momentum twistors are also provided.
\end{itemize}

\bibliographystyle{JHEP}
\bibliography{bibliography}
\end{document}